\documentclass[prd,twocolumn,showpacs,amsmath,amssymb]{revtex4}

\usepackage{graphicx}
\usepackage{dcolumn}
\usepackage{bm}
\usepackage{psfrag}
\usepackage{subfigure}

\newcommand{\be}{\begin{eqnarray}}
\newcommand{\ee}{\end{eqnarray}}
\newcommand{\bn}{\begin{eqnarray*}}
\newcommand{\en}{\end{eqnarray*}}

\newcommand{\nn}{\nonumber \\}
\newcommand{\nl}{\\}

\renewcommand{\vec}[1]{\mbox{\boldmath$#1$}}
\renewcommand{\d}{\mbox{\rm d}}
\newcommand{\gslash}[1]{\mbox{\slash{\hspace{-2mm}}$#1$}}
\newcommand{\kbar}{\mathchar'26\mkern-9mu k}

\renewcommand{\th}{\ensuremath{\theta}}

\newcommand{\Th}{\ensuremath{\Theta}}
\newcommand{\ph}{\ensuremath{\phi}}

\newcommand{\al}{\ensuremath{\alpha}}
\newcommand{\bt}{\ensuremath{\beta}}
\newcommand{\sg}{\ensuremath{\sigma}}
\newcommand{\gm}{\ensuremath{\gamma}}
\newcommand{\dl}{\ensuremath{\delta}}
\newcommand{\lm}{\ensuremath{\lambda}}

\newcommand{\gfive}{\ensuremath{\gm^5}}
\newcommand{\Dl}{\ensuremath{\Delta}}

\newcommand{\Gm}{\ensuremath{\Gamma}}
\newcommand{\Om}{\ensuremath{\Omega}}
\newcommand{\OmK}{\ensuremath{\Omega_{\rm K}}}

\newcommand{\ze}{\ensuremath{\hat{0}}}

\newcommand{\pvec}{\ensuremath{\vec{p}}}
\newcommand{\Pvec}{\ensuremath{\vec{P}}}
\newcommand{\Qvec}{\ensuremath{\vec{Q}}}
\newcommand{\nvec}{\ensuremath{\vec{n}}}

\newcommand{\Rvec}{\ensuremath{\vec{R}}}

\newcommand{\sgvec}{\ensuremath{\vec{\sg}}}

\newcommand{\Avec}{\ensuremath{\vec{A}}}
\newcommand{\Dvec}{\ensuremath{\vec{D}}}
\newcommand{\Jvec}{\ensuremath{\vec{J}}}

\newcommand{\Xvec}{\ensuremath{\vec{X}}}
\newcommand{\Wvec}{\ensuremath{\vec{W}}}
\newcommand{\Gmvec}{\ensuremath{\vec{\Gm}}}
\newcommand{\Omvec}{\ensuremath{\vec{\Omega}}}

\newcommand{\nabvec}{\ensuremath{\vec{\nabla}}}

\newcommand{\hb}{\ensuremath{\hbar}}

\newcommand{\lt}{\ensuremath{\left}}
\newcommand{\rt}{\ensuremath{\right}}

\renewcommand{\d}{\mbox{\rm d}}

\begin{document}

\pagenumbering{arabic}

\title{Breakdown of Lorentz Invariance for Spin-1/2 Particle Motion in \\
Curved Space-Time with Applications to Muon Decay}%

\author{Dinesh Singh}
\email{dinesh.singh@uregina.ca}
\author{Nader Mobed}
\email{nader.mobed@uregina.ca}
\affiliation{%
Department of Physics, University of Regina \\
Regina, Saskatchewan, S4S 0A2, Canada
}%
\date{\today}

\begin{abstract}
This paper explores the properties of the Pauli-Lubanski spin vector for the general motion of \mbox{spin-1/2} particles
in curved space-time.
Building upon previously determined results in flat space-time, it is shown that the associated Casimir scalar for spin
possesses both gravitational contributions and frame-dependent contributions due to non-inertial motion, where the latter
represents a possible quantum violation of Lorentz invariance that becomes significant at the Compton wavelength scale.
When applied to muon decay near the event horizon of a microscopic Kerr black hole, it is shown that its
differential cross section is strongly affected by curvature, with particular sensitivity to changes in the black hole's
spin angular momentum.
In the absence of curvature, the non-inertial contributions to the decay spectrum are also identified
and explored in detail, where its potential for observation is highest for large electron opening angles.
It is further shown how possible contributions to noncommutative geometry can emerge from within this
formalism at some undetermined length scale.
Surprisingly, while the potential exists to identify noncommutative effects in muon decay,
the relevant terms make no contribution to the decay spectrum, for reasons which remain unknown.
\end{abstract}


\maketitle

\section{Introduction}
\label{intro}

The search for a viable theory of quantum gravity has been ongoing for the
past several decades, with many distinct approaches of late.
Arguably, the two leading contenders are string theory \cite{Polchinski,Green} and loop quantum gravity \cite{Ashtekar,Rovelli},
each with their respective strengths and weaknesses.
As well, other approaches to this ultimate goal, such as twistor theory \cite{Penrose} and causal set theory \cite{Sorkin}
among them, create their own unique set of possibilities for how a quantum gravity theory might emerge.
Each of these approaches and others are based on largely inequivalent mathematical and philosophical foundations,
which make for great difficulty in determining their comparative value on purely theoretical grounds.
More importantly, however, with the lack of any concrete observational evidence for guidance, it is unclear which
approach or combination thereof will arise as the most successful one(s) to follow.
While the prediction of Hawking radiation from black holes \cite{Hawking} is widely cited as a standard candle
for competing theories to recover \cite{Carlip} when resolved to length scales consistent with quantum field theory
in curved space-time, this claim may be put into jeopardy if certain technical challenges with
Hawking radiation, such as the presence of trans-Planckian energy modes near the event horizon \cite{Helfer}, remain unresolved.
It is certainly reasonable to assume that these difficulties arise from still unverified assumptions
about the precise nature of space-time ranging from the Compton wavelength scale down to the Planck scale.

A recent paper \cite{Singh-Mobed} raises questions about the applicability of the Poincar\'{e}
group to adequately describe the properties of quantum mechanical objects while in non-inertial motion.
Specifically, it examines the Pauli-Lubanski four-vector in flat space-time for spin-1/2 particles,
where the Poincar\'{e} group generators for the canonical momentum $\Pvec$ are expressed in terms of
curvilinear co-ordinates in a local Lorentz frame to accommodate the symmetries of a discernable classical spatial trajectory
with respect to a static laboratory frame, such as cylindrical co-ordinates for circular motion in a storage ring.
An interesting consequence shown here is that the associated Casimir scalar for particle spin, widely regarded as
a Lorentz invariant, now exhibits a {\em frame dependence}.
This is due to an additive term coupling the Pauli spin operator
$\sgvec$ with a Hermitian three-vector $\Rvec^{\hat{\imath}} = \lt(i \over 2\hbar \rt)
\epsilon^{\hat{\imath}\hat{\jmath}\hat{k}} [\Pvec_{\hat{\jmath}}, \Pvec_{\hat{k}}]$, referred to by
its proponents as the {\em non-inertial dipole operator} \cite{Singh-Mobed,Singh1,Singh2}, since its
coupling to $\sgvec$ is analogous to a magnetic dipole interaction.
The dimensionality of $\Rvec$ is $|\Pvec|/r$, where $r$ is the particle trajectory's local radius of curvature
with respect to the laboratory frame.
For Cartesian co-ordinates to represent rectilinear motion, $\Rvec = \vec{0}$ identically,
while $\Rvec \rightarrow \vec{0}$ for fixed momentum as $r \rightarrow \infty$.

When examining the potential for observing $\Rvec$ within the context of muon decay for free-particle beams
in a circular storage ring, a preliminary computation shows that it is negligibly small for observation by current facilities
like the Brookhaven National Laboratory.
However, in the absence of other interactions it predicts that $\Rvec$ in cylindrical co-ordinates
contributes to the prolongation of the muon's lifetime when $r \rightarrow 1.18 \times 10^{-12}$ cm, the muon's Compton wavelength \cite{Singh-Mobed}.
This is a remarkable outcome if shown to persist within a more realistic treatment of the problem,
since this acceleration-induced effect occurs at a length scale that is over {\em twenty orders of magnitude} larger
than the Planck scale, and conventional wisdom purports that the intersection between quantum mechanics and gravitation
becomes relevant {\em only} near the Planck scale.
Furthermore, there is evidence from other considerations \cite{Rosquist} that suggest potentially significant
consequences of quantum mechanical interactions in curved space-time at the Compton wavelength scale,
apparently unaccounted for in existing treatments of quantum gravity research.
This provides strong motivation for pursuing a more careful treatment of this muon decay problem \cite{Singh-Mobed},
but now in the presence of a strong gravitational field.

The purpose of this paper is to present the analysis of spin-1/2 particle dynamics in a gravitational field
described by Fermi normal co-ordinates, where the spatial components are represented by orthonormal curvilinear co-ordinates,
and explore the consequences when applied to muon decay.
It begins in Sec.~\ref{sec:locality} with a brief description of the hypothesis of locality and its limitations for the Poincar\'{e}
symmetry group, which provides the foundation to motivate this paper.
This is followed by an outline of the covariant Dirac equation in Sec.~\ref{sec:Dirac-equation},
where the formalism of Fermi normal co-ordinates is presented therein.
A presentation of the Pauli-Lubanski four vector $\Wvec$ is next given in Sec.~\ref{sec:Pauli-Lubanski},
where the momentum generators are expressed in terms of both the canonical momentum in curvilinear co-ordinates and
gravitational corrections in the Fermi frame.
This leads to the main expression for the Casimir scalar associated with spin, which shows the results obtained earlier \cite{Singh-Mobed},
along with gravitational corrections.
Following this result, its formal application to the muon decay problem is presented in Sec.~\ref{sec:Muon-Decay}, which shows the
leading order gravitational corrections to its differential decay cross section.
The explicit example of muon decay near the event horizon of a Kerr black hole is then explored in Sec.~\ref{sec:Kerr-BH-application},
which identifies predicted contributions to the decay rate from both gravitational and non-inertial effects
for the full range of possible opening angles.
An interesting and previously unanticipated generalization of the formalism to incorporate noncommutative geometry
is then presented in Sec.~\ref{sec:noncommutative-geometry}, where the possibility of identifying its contribution to
the muon decay rate is explored.
This leads to an overall discussion of the main results in Sec.~\ref{sec:discussion} with comments on other avenues of exploration
and relevant issues for consideration, followed by a conclusion in Sec.~\ref{sec:conclusion}.

\section{The Hypothesis of Locality and Its Limitations for the Poincar\'{e} Group}
\label{sec:locality}

One of the fundamental principles of physics often taken for granted is the {\em hypothesis of locality}
for describing a physical object's motion in space.
The essential idea \cite{Mashhoon} is the suggestion that an accelerated observer can always be identified
with a continuous set of instantaneously comoving inertial observers at each given moment of proper
time throughout the accelerated observer's worldline.
Mathematically, this corresponds to defining a family of velocity vectors tangent to the worldline that have a one-to-one
relationship with the comoving inertial frames, with intrinsic length and time scales for the accelerated observer.
For classical objects of arbitrary composition, the hypothesis of locality is undoubtedly in agreement with existing observations.
However, matters become more complicated when considering the motion of quantum mechanical objects.
This is because of the wave-like nature of quantum systems, where quantum interference considerations become relevant.
In particular, the notion of localization for a quantum object is subject to the Heisenberg uncertainty relations \cite{Singh-Mobed},
which makes for conceptual difficulties in terms of quantifying its transport in space-time.
It is especially unclear whether a manifold structure even makes any physical sense at a certain threshold of length scale
compared to the typical length scale describing quantum fluctuations of the quantum object.
With these issues in mind, it is certainly possible to consider a breakdown of the hypothesis of locality
for an accelerated quantum system with sufficiently large wavelengths, when a {\em region} of space-time
is then required to properly identify its location from one moment to the next.

When considering the boundary layer between classical and quantum effects as it concerns gravitation, it is useful
to consider the intrinsic spin properties of quantum particles propagating in curved space-time.
In flat space-time, the Poincar\'{e} symmetry group \cite{Ryder} is unquestionably invaluable for
determining the classification of subatomic particles, based on their mass and spin angular momentum.
However, the viability of using the Poincar\'{e} group to describe non-inertial motion, even in the absence of space-time curvature,
is worthy of considerable scrutiny.
This is because the Poincar\'{e} group can only be applied for particles in strictly inertial motion,
which is an impossibility, considering the idealized nature of truly inertial reference frames \cite{Singh-Mobed}
and given the understanding from general relativity that {\em any} source of mass-energy produces space-time
curvature, no matter how weak it may be.
In curved space-time, the best approximation to an inertial frame is the {\it freely falling frame}
for propagation along a geodesic, corresponding to force-free motion consistent with the weak principle of equivalence
at a mathematical point on the manifold.
Even this approach does not negate these challenges posed to the Poincar\'{e} group when applied to a quantum system
with some spatial extension, since its internal structure would then be sensitive to the tidal forces
generated by space-time curvature, according to measurements from observers on neighbouring geodesics.

There exists a limited but useful generalization of the Poincar\'{e} group that is applicable to
conformally flat space-times in four dimensions, known as the {\it de~Sitter group} \cite{Gursey}.
Given that the de~Sitter space-time metric can exist as an embedding in a flat five-dimensional background
with an overall radius of curvature that defines the length scale of the expanding space, the momentum operators
are then represented as elements for a ten-parameter set of operators that generate rotation.
Nonetheless, the de~Sitter group also shares the same set of limitations as the Poincar\'{e} group when
applied to non-inertial motion.

Upon reflection, the problem appears to come from presupposing local Cartesian co-ordinates for the description
of the co-ordinates and derivative operators that determine the group generators.
If curvilinear co-ordinates are used instead to better reflect the symmetries of the quantum object's motion in space,
would such a consideration be physically relevant?
This is the primary motivation for considering a re-assessment of the Poincar\'{e} and de~Sitter groups within
the context of non-inertial motion as initially put forward \cite{Singh-Mobed}, now with the generalization of
adding space-time curvature in terms of Fermi normal co-ordinates.
Choosing this approach to represent local curvature is particularly useful because it naturally introduces
a local $3+1$ structure with mutually orthogonal directions for the basis vectors, where a proper time co-ordinate
unambiguously appears in relation to the three spatial directions.

\section{Covariant Dirac Equation in Fermi Normal Co-ordinates}
\label{sec:Dirac-equation}

The first step is to work with the covariant Dirac equation in Fermi normal co-ordinates $X^\mu = \lt(T, X^i\rt)$,
where the spatial components $X^i$ are expressed in terms of local Cartesian co-ordinates corresponding to rectilinear motion
and $T$ is the proper time along the worldline.
For the purposes of this paper, the standard perturbative expansion of second-order in $X^i$ away from the worldline is sufficient
\cite{Manasse,Synge,Mashhoon1,Mashhoon2,Poisson}, though it is certainly possible to either use exact expressions for specific
space-time backgrounds \cite{Chicone} or go to third-order and higher in the perturbation to incorporate additional curvature
and inertial contributions \cite{Ni,Li,Huang}.
Geometric units of $G = c = 1$ are assumed throughout, where the Riemann and Ricci tensors satisfy the conventions of
Misner, Thorne, and Wheeler \cite{MTW} but with $-2$ metric signature.

To begin, consider a worldline ${\cal C}$ in a general space-time background parametrized by proper time $\tau$.
At some event $P_0$ on ${\cal C}$, the Fermi frame \cite{Manasse,Synge,Mashhoon1,Mashhoon2} is determined by
constructing a local orthonormal vierbein set $\lt\{ \lm^{\bar{\alpha}}{}_\mu \rt\} \,$ and its inverse set $\lt\{ \lm^\mu{}_{\bar{\alpha}} \rt\}$,
such that $\lm^{\bar{\mu}}{}_0 = \d x^{\bar{\mu}}/\d \tau$ and $\lm^{\bar{\mu}}{}_a$ define the local spatial axes.
If the unit spatial tangent vector from $P_0$ to a neighbouring event $P$ is described by $\xi^{\bar{\mu}} = \lt(\d x^{\bar{\mu}}/\d \sg\rt)_0$,
where $\sg$ is the proper length of a unique spacelike geodesic orthogonal to ${\cal C}$, then the Fermi normal co-ordinates at $P$ are described by
$T = \tau$ and $X^i = \sg \, \xi^{\bar{\mu}} \, \lm^i{}_{\bar{\mu}}$.
It follows that the space-time metric in Fermi normal co-ordinates is described by
\be
\d s^2 & = & {}^F{}g_{\mu \nu}(X) \, \d X^\mu \, \d X^\nu \, ,
\label{ds^2-symm}
\ee
where
\begin{subequations}
\label{F-g}
\be
{}^F{}g_{00}(X) & = & 1 - {}^F{}R_{l00m}(T) \, X^l \, X^m + \cdots \, ,
\label{F-g00}
\nl
{}^F{}g_{0j}(X) & = & -{2 \over 3} \, {}^F{}R_{l0jm}(T) \, X^l \, X^m + \cdots \, ,
\label{F-g0j}
\nl
{}^F{}g_{ij}(X) & = & \eta_{ij} - {1 \over 3} \, {}^F{}R_{lijm}(T) \, X^l \, X^m + \cdots \, ,
\label{F-gij}
\ee
\end{subequations}
$\eta_{\mu \nu}$ is the Minkowski metric and
\be
{}^F{}R_{\mu \al \bt \nu}(T) & = & R_{\bar{\mu} \bar{\al} \bar{\bt} \bar{\nu}} \,
\lm^{\bar{\mu}}{}_\mu \, \lm^{\bar{\al}}{}_\al \, \lm^{\bar{\bt}}{}_\bt \, \lm^{\bar{\nu}}{}_\nu
\label{Riemann-fermi}
\ee
is the projection of the Riemann tensor onto the Fermi frame \cite{Mashhoon2}.

The covariant Dirac equation for a spin-1/2 particle with mass $m$ and $\partial_\mu = \partial/\partial X^\mu$ is
\be
\lt[i \gm^\mu(X) \lt(\partial_\mu + i \, \Gamma_\mu(X) \rt) - m/\hb\rt]\psi(X) & = & 0 \, ,
\label{Dirac-eq}
\ee
where the set of gamma matrices $\lt\{ \gm^\mu(X) \rt\}$ satisfy
$\lt\{ \gm^\mu(X), \gm^\nu(X) \rt\} = 2 \, g_F^{\mu \nu}(X)$ and $\Gamma_\mu(X)$ is the spin connection.
An orthonormal vierbein set $\lt\{ \bar{e}^\mu{}_{\hat{\alpha}}(X) \rt\}$ and its inverse
$\lt\{ \bar{e}^{\hat{\alpha}}{}_\mu (X)\rt\} \,$ can be obtained to define a local Lorentz frame \cite{Poisson},
denoted by co-ordinates with hatted indices, which satisfy
${}^F{}g_{\mu \nu}(X) = \eta_{\hat{\alpha}\hat{\beta}} \, \bar{e}^{\hat{\alpha}}{}_\mu (X) \, \bar{e}^{\hat{\beta}}{}_\nu (X)$,
where
\begin{subequations}
\be
\bar{e}^{\hat{0}}{}_0(X) & = & 1 + {1 \over 2} \, {}^F{}R^0{}_{i0j}(T) \, X^i \, X^j \, ,
\label{inv-tetrad-00}
\nl
\bar{e}^{\hat{0}}{}_j(X) & = & {1 \over 6} \, {}^F{}R^0{}_{ijk}(T) \, X^i \, X^k \, ,
\label{inv-tetrad-0j}
\nl
\bar{e}^{\hat{\imath}}{}_0(X) & = & -{1 \over 2} \, {}^F{}R^i{}_{j0k}(T) \, X^j \, X^k \, ,
\label{inv-tetrad-i0}
\nl
\bar{e}^{\hat{\imath}}{}_j(X) & = & \dl^i{}_j - {1 \over 6} \, {}^F{}R^i{}_{kjl}(T) \, X^k \, X^l \, ,
\label{inv-tetrad-ij}
\ee
\end{subequations}
and
\begin{subequations}
\be
\bar{e}^0{}_{\hat{0}}(X) & = & 1 - {1 \over 2} \, {}^F{}R^0{}_{i0j}(T) \, X^i \, X^j \, ,
\label{tetrad-00}
\nl
\bar{e}^i{}_{\hat{0}}(X) & = & {1 \over 2} \, {}^F{}R^i{}_{j0k}(T) \, X^j \, X^k \, ,
\label{tetrad-i0}
\nl
\bar{e}^0{}_{\hat{\jmath}}(X) & = & -{1 \over 6} \, {}^F{}R^0{}_{ijk}(T) \, X^i \, X^k \, ,
\label{tetrad-0j}
\nl
\bar{e}^i{}_{\hat{\jmath}}(X) & = & \dl^i{}_j + {1 \over 6} \, {}^F{}R^i{}_{kjl}(T) \, X^k \, X^l \, .
\label{tetrad-ij}
\ee
\end{subequations}
The spin connection is expressed as
\be
\Gm_\mu(X) & = & -{1 \over 4} \, \sg^{\hat{\al} \hat{\bt}} \, \eta_{\hat{\bt} \hat{\gm}} \, \bar{e}^\al{}_{\hat{\al}}
\lt(\nabla_\mu \, \bar{e}^{\hat{\gm}}{}_\al \rt),
\label{spin-connection}
\ee
where $\nabla_\mu$ is the covariant derivative operator, $\lt\{ \gm^{\hat{\al}} \rt\}$ is the set of flat space-time gamma matrices \cite{Itzykson}
satisfying $\lt\{ \gm^{\hat{\al}}, \gm^{\hat{\bt}} \rt\} = 2 \, \eta^{\hat{\al} \hat{\bt}}$
and $\sg^{\hat{\al} \hat{\bt}} = {i \over 2} \lt[\gm^{\hat{\al}}, \gm^{\hat{\bt}}\rt]$.
Retaining only contributions to first-order in the Riemann tensor throughout this paper,
it therefore follows from (\ref{spin-connection}) that
\be
\Gm_0(X) & = & i \, \gm^{\hat{0}} \, \gm^{\hat{\imath}} \lt[{1 \over 2} \, {}^F{}R_{i00k}(T) + {1 \over 3} \, {}^F{}R_{ij0k,0}(T) \, X^j \rt] X^k \, ,
\nn
\label{Gm-0}
\nl
\Gm_l(X) & = & i \, \gm^{\hat{0}} \, \gm^{\hat{\imath}} \lt[{1 \over 3} \, \lt({}^F{}R_{il0k}(T) + {}^F{}R_{i[k0]l}(T)\rt) \rt.
\nn
& &{} \lt. - {1 \over 12} \, {}^F{}R_{ijkl,0}(T) \, X^j \rt] X^k \, .
\label{Gm-l}
\ee

Conversion of (\ref{Dirac-eq}) to curvilinear co-ordinates is straightforward.
In what follows, it is assumed that $X^i = X^i(u^1, u^2, u^2)$, where $u^j$ represent
mutually orthogonal curvilinear co-ordinates.
This results in a new set of Fermi normal co-ordinates $U^\mu$, where $U^0 = T$ and $U^i = u^i$.
A corresponding set of orthonormal vierbeins is then obtained, where
$e^\bt{}_{\hat{\al}}(U) = {\partial U^\bt \over \partial X^\al} \, \bar{e}^\al{}_{\hat{\al}}(X)$,
$e^{\hat{\al}}{}_\bt(U) = {\partial X^\al \over \partial U^\bt} \, \bar{e}^{\hat{\al}}{}_\al(X)$
and $\Gm_\mu(U) = {\partial X^\al \over \partial U^\mu} \, \Gm_\al(X)$.
By projecting onto the local Lorentz frame, it follows that (\ref{Dirac-eq}) becomes
\be
\lt[i \gm^{\hat{\mu}} \lt(\hat{\nabla}_{\hat{\mu}} + i \, \Gm_{\hat{\mu}}(U) \rt) - m/\hb\rt]\psi(U) & = & 0 \, ,
\label{Dirac-eq-curvilinear}
\ee
where
$\hat{\nabla}_{\hat{\mu}} = \nabvec_{\hat{\mu}} + i \, \hat{\Gm}_{\hat{\mu}}(U)$ is the covariant derivative operator
defined with respect to Minkowski space-time in curvilinear co-ordinate form, with
$\nabvec_{\hat{0}} \equiv {\partial \over \partial T}$ and
$\nabvec_{\hat{\jmath}} \equiv {1 \over \lm^{\hat{\jmath}}(u)} \, {\partial \over \partial u^{\hat{\jmath}}}$,
where $\lm^{\hat{k}}(u)$ are dimensional scale functions \cite{Hassani} and $\hat{\Gm}_{\hat{\mu}}(U)$ is the spin connection
due to curvilinear co-ordinates.
It follows from (\ref{Dirac-eq-curvilinear}) that
\be
\lt[\gm^{\hat{\mu}} \lt(\Pvec_{\hat{\mu}} - \hb \, \Gmvec_{\hat{\mu}}(U) \rt) - m\rt]\psi(U) & = & 0 \, ,
\label{Dirac-eq-curvilinear-1}
\ee
where
\be
\Pvec_{\hat{\mu}} & = & \pvec_{\hat{\mu}} + \Omvec_{\hat{\mu}}
\label{Pvec}
\ee
is the momentum operator in curvilinear co-ordinates \cite{Singh-Mobed},
in terms of
\begin{subequations}
\be
\pvec_{\hat{\mu}} & = & i \hbar \, \nabvec_{\hat{\mu}} \, ,
\label{pvec}
\nl
\Omvec_{\hat{\mu}} & = & i \hbar \lt[\nabvec_{\hat{\mu}} \ln \lt(\lm^{\hat{1}}(u) \, \lm^{\hat{2}}(u) \, \lm^{\hat{3}}(u)\rt)^{1/2} \rt] \, ,
\label{Ovec}
\nl
i \, \Gmvec_{\hat{\mu}} & = & \bar{\Gmvec}^{(\rm S)}_{\hat{\mu}}
+ \gm^{\hat{l}} \, \gm^{\hat{m}} \, \bar{\Gmvec}^{(\rm T)}_{\ze[\hat{l}\hat{m}]} \, \dl^{\ze}{}_{\hat{\mu}} \, ,
\label{Spin-Connection}
\ee
\end{subequations}
where
\begin{subequations}
\be
\bar{\Gmvec}^{(\rm S)}_{\ze} & = & {1 \over 12} {}^F{}R^m{}_{jmk,0}(T) \, X^j \, X^k \, ,
\label{Spin-Connection-0-S}
\nl
\bar{\Gmvec}^{(\rm S)}_{\hat{\jmath}} & = & - {1 \over 2} {}^F{}R_{j00m}(T) \, X^m + {1 \over 3} {}^F{}R_{lj0m,0}(T) \, X^l \, X^m \, ,
\nn
\label{Spin-Connection-j-S}
\nl
\bar{\Gmvec}^{(\rm T)}_{\ze[\hat{l}\hat{m}]} & = & {1 \over 2} \, {}^F{}R_{lm0k}(T) \, X^k
+ {1 \over 12} \, {}^F{}R_{k[lm]j,0}(T) \, X^j \, X^k \, .
\nn
\label{Spin-Connection-0-T}
\ee
\end{subequations}
%
By defining $\Dvec_{\hat{\mu}} = \Pvec_{\hat{\mu}} - \hb \, \Gmvec_{\hat{\mu}}$ and making use of the identity \cite{Aitchison}
\be
\gm^{\hat{\mu}} \, \gm^{\hat{\nu}} \, \gm^{\hat{\rho}} & = & \eta^{\hat{\nu} \hat{\rho}} \, \gm^{\hat{\mu}}
- 2 \, \gm^{[\hat{\nu}} \eta^{\hat{\rho}]\hat{\mu}} - i \, \gm^5 \, \gm^{\hat{\sg}} \, \varepsilon^{\hat{\mu} \hat{\nu} \hat{\rho}}{}_{\hat{\sg}} \, ,
\label{gm-identity}
\ee
where $\varepsilon^{\hat{\mu} \hat{\nu} \hat{\rho} \hat{\sg}}$ is the Levi-Civita symbol with
$\varepsilon^{\hat{0} \hat{1} \hat{2} \hat{3}} = 1$, it is shown that the covariant momentum
operator with curvature exists in the very useful form
\be
\Dvec_{\hat{\mu}} & = & \Pvec_{\hat{\mu}} -
\hb \lt(\gm^5 \, \bar{\Gmvec}^{(\rm C)}_{\hat{\mu}} - i \, \bar{\Gmvec}^{(\rm S)}_{\hat{\mu}}\rt) \, ,
\label{D}
\nl
\bar{\Gmvec}^{(\rm C)}_{\hat{\mu}} & = & \varepsilon^{\hat{0} \hat{l} \hat{m}}{}_{\hat{\mu}} \, \bar{\Gmvec}^{(\rm T)}_{\ze[\hat{l}\hat{m}]} \, ,
\label{Spin-Connection-C}
\ee
where the ``C'' refers to the chiral-dependent part of the spin connection and the ``S'' denotes the symmetric part under chiral symmetry.

\section{Pauli-Lubanski Vector and Casimir Properties}
\label{sec:Pauli-Lubanski}

The Pauli-Lubanski vector defined with respect to the local Lorentz frame and in terms of $\Dvec_{\hat{\mu}}$ is
\be
\Wvec^{\hat{\mu}} & = & -{1 \over 4} \, \varepsilon^{\hat{\mu}}{}_{\hat{\al}\hat{\bt}\hat{\gm}} \,
\sg^{\hat{\al}\hat{\bt}} \, \Dvec^{\hat{\gm}} \, .
\label{Wvec-grav-defn}
\ee
By making use of (\ref{gm-identity}) and the identity
$\varepsilon_{\hat{\mu} \hat{\nu} \hat{\rho} \hat{\sg}} \, \sg^{\hat{\rho} \hat{\sg}}
= -2 \, i \, \gm^5 \, \sg_{\hat{\mu} \hat{\nu}}$ \cite{Itzykson}, it is straightforward to show that
the squared magnitude of the Pauli-Lubanski vector is
\be
\Wvec^{\hat{\al}} \, \Wvec_{\hat{\al}} & = &
-{3 \over 4} \, \Dvec^{\hat{\al}} \, \Dvec_{\hat{\al}} 
+ {i \over 4} \, \sg^{\hat{\al}\hat{\bt}} \lt[\Dvec_{\hat{\al}} , \Dvec_{\hat{\bt}}\rt] \, , \hspace{1mm}
\label{W^2-1}
\ee
where the first term of (\ref{W^2-1}) corresponds to $-{1 \over 2}\lt({1 \over 2} + 1 \rt) m^2$ for spin-1/2 particles
in the absence of external fields and the second term \cite{Pagels} is given by
\be
i \lt[\Dvec_{\hat{\al}} , \Dvec_{\hat{\bt}}\rt] & = & {\hbar^2 \over 4} \, \sg^{\hat{\mu}\hat{\nu}}
\lt( {}^F{}R_{\hat{\mu}\hat{\nu}\hat{\al}\hat{\bt}} \rt)
+ i \lt[\Pvec_{\hat{\al}} , \Pvec_{\hat{\bt}}\rt]
\nn
& &{} - \hbar^2 \, C^{\hat{\mu}}{}_{\hat{\al}\hat{\bt}} \, \Gmvec_{\hat{\mu}} \, ,
\label{D-commutator}
\ee
where
\be
C^{\hat{\mu}}{}_{\hat{\al}\hat{\bt}} & = & 
2 \, \bar{e}^{\hat{\mu}}{}_\sg \lt( \nabla_\lm \, \bar{e}^\sg{}_{[{\hat{\al}}} \rt) \bar{e}^\gm{}_{{\hat{\bt}}]} \,
{\partial U^\lm \over \partial X^\gm} \, ,
\label{C-defn}
\ee
is the object of anholonomicity for the local Lorentz frame, and is exclusively first-order
in ${}^F{}R_{\hat{\mu}\hat{\nu}\hat{\al}\hat{\bt}}$.
It can be shown that the second term in (\ref{D-commutator}) is given by
\be
i \lt[\Pvec_{\hat{\al}} , \Pvec_{\hat{\bt}}\rt] & = & \hbar \, C^{\hat{\mu}}{}_{\hat{\al}\hat{\bt}} \, \Pvec_{\hat{\mu}}
\nn
&  &{} + 2 \, \hbar \, \bar{e}^\sg{}_{[\hat{\al}} \, \bar{e}^\gm{}_{\hat{\bt}]} \lt(\nabvec_{\hat{\sg}} \, \ln \lm^{(\gm)}\rt) \Pvec_{\hat{\gm}}
 \, ,
\qquad \hspace{2mm}
\label{P-commutator}
\ee
where the second term in (\ref{P-commutator}) is the source for the non-inertial effect now considered \cite{Singh-Mobed,Singh1,Singh2}.
In particular, the general expression for the non-inertial dipole operator $\Rvec$ is embedded in (\ref{P-commutator})
as the limit of vanishing curvature, such that
\be
\Rvec^{\hat{k}} & = & \lt. {i \over 2\hbar} \, \varepsilon^{\hat{0} \hat{\imath} \hat{\jmath} \hat{k}} \,
\lt[\Pvec_{\hat{\imath}}, \Pvec_{\hat{\jmath}}\rt] \rt|_{{}^F{}R_{\hat{\mu}\hat{\nu}\hat{\al}\hat{\bt}} \rightarrow 0}
\nn
& = & \varepsilon^{\hat{0} \hat{\imath} \hat{\jmath} \hat{k}} \,
\lt(\nabvec_{\hat{\imath}} \, \ln \lm^{(j)}\rt) \Pvec_{\hat{\jmath}} \, .
\label{R-defn}
\ee
It is important to note that (\ref{P-commutator}) is not merely the consequence of a co-ordinate transformation,
which would otherwise lead to the standard result
$i \lt[\Pvec_{\hat{\al}} , \Pvec_{\hat{\bt}}\rt] = \hbar \, C^{\hat{\mu}}{}_{\hat{\al}\hat{\bt}} \, \Pvec_{\hat{\mu}}$
that vanishes as ${}^F{}R_{\hat{\mu}\hat{\nu}\hat{\al}\hat{\bt}} \rightarrow 0$.
The presence of the second term in (\ref{P-commutator}) is a subtle feature that follows from preserving the operator status of
$\Pvec_{\hat{\mu}}$ in curvilinear co-ordinates, and is the defining structure illustrating the limitations of the hypothesis of locality
for a quantum system in general motion.
A straightforward derivation of (\ref{P-commutator}) can be found in Appendix~\ref{sec:momentum-comm} of this paper.

By substitution of (\ref{D-commutator})--(\ref{P-commutator}) into (\ref{W^2-1}), it is shown that the Casimir scalar
for spin is
\be
\Wvec^{\hat{\al}} \, \Wvec_{\hat{\al}} & = & -{3 \over 4} \lt[m_0^2 + {\hbar^2 \over 6} \lt({}^F{}R^{\hat{\al}}{}_{\hat{\al}}\rt) \rt]
+ {\hbar \over 2} \lt(\sgvec \cdot \Rvec\rt)
\nn
&  &{} -{\hbar \over 4} \, \sg^{\hat{\al} \hat{\bt}} \, Q_{\hat{\al}\hat{\bt}}
+ {3 \over 2} \, \hbar \lt(\gm^5 \, \bar{\Gmvec}^{(\rm C)}_{\hat{\al}} - i \, \bar{\Gmvec}^{(\rm S)}_{\hat{\al}}\rt) \Pvec^{\hat{\al}}
\nn
&  &{} + {3 \over 4} \, \hbar^2 \, \nabvec^{\hat{\al}}
\lt(\bar{\Gmvec}^{(\rm S)}_{\hat{\al}} + i \, \gm^5 \, \bar{\Gmvec}^{(\rm C)}_{\hat{\al}}\rt) \, ,
\label{W^2-2}
\ee
where $m_0^2 = \Pvec^{\hat{\al}} \, \Pvec_{\hat{\al}}$, the Casimir scalar for momentum, is a Lorentz invariant and
\be
Q_{\hat{\al}\hat{\bt}} & = &
{i \over \hbar} \lt[\Pvec_{\hat{\al}} , \Pvec_{\hat{\bt}}\rt]
- \dl^{\hat{\imath}}{}_{\hat{\al}} \, \dl^{\hat{\jmath}}{}_{\hat{\bt}} \,
\varepsilon_{\ze {\hat{\imath}} {\hat{\jmath}} {\hat{k}}} \, \Rvec^{\hat{k}} \,
\nn
& = & 2 \lt(\bar{e}^\sg{}_{[\hat{\al}} \, \bar{e}^\gm{}_{\hat{\bt}]} - \dl^\sg{}_{[\hat{\al}} \, \dl^\gm{}_{\hat{\bt}]}\rt)
\lt(\nabvec_{\hat{\sg}} \, \ln \lm^{(\gm)}\rt) \Pvec_{\hat{\gm}}
\nn
&  &{} + C^{\hat{\mu}}{}_{\hat{\al}\hat{\bt}} \, \Pvec_{\hat{\mu}} \, ,
\label{Q-defn}
\ee
%
%
is exclusively first-order in the Riemann tensor.
With the exception of the effective mass term due to the Ricci scalar, the first line of
(\ref{W^2-2}) corresponds exactly with the expression obtained in flat space-time \cite{Singh-Mobed}.
The presence of $\Rvec$ in (\ref{W^2-2}) illustrates the frame-dependent quantity that represents
the predicted quantum violation of Lorentz invariance in the Casimir scalar, though it is important
to note that it does not appear whenever $\Wvec^{\hat{\al}} \, \Wvec_{\hat{\al}}$ acts on an {\em eigenstate}
\cite{Singh-Mobed}.
It is also interesting to note the prediction of a spin-gravity interaction due to (\ref{Q-defn}),
as well as contributions due to the spin connection.

Finally, it is very straightforward to generalize this expression to include an external electromagnetic (EM) field.
By introducing minimal coupling into (\ref{W^2-2}), such that
\be
\Pvec^{\hat{\al}} & \rightarrow & \bar{\Pvec}^{\hat{\al}} \ = \ \Pvec^{\hat{\al}} - e \, \Avec^{\hat{\al}} \, ,
\label{P-bar-defn}
\ee
it follows naturally that
\be
\Wvec^{\hat{\al}} \, \Wvec_{\hat{\al}} (\Pvec) & \rightarrow & \Wvec^{\hat{\al}} \, \Wvec_{\hat{\al}} (\bar{\Pvec})
+ {e \hbar \over 4} \, \sg^{\hat{\al} \hat{\bt}}
\lt( {}^F{}F_{\hat{\al}\hat{\bt}} \rt) \, , \qquad
\label{W^2-2+EM}
\ee
with a spin-EM interaction term emerging for $\Wvec^{\hat{\al}} \, \Wvec_{\hat{\al}}$,
where ${}^F{}F_{\hat{\al}\hat{\bt}}$ is the electromagnetic field tensor in the Fermi frame.

\section{Applications to Muon Decay in Curved Space-Time}
\label{sec:Muon-Decay}

Given (\ref{Wvec-grav-defn}) and the interesting results obtained for (\ref{W^2-2}),
it is worthwhile to consider its application to muon decay in curved space-time,
following the approach taken earlier \cite{Singh-Mobed}.
With this in mind, consider the reaction $\mu^- \rightarrow e^- + \bar{\nu}_e + \nu_\mu$ in an arbitrary gravitational background.
Then the matrix element \cite{Singh-Mobed} associated with muon decay is
\be
|{\cal M}|^2 & = &
{G_{\rm F}^2 \over 2} \, L^{(\mu)}_{\hat{\mu} \hat{\nu}} \, M^{\hat{\mu} \hat{\nu}}_{(e)},
\label{M^2}
\ee
where
\begin{subequations}
\label{L-M}
\be
\hspace{-0.8cm}
L^{\hat{\mu} \hat{\nu}}_{(\mu)} & = & {\rm Tr} \lt[\gslash{\pvec}_{\nu_\mu} \, \gm^{\hat{\mu}} \lt(\gslash{\Dvec}_{\mu}
+ m_{\mu} \, \gfive \, \gslash{\nvec}_{\mu}\rt)\gm^{\hat{\nu}} \lt(1 - \gfive\rt)\rt] \, ,
\label{L_uv}
\nl
\hspace{-0.8cm}
M^{\hat{\mu} \hat{\nu}}_{(e)} & = & {\rm Tr} \lt[\lt(\gslash{\Dvec}_{e} + m_{e} \, \gfive \, \gslash{\nvec}_e\rt)
\gm^{\hat{\mu}} \, \gslash{\pvec}_{\nu_e} \, \gm^{\hat{\nu}} \lt(1 - \gfive\rt)\rt] \, ,
\label{M_uv}
\ee
\end{subequations}
and $\nvec^{\hat{\mu}}$ is the polarization vector for the charged lepton.
By defining $\nvec^{\hat{\mu}}$ such that the orthogonality condition $\nvec^{\hat{\mu}} \, \Dvec_{\hat{\mu}} = 0$
is satisfied, where $\nvec^{\rm (0)}_{\hat{\mu}} = {E \over m_0 |\Pvec|} \, \Pvec_{\hat{\mu}} - {m_0 \over |\Pvec|}
 \, \dl^{\hat{0}}{}_{\hat{\mu}}$ is the polarization vector in flat space-time and
$|\Pvec| = \sqrt{-\Pvec^{\hat{\jmath}} \, \Pvec_{\hat{\jmath}}}$ ,
it is possible to express $\gfive \, \gslash{\nvec}$ in terms of $\Wvec^{\hat{\mu}}$.
That is,
\be
\lefteqn{
\gfive \, \gslash{\nvec} \ = \  -{2 \over m_0^2} \lt[\lt(\nvec^{\hat{\al}} \, \Wvec_{\hat{\al}}\rt)\gslash{\Dvec}
+ \gslash{\nvec} \lt(\Dvec^{\hat{\al}} \, \Wvec_{\hat{\al}}\rt) \rt] }
\nn
&&{} \times \lt[1 - {2 \hbar \over m_0^2} \lt(i \, \bar{\Gmvec}_{(\rm S)}^{\hat{\al}} - \gm^5 \, \bar{\Gmvec}_{(\rm C)}^{\hat{\al}} \rt)
\Pvec_{\hat{\al}} \rt.
\nn
&&{} + \lt. {\hbar^2 \over m_0^2} \, \nabvec_{\hat{\al}} \lt(\bar{\Gmvec}_{(\rm S)}^{\hat{\al}} + i \, \gm^5 \, \bar{\Gmvec}_{(\rm C)}^{\hat{\al}} \rt)
- {2 \hbar \over m_0^2} \, \gm^5 \, \gm^{\hat{\al}} \, \gm^{\hat{\bt}} \, \bar{\Gmvec}^{(\rm C)}_{\hat{\al}} \, \Pvec_{\hat{\bt}} \rt]
\nn
&&{} + {2 \hbar \over m_0 |\Pvec|} \lt[E \, \gslash{\bar{\Gmvec}_{(\rm C)}} - \gm^{\hat{0}}
\lt(2 \, \bar{\Gmvec}_{(\rm C)}^{\hat{\al}} \, \Pvec_{\hat{\al}} + i \hbar \, \nabvec_{\hat{\al}} \bar{\Gmvec}_{(\rm C)}^{\hat{\al}} \rt)  \rt]
\nn
&&{} \times \lt[1 - {2 \over m_0^2} \, \gm^5 \lt(\Dvec^{\hat{\al}} \, \Wvec_{\hat{\al}}\rt) \rt] \, ,
\label{gm5-nslash}
\ee
where $E = |\Pvec^{\hat{0}}|$.
When expanded out and simplified using (\ref{gm-identity}), it follows from (\ref{gm5-nslash}) that
$\gslash{\nvec} = \lt[\gslash{\nvec^{\rm (S, Re)}} + i \, \gslash{\nvec^{\rm (S, Im)}}\rt]
+ \gm^5 \lt[\gslash{\nvec^{\rm (C, Re)}} + i \, \gslash{\nvec^{\rm (C, Im)}}\rt]$, where the
contributions up to leading-order in curvature are
\be
\nvec^{\rm (S, Re)}_{\hat{\mu}} & \approx & \nvec^{\rm (0)}_{\hat{\mu}}
+ {4 \hbar \, E \over m_0^3 |\Pvec|} \, \dl^{\hat{0}}{}_{[\hat{\mu}} \, \dl^{\hat{\jmath}}{}_{\hat{\nu}]} \, \Rvec_{\hat{\jmath}} \, \Pvec^{\hat{\nu}}
\nn
& &{}
- {\hbar^2 \, E \over m_0^3 |\Pvec|} \lt(\nabvec^{\hat{\nu}} Q_{\hat{\mu} \hat{\nu}}\rt) \, ,
\label{n-S,Re}
\ee
\be
\lefteqn{
\nvec^{\rm (S, Im)}_{\hat{\mu}} \ \approx \
{2 \hbar^2 \, E \over m_0^3 |\Pvec|} \, \dl^{\hat{0}}{}_{[\hat{\mu}} \, \dl^{\hat{\jmath}}{}_{\hat{\nu}]}
\lt(\nabvec^{\hat{\nu}} \Rvec_{\hat{\jmath}}\rt) + {\hbar \, E \over m_0 |\Pvec|} \, \bar{\Gmvec}^{\rm (S)}_{\hat{\mu}} }
\nn
&&{} - {4 \hbar \over m_0 |\Pvec|} \lt[\dl^{\hat{0}}{}_{\hat{\mu}}
\lt(\bar{\Gmvec}_{\rm (S)}^{\hat{\lm}} \, \Pvec_{\hat{\lm}} \rt)
+ {1 \over 2} \, \varepsilon^{\hat{0} \hat{\al} \hat{\bt}}{}_{\hat{\mu}} \, \bar{\Gmvec}^{\rm (C)}_{\hat{\al}} \, \Pvec_{\hat{\bt}} \rt]
\nn
&&{} - {2 \hbar \over m_0^2} \, \nvec^{\rm (0)}_{\hat{\nu}} \lt[\dl^{\hat{\nu}}{}_{\hat{\mu}}
\lt(\bar{\Gmvec}_{\rm (S)}^{\hat{\lm}} \, \Pvec_{\hat{\lm}} \rt)
+ \varepsilon^{\hat{\nu} \hat{\al} \hat{\bt}}{}_{\hat{\mu}} \, \bar{\Gmvec}^{\rm (C)}_{\hat{\al}} \, \Pvec_{\hat{\bt}} \rt]
\nn
&&{} + {\hbar \, E \over m_0^3 |\Pvec|}\lt[
2 \lt(\bar{\Gmvec}_{(\rm S)}^{\hat{\lm}} \, \Pvec_{\hat{\lm}} \rt) \Pvec_{\hat{\mu}}
- {\hbar^2 \over 4} \, \nabvec_{\hat{\mu}} \lt({}^F{}R^{\hat{\al}}{}_{\hat{\al}}\rt) \rt]
\nn
&&{} + {2 \hbar \, E \over m_0^3 |\Pvec|} \,  Q_{\hat{\mu} \hat{\nu}} \, \Pvec^{\hat{\nu}} \, ,
\label{n-S,Im}
\ee
\be
\lefteqn{
\nvec^{\rm (C, Re)}_{\hat{\mu}} \ \approx \
-{\hbar^2 \, E \over m_0^3 |\Pvec|} \, \varepsilon^{\hat{\sg} \hat{\al} \hat{\bt}}{}_{\hat{\mu}} \,
\dl^{\hat{0}}{}_{[\hat{\al}} \, \dl^{\hat{\jmath}}{}_{\hat{\bt}]}
\lt(\nabvec_{\hat{\sg}} \Rvec_{\hat{\jmath}}\rt) }
\nn
&&{} + {3 \hbar \, E \over m_0 |\Pvec|} \, \bar{\Gmvec}^{\rm (C)}_{\hat{\mu}}
- {2 \hbar \over m_0 |\Pvec|} \, \dl^{\hat{0}}{}_{\hat{\mu}} \, \lt(\bar{\Gmvec}_{\rm (C)}^{\hat{\lm}} \, \Pvec_{\hat{\lm}} \rt)
\nn
&&{} - {2 \hbar \, E \over m_0^3 |\Pvec|}\lt[
\eta_{\hat{\mu} \hat{\nu}} \lt(\bar{\Gmvec}_{\rm (S)}^{\hat{\lm}} \, \Pvec_{\hat{\lm}} \rt)
+ 2 \, \bar{\Gmvec}^{\rm (C)}_{[\hat{\mu}} \, \Pvec_{\hat{\nu}]} \rt] \Pvec^{\hat{\nu}} \, ,
\label{n-C,Re}
\ee
\be
\nvec^{\rm (C, Im)}_{\hat{\mu}} & \approx &
-{\hbar^2 \, E \over 2 m_0^3 |\Pvec|} \,
\varepsilon^{\hat{\sg} \hat{\al} \hat{\bt}}{}_{\hat{\mu}} \lt(\nabvec_{\hat{\sg}} Q_{\hat{\al} \hat{\bt}}\rt) \, .
\label{n-C,Im}
\ee
%
%

Substitution of (\ref{L-M})--(\ref{n-C,Im}) into (\ref{M^2}) leads to
\be
|{\cal M}|^2 & = & 32 \, G_{\rm F}^2 \lt(\Pvec_{\nu_e}^{\hat{\al}} \, \bar{\Dvec}^{\mu}_{\hat{\al}} \rt)
\lt(\Pvec_{\nu_\mu}^{\hat{\bt}} \, \bar{\Dvec}^{e}_{\hat{\bt}} \rt) \, ,
\label{M^2-eval}
\ee
where
\be
\bar{\Dvec}_{\hat{\al}} = \Pvec_{\hat{\al}} - \hbar \lt(\bar{\Gmvec}^{\rm (C)}_{\hat{\al}} - i \, \bar{\Gmvec}^{\rm (S)}_{\hat{\al}} \rt)
+ m_0 \lt(\nvec_{\hat{\al}}^{\rm (C)} - \nvec_{\hat{\al}}^{\rm (S)}\rt) \, . \quad
\label{D-bar}
\ee
Apart from (\ref{n-C,Im}), each of the contributions to $\nvec_{\hat{\mu}}$ has a component that
is dependent on $\Rvec$ or its gradient that survives in the limit of vanishing curvature.
It is also worthwhile to note that $|{\cal M}|^2$ in (\ref{M^2-eval}) is {\em complex-valued},
even when the notation for the matrix element suggests a {\em real-valued} result.
Given that the imaginary terms in $|{\cal M}|^2$ come exclusively from the non-Hermiticity properties of quantum systems
in curved space-time \cite{Parker}, this is not a surprising outcome and not a cause for concern.

In a similar fashion as shown by (\ref{W^2-2+EM}), it is straightforward
to generalize (\ref{n-S,Re})--(\ref{n-C,Im}) to incorporate external electromagnetic fields.
By again introducing (\ref{P-bar-defn}) and defining
\be
\bar{F}_{\hat{\al}\hat{\bt}} & = & {}^F{}F_{\hat{\al}\hat{\bt}} + C^{\hat{\mu}}{}_{\hat{\al}\hat{\bt}} \, \Avec_{\hat{\mu}} \, ,
\label{F-bar-EM-field}
\ee
it follows that the contributions to the polarization vector change as
\be
\lefteqn{\nvec^{\rm (S, Re)}_{\hat{\mu}}(\Pvec) \ \rightarrow \ \nvec^{\rm (S, Re)}_{\hat{\mu}}(\bar{\Pvec})
+ {2 e \hbar^2 \over m_0^3 |\bar{\Pvec}|} }
\nn
&&{} \times \bar{F}_{\hat{\al}\hat{\bt}} \lt\{
\lt[\eta^{\hat{\al} \hat{0}} \, \dl^{\hat{\bt}}{}_{\hat{\mu}}
\lt(\bar{\Gmvec}_{\rm (S)}^{\hat{\lm}} \, \bar{\Pvec}_{\hat{\lm}} \rt)
+ {1 \over 2} \, \varepsilon^{\hat{0} \hat{\al} \hat{\bt}}{}_{\hat{\mu}} \, \lt(\bar{\Gmvec}_{\rm (C)}^{\hat{\lm}} \, \bar{\Pvec}_{\hat{\lm}} \rt) \rt]
\rt.
\nn
&&{} + \lt. E \, \lt[\dl^{\hat{\al}}{}_{\hat{\mu}} \, \bar{\Gmvec}_{\rm (S)}^{\hat{\bt}}
- {1 \over 2} \, \varepsilon^{\hat{\nu} \hat{\al} \hat{\bt}}{}_{\hat{\mu}} \, \bar{\Gmvec}^{\rm (C)}_{\hat{\nu}} \rt] \rt\}
\nn
&&{} + {e \hbar^2 \, E \over m_0^3 |\bar{\Pvec}|} \lt(\nabvec^{\hat{\nu}} \bar{F}_{\hat{\mu} \hat{\nu}}\rt)
- {2 e \hbar^2 \over m_0^4 |\bar{\Pvec}|} \, \varepsilon^{\hat{\al} \hat{\bt} \hat{\lm} \hat{\nu}} \,
\bar{F}_{\hat{\al}\hat{\bt}} \, \bar{\Pvec}_{\hat{\lm}} \, \bar{\Gmvec}^{\rm (C)}_{[\hat{\mu}} \, \bar{\Pvec}_{\hat{\nu}]}
\nn
&&{} - {4 e \hbar^2 \, E \over m_0^5 |\bar{\Pvec}|} \, \bar{F}_{\hat{\sg}\hat{\gm}} \, \bar{\Pvec}^{\hat{\gm}}
\lt[\dl^{\hat{\sg}}{}_{\hat{\mu}} \lt(\bar{\Gmvec}_{\rm (S)}^{\hat{\lm}} \, \bar{\Pvec}_{\hat{\lm}} \rt)
+ \varepsilon^{\hat{\sg} \hat{\al} \hat{\bt}}{}_{\hat{\mu}} \, \bar{\Gmvec}^{\rm (C)}_{\hat{\al}} \, \bar{\Pvec}_{\hat{\bt}} \rt] \, ,
\nn
\label{n-S,Re+EM}
\ee
\be
\nvec^{\rm (S, Im)}_{\hat{\mu}}(\Pvec) & \rightarrow & \nvec^{\rm (S, Im)}_{\hat{\mu}}(\bar{\Pvec})
- {2 e \hbar \, E \over m_0^3 |\bar{\Pvec}|} \,  \bar{F}_{\hat{\mu} \hat{\nu}} \, \bar{\Pvec}^{\hat{\nu}} \, ,
\label{n-S,Im+EM}
\nl
\nvec^{\rm (C, Re)}_{\hat{\mu}}(\Pvec) & \rightarrow & \nvec^{\rm (C, Re)}_{\hat{\mu}}(\bar{\Pvec}) \, ,
\label{n-C,Re+EM}
\ee
\be
\lefteqn{\nvec^{\rm (C, Im)}_{\hat{\mu}}(\Pvec) \ \rightarrow \ \nvec^{\rm (C, Im)}_{\hat{\mu}}(\bar{\Pvec})
+ {2 e \hbar^2 \over m_0^3 |\bar{\Pvec}|} }
\nn
&&{} \times \bar{F}_{\hat{\al}\hat{\bt}}
\lt[\eta^{\hat{\al} \hat{0}} \, \dl^{\hat{\bt}}{}_{\hat{\mu}}
\lt(\bar{\Gmvec}_{\rm (C)}^{\hat{\lm}} \, \bar{\Pvec}_{\hat{\lm}} \rt)
- {1 \over 2} \, \varepsilon^{\hat{0} \hat{\al} \hat{\bt}}{}_{\hat{\mu}} \, \lt(\bar{\Gmvec}_{\rm (S)}^{\hat{\lm}} \, \bar{\Pvec}_{\hat{\lm}} \rt) \rt]
\nn
&&{} + {e \hbar^2 \, E \over 2 m_0^3 |\bar{\Pvec}|}
\varepsilon^{\hat{0} \hat{\al} \hat{\bt}}{}_{\hat{\mu}} \lt(\nabvec_{\hat{\sg}} \bar{F}_{\hat{\al} \hat{\bt}}\rt)
- {e \hbar^2 \over m_0^4 |\bar{\Pvec}|}
\nn
&&{} \times \varepsilon^{\hat{\al} \hat{\bt} \hat{\gm}}{}_{\hat{\sg}} \, \bar{F}_{\hat{\al}\hat{\bt}} \, \bar{\Pvec}_{\hat{\gm}}
\lt[\dl^{\hat{\sg}}{}_{\hat{\mu}} \lt(\bar{\Gmvec}_{\rm (S)}^{\hat{\lm}} \, \bar{\Pvec}_{\hat{\lm}} \rt)
+ \varepsilon^{\hat{\sg} \hat{\lm} \hat{\nu}}{}_{\hat{\mu}} \, \bar{\Gmvec}^{\rm (C)}_{\hat{\lm}} \, \bar{\Pvec}_{\hat{\nu}} \rt]
\nn
&&{} + {8 e \hbar^2 \, E \over m_0^5 |\bar{\Pvec}|} \,
\bar{F}_{\hat{\al}\hat{\bt}} \, \bar{\Pvec}^{\hat{\bt}}
\lt[\eta^{\hat{\al} \hat{\nu}} \, \bar{\Gmvec}^{\rm (C)}_{[\hat{\mu}} \, \bar{\Pvec}_{\hat{\nu}]}
- \dl^{\hat{\al}}{}_{\hat{\mu}} \lt(\bar{\Gmvec}_{\rm (C)}^{\hat{\lm}} \, \bar{\Pvec}_{\hat{\lm}} \rt) \rt] \, .
\nn
\label{n-C,Im+EM}
\ee
%

Clearly, there exist numerous contributions involving the coupling of the EM field with space-time curvature,
as well as exclusively electromagnetic contributions that survive as ${}^F{}R_{\hat{\mu}\hat{\nu}\hat{\al}\hat{\bt}} \rightarrow 0$.

\section{Muon Decay near the Event Horizon of a Kerr Black Hole}
\label{sec:Kerr-BH-application}


%
\begin{figure}
\psfrag{al}[tr][][1.8][0]{\Large $a = \al \, r_0$}
\psfrag{x}[bc][][2.0][0]{\huge $\rm x$}
\psfrag{y}[tc][][2.0][0]{\huge $\rm y$}
\psfrag{z}[bl][][2.0][0]{\huge $\rm z$}
\psfrag{mu}[cr][][1.8][0]{\Large $\mu^-$}
\psfrag{e}[cc][][1.8][0]{\Large $e^-$}
\psfrag{r0}[cc][][1.5][0]{\Large $r_0$}
\psfrag{th0}[cc][][1.5][0]{\Large $\th_0$}
\psfrag{Omt}[tc][][1.5][0]{\Large $\ph_0$}
\psfrag{r}[bl][][1.5][0]{\large $r$}
\psfrag{th}[cc][][1.5][0]{\large $\th$}
\psfrag{ph}[cc][][1.5][0]{\large $\ph$}
\psfrag{nue}[cc][][1.8][0]{\Large $\bar{\nu}_e$}
\psfrag{num}[cc][][1.5][0]{\Large $\nu_\mu$}
\begin{minipage}[t]{0.3 \textwidth}
\centering
\rotatebox{0}{\includegraphics[width = 6.0cm, height = 4.0cm, scale = 1]{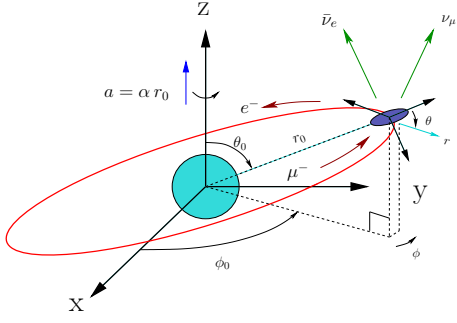}}
\end{minipage}%
\caption{\label{fig:black-hole-orbit} Muon in circular orbit around a Kerr black hole, subject to electroweak decay.
The classical orbit is defined by $(r_0, \th_0, \ph_0)$, where quantum fluctuations about it are defined by $(r, \th, \ph)$.
While $\th_0 = \pi/2$ for this paper, the inclined orbit here emphasizes the point that the decaying muon is
subject to polar variations in the curvature that need to be accounted for properly.}
\end{figure}
Having now obtained the formal expression for the muon decay matrix element in a general curved space-time background,
it is useful to explore a concrete example.
An interesting test case is that of muon decay in circular orbit near the event horizon of a Kerr black hole,
which is especially important for identifying any signatures in the muon decay spectrum that may result from strong
curvature effects, particularly where it concerns changes in the black hole's spin angular momentum.

To begin, consider an orthonormal tetrad frame situated as shown in Figure~\ref{fig:black-hole-orbit}, where it defines
a classical circular orbit around a black hole of mass $M$ and specific spin angular momentum
$a = J/M$ in standard Boyer-Lindquist co-ordinates $x^{\bar{\mu}} = (t, r_0, \th_0, \ph_0)$, such that \cite{Mashhoon2,Chicone1}
\be
t & = & {1 \over N_0} \lt(1 + a \, \OmK\rt) T \, ,
\label{t=}
\nl
\ph_0 & = & {\OmK \, T \over N_0 \, \sin \th_0}   \, ,
\label{phi=}
\ee
are expressed in terms of the Keplerian frequency
\be
\OmK & = & \sqrt{M \over r_0^3} \, ,
\label{Om-Kepler}
\ee
and
\be
N_0 & = & \sqrt{1 - {3 M \over r_0} + 2 \, a \, \OmK} \, ,
\label{N0=}
\ee
with boundary conditions chosen such that $t = \ph_0 = 0$ at $T = 0$.
For all computations considered in this paper, $\th_0 = \pi/2$ and $r_0$ is strictly constant.
Given that the orbital energy and angular momentum expressions are
\be
E_0 & = & {1 \over N_0} \lt(1 - {2M \over r_0} + a \, \OmK\rt) \, ,
\label{E0=}
\nl
L_0 & = & {r_0^2 \OmK \over N_0} \lt(1 - 2 \, a \, \OmK + {a^2 \over r_0^2} \rt) \, ,
\label{L0=}
\ee
respectively, it is shown that the orthonormal tetrad frame corresponding to Figure~\ref{fig:black-hole-orbit}
is described by
\be
\lm^{\bar{\mu}}{}_{0} & = & \lt({1 + a \, \OmK \over N_0}, 0, 0, {\OmK \over N_0 \, \sin \th_0} \rt),
\label{tetrad-0}
\nl
\lm^{\bar{\mu}}{}_{1} & = & \lt(-{L_0 \over r_0 \, A_0} \, \sin \lt(\OmK \, T\rt), \, A_0 \, \cos \lt(\OmK \, T\rt), \, 0, \rt.
\nn
&  &{} \lt.
 \, -{E_0 \over r_0 \, A_0 \, \sin \th_0} \, \sin \lt(\OmK \, T\rt) \rt),
\label{tetrad-1}
\nl
\lm^{\bar{\mu}}{}_{2} & = & \lt(0, 0, {1 \over r_0}, 0\rt),
\label{tetrad-2}
\nl
\lm^{\bar{\mu}}{}_{3} & = & \lt({L_0 \over r_0 \, A_0} \, \cos \lt(\OmK \, T\rt), \, A_0 \, \sin \lt(\OmK \, T\rt), \, 0, \rt.
\nn
&  &{} \lt.
 \, {E_0 \over r_0 \, A_0 \, \sin \th_0} \, \cos \lt(\OmK \, T\rt) \rt),
\label{tetrad-3}
\ee
where
\be
A_0 & = & \sqrt{1 - {2M \over r_0} + {a^2 \over r_0^2}} \, .
\label{A0=}
\ee
%
%
%
%
\begin{figure}
\psfrag{x}[tc][][1.8][0]{\Large $x$}
\psfrag{dG}[bc][][1.8][0]{\Large ${1 \over \Gm_0} \, \hat{\Gm}\lt(x,\cos \Th\rt)$}
\begin{minipage}[t]{0.3 \textwidth}
\centering
\subfigure[\hspace{0.2cm} ($\Th = 0^\circ \, , \ \al = 1$)]{
\label{fig:dG-th=000-al=100}
\rotatebox{0}{\includegraphics[width = 6.0cm, height = 4.0cm, scale = 1]{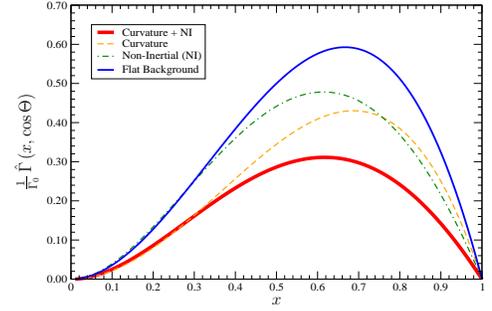}}}
\vspace{5mm}
\end{minipage}
\begin{minipage}[t]{0.3 \textwidth}
\centering
\subfigure[\hspace{0.2cm} ($\Th = 0^\circ \, , \ \al = 0.99$)]{
\label{fig:dG-th=000-al=099}
\rotatebox{0}{\includegraphics[width = 6.0cm, height = 4.0cm, scale = 1]{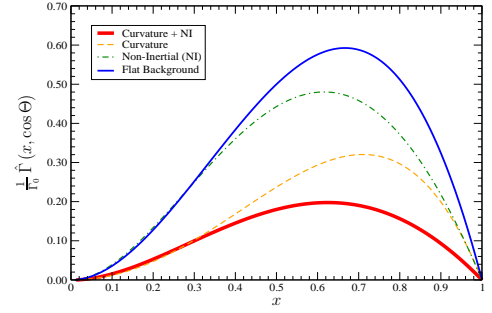}}}
\end{minipage}
\caption{\label{fig:dG-th=000} The muon differential decay rate as a function of electron energy fraction
for the muon near the event horizon of a Kerr black hole with mass $M = 2 \times 10^{-11}$~cm and orbital radius $r_0 \sim M$.
Figs.~\ref{fig:dG-th=000-al=100} and \ref{fig:dG-th=000-al=099} display comparative plots of the decay spectrum due
to curvature and non-inertial effects, for $\al = 1$ and $\al = 0.99$, respectively.
It is clear that the plots are very sensitive to changes in $\al$ for fixed $r_0$.}
\end{figure}

For future reference, it is useful to express $a$ in terms of the dimensionless parameter $\al = a/r_0$
with the range of $-1/4 \leq \al \leq 1$ to accommodate for both co-rotating and counter-rotating black holes
with respect to the orbital direction.
The range of $\al$ corresponds to $-M \leq a \leq M$ for the innermost (photon) radius of $r_{0+} = M$ for $\al =~1$
and $r_{0-} = 4 \, M$ for $\al = -1/4$ \cite{Chandra}.
By defining all dimensional quantities in terms of $M$, it is shown that the orbital radius near the black hole event horizon
is described by
\be
r_0 & = & 9 \, M \lt[\al + \sqrt{\al^2 + 3 \lt(1 - N_0^2\rt)}\rt]^{-2} \, ,
\label{r0=}
\ee
where $N_0 \gtrsim 0$ denotes the separation away from the innermost circular orbit.
As indicated by Figure~\ref{fig:black-hole-orbit}, the muon undergoes a decay into an electron and two neutrinos
within some region of space-time in the viscinity of the classical orbit, and is subject to quantum fluctuations
described by $(r, \th, \ph)$ with respect to the orthonormal frame during the decay process.
This requirement suggests that the muon is sensitive to polar variations in the Riemann curvature within the time frame
of the decay, in accordance with the Heisenberg uncertainty principle.

\begin{figure}
\psfrag{x}[tc][][1.8][0]{\Large $x$}
\psfrag{dG}[bc][][1.8][0]{\Large ${1 \over \Gm_0} \, \Dl \hat{\Gm}\lt(x,\cos \Th\rt)$}
\psfrag{th = 180d}[cc][][1.8][0]{$\Th = 180^\circ$}
\psfrag{th = 150d}[cc][][1.8][0]{$\Th = 150^\circ$}
\psfrag{th = 120d}[cc][][1.8][0]{$\Th = 120^\circ$}
\psfrag{th = 90d}[cc][][1.8][0]{$\Th = 90^\circ$}
\psfrag{th = 60d}[cc][][1.8][0]{$\Th = 60^\circ$}
\psfrag{th = 30d}[cc][][1.8][0]{$\Th = 30^\circ$}
\psfrag{th = 0d}[cc][][1.8][0]{$\Th = 0^\circ$}
\begin{minipage}[t]{0.3 \textwidth}
\centering
\rotatebox{0}{\includegraphics[width = 6.0cm, height = 4.0cm, scale = 1]{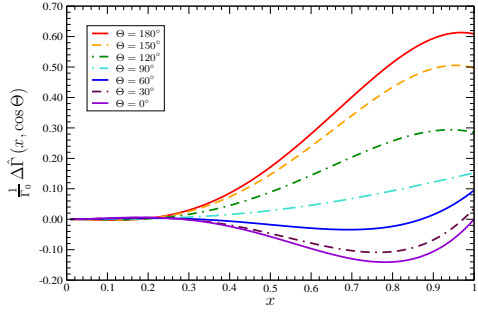}}
\end{minipage}%
\caption{\label{fig:rel-NI-angles} Net differential decay rate due to the non-inertial dipole operator
for various decay angles, after subtracting the flat space-time background contribution.
It is evident that the contribution of $\Rvec$ to the decay spectrum becomes identifiable for large emission angles.}
\end{figure}
With the necessary framework established, it is now possible to determine the differential decay rate for the muon in the Kerr background.
This is given by \cite{Kaku,Arbuzov}
\be
\lefteqn{ {\d^2 \Gm \over \d E_e \, \d \lt(\cos \Th\rt)} \ \equiv \ \hat{\Gm}\lt(E_e, \cos \Th\rt) }
\nn
&&{} = \Gm_0 \lt[{8 \, |\Pvec_e| \over m_\mu^5 \, E_\mu} \, \bar{\Dvec}_\mu^{\hat{\al}} \, \bar{\Dvec}_e^{\hat{\bt}}
\lt(\eta_{\hat{\al}\hat{\bt}} \, Q^2 + 2 \, \Qvec_{\hat{\al}} \, \Qvec_{\hat{\bt}} \rt) \rt] \, , \quad
\label{diff-cross-section=}
\ee
where $\Th$ is the scattering angle for the outgoing electron,
\be
\Qvec^{\hat{\al}} & = & \lt(\Pvec_\mu - \Pvec_e \rt)^{\hat{\al}}
\label{Q=}
\ee
is the momentum transfer and
\be
\Gm_0 & \approx & {G_{\rm F}^2 \, m_\mu^5 \over 192 \, \pi^3} \ \approx \ 2.965 \times 10^{-16} \ \ {\rm MeV}
\label{Gm0=}
\ee
is the total decay rate in flat space-time.
All computations for the various plots shown in this paper are performed in the local rest frame of the muon,
such that
\be
E_e = {m_\mu \over 2} \, x \, ,
\label{Ee-rest-frame=}
\ee
where $x$ is the electron energy fraction in the decay process.

It is not surprising to note that there is no discernable contribution of space-time curvature to the
differential decay rate for astrophysical black holes.
However, matters change considerably for the case of microscopic black holes in the range of $M \sim 10^{-10} - 10^{-11}$~cm,
with interesting results to report.
Unless otherwise specified, all plots presented in this paper are generated for $M = 2 \times 10^{-11}$ cm and $N_0 = 10^{-2}$
corresponding to $r_0 \sim M$, the event horizon of an extreme Kerr black hole $(\al = 1)$ co-rotating with the muon's orbital direction.
\begin{figure}
\psfrag{x}[tc][][1.8][0]{\Large $x$}
\psfrag{dG}[bc][][1.8][0]{\Large ${1 \over \Gm_0} \, \Dl \hat{\Gm}\lt(x,\cos \Th\rt)$}
\begin{minipage}[t]{0.3 \textwidth}
\centering
\subfigure[\hspace{0.2cm} ($\Th = 0^\circ \, , \ \al = 1$)]{
\label{fig:rel-dG-th=000-al=100}
\rotatebox{0}{\includegraphics[width = 6.0cm, height = 4.0cm, scale = 1]{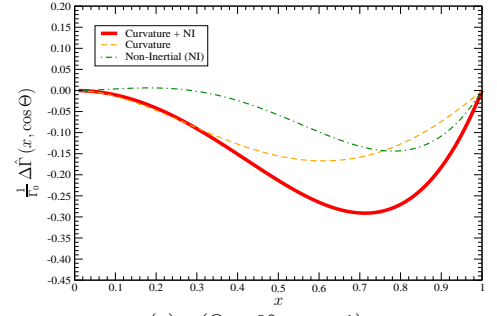}}}
\vspace{5mm}
\end{minipage}
\begin{minipage}[t]{0.3 \textwidth}
\centering
\subfigure[\hspace{0.2cm} ($\Th = 0^\circ \, , \ \al = 0.99$)]{
\label{fig:rel-dG-th=000-al=099}
\rotatebox{0}{\includegraphics[width = 6.0cm, height = 4.0cm, scale = 1]{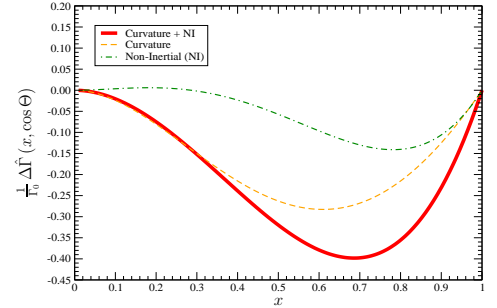}}}
\end{minipage}
\caption{\label{fig:rel-dG-th=000} Net differential decay rate due to various combinations of curvature and the
non-inertial dipole operator after subtracting the flat space-time background contribution.}
\end{figure}

Figure~\ref{fig:dG-th=000} displays the contributions of curvature and the non-inertial dipole operator $\Rvec$
to the overall differential cross section due to forward emission $(\Th = 0)$.
A complete list of all plots illustrating the full range of scattering angles can be found in
Appendix~\ref{sec:diff-cross-section-plots} of this paper.
It is clear from examining Figs.~\ref{fig:dG-th=000-al=100} and \ref{fig:dG-th=000-al=099}
that the contributions of curvature and non-inertial interactions make for a considerable
adjustment in the decay spectrum compared with the flat background contribution.
Specifically, the effect of these contributions is to reduce the decay spectrum, contributing to
a lifetime enhancement for the muon at this angle.
It is also interesting to note that the contribution of $\Rvec$ to the decay cross section enhances
the contribution due to curvature alone.

For the effect due to $\Rvec$ in Figure~\ref{fig:dG-th=000-al=100}, its peak value at $x = 0.60$ corresponds
to a reduction in the decay profile by about 17\% relative to the flat space-time profile.
For the contribution due to curvature alone, which peaks at $x = 0.66$, the reduction is about 28\%, while
the total contribution results in a reduction of 46\% at $x = 0.60$.
It is also clear, based on Figure~\ref{fig:dG-th=000-al=099}, that the decay spectrum is very sensitive to changes
in $a$, since even a 1\% decrease in $\al$ for fixed $r_0$ leads to a curvature-induced reduction of its peak by 46\% at $x = 0.70$,
with the total contribution due to curvature and non-inertial effects reduced by about 66\% at $x = 0.60$.

It is worthwhile to determine the net differential decay rate due to non-inertial and curvature effects after
subtracting off the flat space-time background contribution.
Figure~\ref{fig:rel-NI-angles} displays the distinctive contribution of $\Rvec$ to the decay cross section
as a function of emission angle, where
\be
\Dl \hat{\Gm}\lt(x,\cos \Th\rt) & \equiv & \hat{\Gm}\lt(x,\cos \Th\rt) - \hat{\Gm}\lt(x,\cos \Th\rt)_{\rm Flat}
\label{Dl-Gm1=}
\ee
reaches a maximum value of $0.60 \, \Gm_0$ at $x = 1$ for backward emission ($\Th = 180^\circ$).
It is useful to note that (\ref{Dl-Gm1=}) is negative-valued for $0^\circ \leq \Th \lesssim 60^\circ$, while it becomes
increasingly positive-valued for $60^\circ \lesssim \Th \leq 180^\circ$.
In particular, all of the curves for $\Th = 90^\circ - 180^\circ$ become nonzero at around $x = 0.22$, which may be suggestive
of an interesting kinematic region to explore in parameter space.

The corresponding comparison due to curvature also yields some interesting features, as shown in Figure~\ref{fig:rel-dG-th=000}.
For the case of $\al = 1$, Figure~\ref{fig:rel-dG-th=000-al=100} shows a $0.17 \, \Gm_0$ reduction of the decay cross section
at $x = 0.60$ due to curvature alone, while the combination of curvature and non-inertial effects for $x = 0.72$
leads to a $0.29 \, \Gm_0$ reduction in the profile.
The corresponding analysis for Figure~\ref{fig:rel-dG-th=000-al=099} at $x = 0.60$ shows a reduction of $0.29 \, \Gm_0$ due to curvature alone,
and an overall reduction of $0.40 \, \Gm_0$ at $x = 0.70$ for $\al = 0.99$,
again indicating an extreme sensitivity of the decay cross section due to changes in $a$.

\begin{figure}
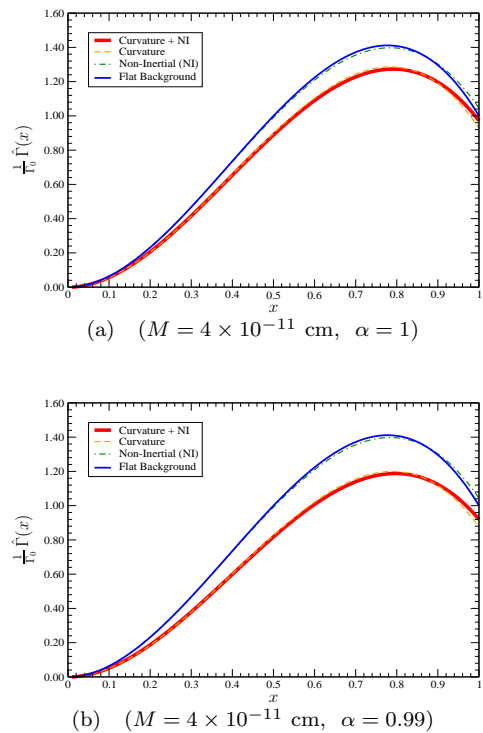

\psfrag{x}[tc][][1.8][0]{\Large $x$}
\psfrag{dG}[bc][][1.8][0]{\Large ${1 \over \Gm_0} \, \hat{\Gm}(x)$}
\begin{minipage}[t]{0.3 \textwidth}
\centering
\subfigure[\hspace{0.2cm} ($M = 4 \times 10^{-11}$ cm, \ $\al = 1$)]{
\label{fig:Michel-r0=4e-11-al=100}
\rotatebox{0}{\includegraphics[width = 6.0cm, height = 4.0cm, scale = 1]{5a}}}
\vspace{5mm}
\end{minipage}
\begin{minipage}[t]{0.3 \textwidth}
\centering
\subfigure[\hspace{0.2cm} ($M = 4 \times 10^{-11}$ cm, \ $\al = 0.99$)]{
\label{fig:Michel-r0=4e-11-al=099}
\rotatebox{0}{\includegraphics[width = 6.0cm, height = 4.0cm, scale = 1]{5b}}}
\end{minipage}
\caption{\label{fig:Michel-r0=4e-11} Michel spectrum with contributions due to curvature and the non-inertial dipole operator
for $M = 4 \times 10^{-11}$ cm and $r_0 \sim M$.
While the combined curvature and non-inertial effect makes a larger change to the profile compared to the non-inertial effect alone,
the fact that there exists a slight but discernable change in its curve compared to the flat background contribution is noteworthy.}
\end{figure}
\begin{figure}
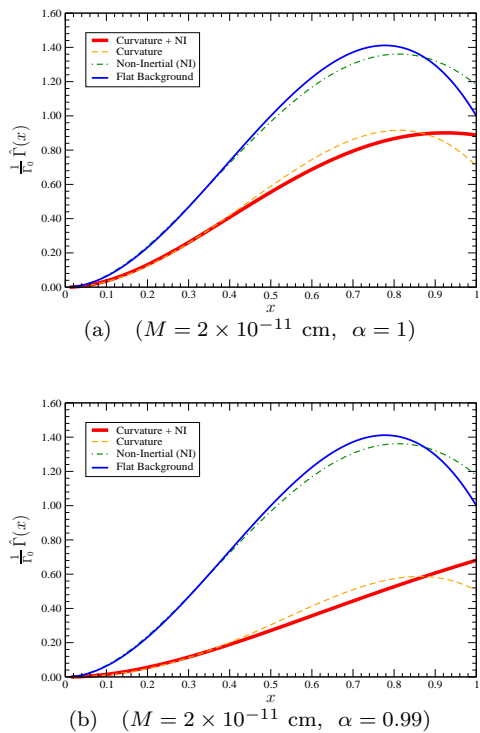

\psfrag{x}[tc][][1.8][0]{\Large $x$}
\psfrag{dG}[bc][][1.8][0]{\Large ${1 \over \Gm_0} \, \hat{\Gm}(x)$}
\begin{minipage}[t]{0.3 \textwidth}
\centering
\subfigure[\hspace{0.2cm} ($M = 2 \times 10^{-11}$ cm, \ $\al = 1$)]{
\label{fig:Michel-r0=2e-11-al=100}
\rotatebox{0}{\includegraphics[width = 6.0cm, height = 4.0cm, scale = 1]{6a}}}
\vspace{5mm}
\end{minipage}
\begin{minipage}[t]{0.3 \textwidth}
\centering
\subfigure[\hspace{0.2cm} ($M = 2 \times 10^{-11}$ cm, \ $\al = 0.99$)]{
\label{fig:Michel-r0=2e-11-al=099}
\rotatebox{0}{\includegraphics[width = 6.0cm, height = 4.0cm, scale = 1]{6b}}}
\end{minipage}
\caption{\label{fig:Michel-r0=2e-11} Michel spectrum with contributions due to curvature and the non-inertial dipole operator
for $M = 2 \times 10^{-11}$ cm and $r_0 \sim M$.
Comparison between Figs.~\ref{fig:Michel-r0=2e-11-al=100} and \ref{fig:Michel-r0=2e-11-al=099} shows that
the profile is extremely sensitive to changes in $\al$, while the increase in $r_0$ leads to greater distinction
between the non-inertial curve compared to the flat background.}
\end{figure}

Integration of the scattering angle over the unit sphere results in the Michel spectrum described by
\be
\hat{\Gm}(x) & = & {\d \Gm \over \d x} \, .
\label{Michel-defn}
\ee
Figures~\ref{fig:Michel-r0=4e-11} and \ref{fig:Michel-r0=2e-11} display the muon decay Michel spectrum for
$M = 4 \times 10^{-11}$~cm and $M = 2 \times 10^{-11}$~cm, respectively, where $r_0 \sim M$.
It is clear from Figure~\ref{fig:Michel-r0=4e-11} that curvature has the effect of measurably reducing $\hat{\Gm}(x)$
for this choice of $M$, and comparison between Figs.~\ref{fig:Michel-r0=4e-11-al=100} and \ref{fig:Michel-r0=4e-11-al=099}
reveals that the overall reduction is enhanced by a decrease in $\al$.
At around $x = 0.77$ and $\al = 1$, the Michel spectrum due to both curvature and $\Rvec$ is reduced by about 9\% from
its peak value due to the flat background alone, while the corresponding curve for $x = 0.77$ and $\al = 0.99$
is reduced by about 17\%.
Comparison between Figs.~\ref{fig:Michel-r0=2e-11-al=100} and \ref{fig:Michel-r0=2e-11-al=099} indicates an increased
reduction in the Michel spectrum due to curvature and non-inertial effects when $r_0 \rightarrow 0$.
Consistent with the differential decay cross section in Figure~\ref{fig:dG-th=000}, the contribution of
$\Rvec$ to the overall Michel spectrum has the effect of slightly counteracting the curvature contribution.

Although this reduction in $\hat{\Gm}(x)$ is overwhelmingly determined by the curvature, the contribution solely due to
the non-inertial dipole operator is nonetheless still apparent in both Figures~\ref{fig:Michel-r0=4e-11} and \ref{fig:Michel-r0=2e-11}.
When $r_0 \sim M = 4 \times 10^{-11}$~cm, the appearance of $\Rvec$ in the Michel spectrum is slight, but still discernable.
For the slightly smaller choice of $r_0 \sim M = 2 \times 10^{-11}$~cm, however, the Michel spectrum with the non-inertial contribution
becomes more pronounced, as shown in Figure~\ref{fig:Michel-r0=2e-11}, with its peak slightly displaced and broadened around $x = 0.80$.
This extreme sensitivity in the profile for $r_0 \sim 10^{-11}$ cm suggests an intriguing possibility of observing a signature due to
$\Rvec$ in the Michel spectrum.
Since this contribution is totally independent of space-time curvature, such a signature may become identifiable for a sufficiently
large data set of muon decay counts, and is worthy of a more detailed examination.

\begin{figure*}
\psfrag{x}[tc][][1.8][0]{\Large $x$}
\psfrag{alpha}[tc][][1.8][0]{\Large $\al$}
\psfrag{dG}[bc][][1.8][90]{\Large ${1 \over \Gm_0} \, \hat{\Gm}(x)$}
\psfrag{dG1}[bc][][1.8][90]{\Large ${1 \over \Gm_0} \, \Dl \hat{\Gm}(x)$}
\begin{minipage}[t]{0.3 \textwidth}
\centering
\subfigure[\hspace{0.2cm} $M = 2 \times 10^{-10}$ cm]{
\label{fig:3d-curv+ni-Michel-r=2e-10}
\rotatebox{0}{\includegraphics[width = 5.0cm, height = 3.5cm, scale = 1]{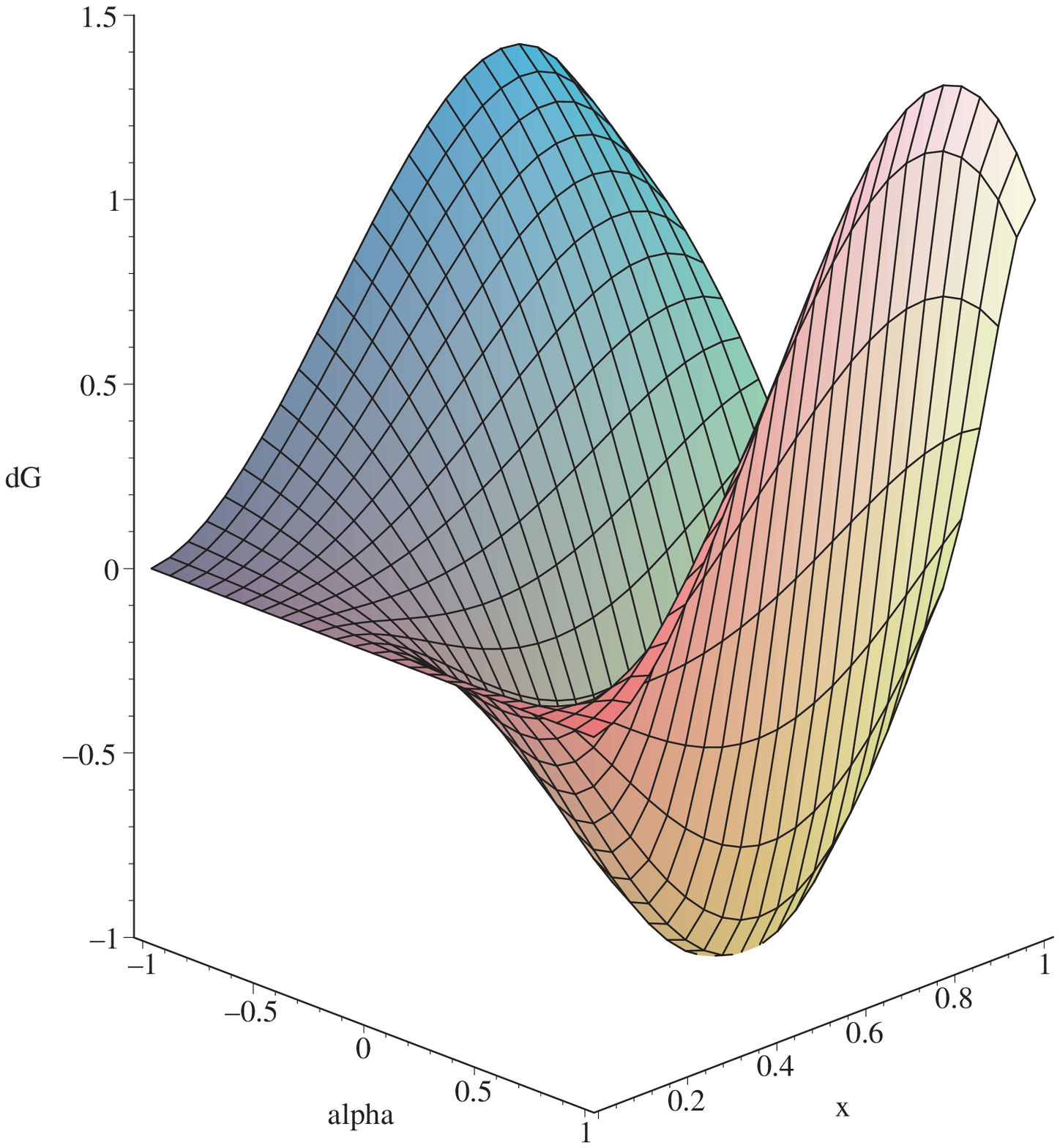}}}
\end{minipage}%
\hspace{0.5cm}
\begin{minipage}[t]{0.3 \textwidth}
\centering
\subfigure[\hspace{0.2cm} $M = 2 \times 10^{-10}$ cm]{
\label{fig:3d-rel-Michel-r=2e-10}
\rotatebox{0}{\includegraphics[width = 5.0cm, height = 3.5cm, scale = 1]{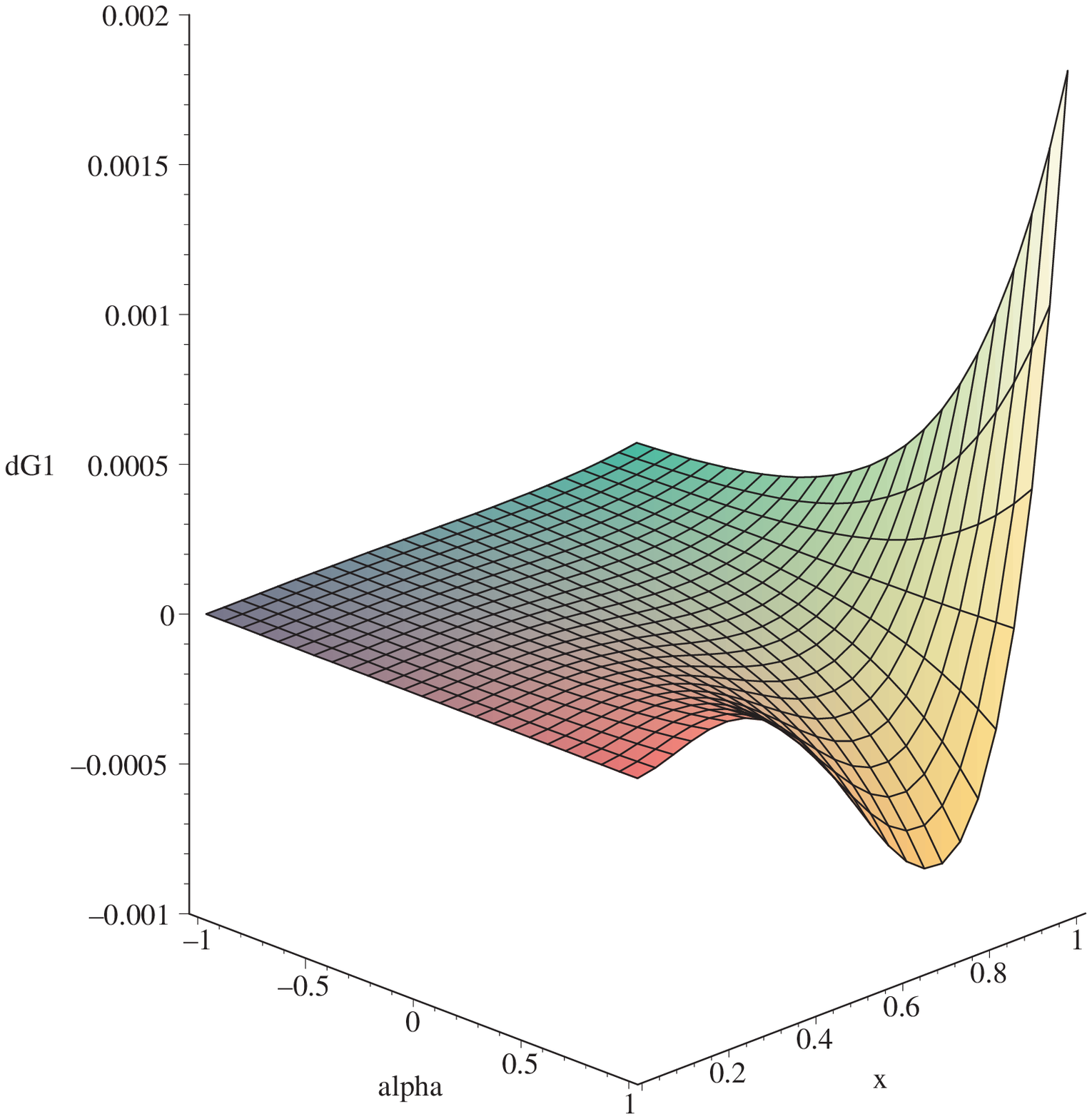}}}
\end{minipage}%
\hspace{0.5cm}
\begin{minipage}[t]{0.3 \textwidth}
\centering
\vspace{-3.5cm}
\caption{\label{fig:3d-Michel-r=2e-10} Three-dimensional plot of the Michel spectrum for varying $\al$ and $x$, with $M = 2 \times 10^{-10}$~cm.
Fig.~\ref{fig:3d-curv+ni-Michel-r=2e-10} starts to develop a minimum value due to curvature and non-inertial contributions,
while Fig.~\ref{fig:3d-rel-Michel-r=2e-10} shows the net contribution after subtracting off the exclusively curvature-dependent
contribution.
\vspace{1mm}}
\end{minipage}
\end{figure*}
\begin{figure*}
\psfrag{x}[tc][][1.8][0]{\Large $x$}
\psfrag{alpha}[tc][][1.8][0]{\Large $\al$}
\psfrag{dG}[bc][][1.8][90]{\Large ${1 \over \Gm_0} \, \hat{\Gm}(x)$}
\psfrag{dG1}[bc][][1.8][90]{\Large ${1 \over \Gm_0} \, \Dl \hat{\Gm}(x)$}
\begin{minipage}[t]{0.3 \textwidth}
\centering
\subfigure[\hspace{0.2cm} $M = 4 \times 10^{-11}$ cm]{
\label{fig:3d-curv+ni-Michel-r=4e-11}
\rotatebox{0}{\includegraphics[width = 5.0cm, height = 3.5cm, scale = 1]{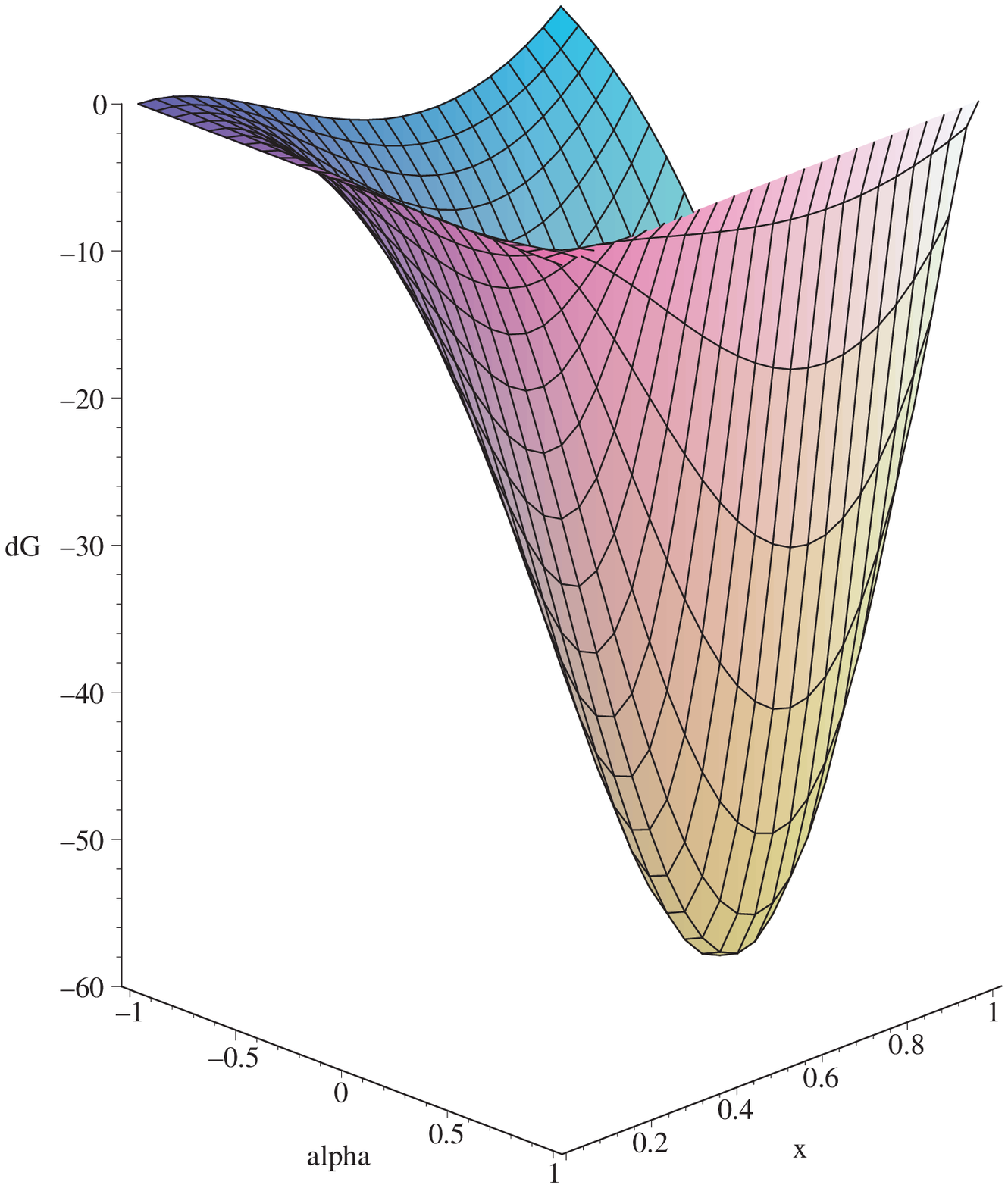}}}
\end{minipage}%
\hspace{0.5cm}
\begin{minipage}[t]{0.3 \textwidth}
\centering
\subfigure[\hspace{0.2cm} $M = 4 \times 10^{-11}$ cm]{
\label{fig:3d-rel-Michel-r=4e-11}
\rotatebox{0}{\includegraphics[width = 5.0cm, height = 3.5cm, scale = 1]{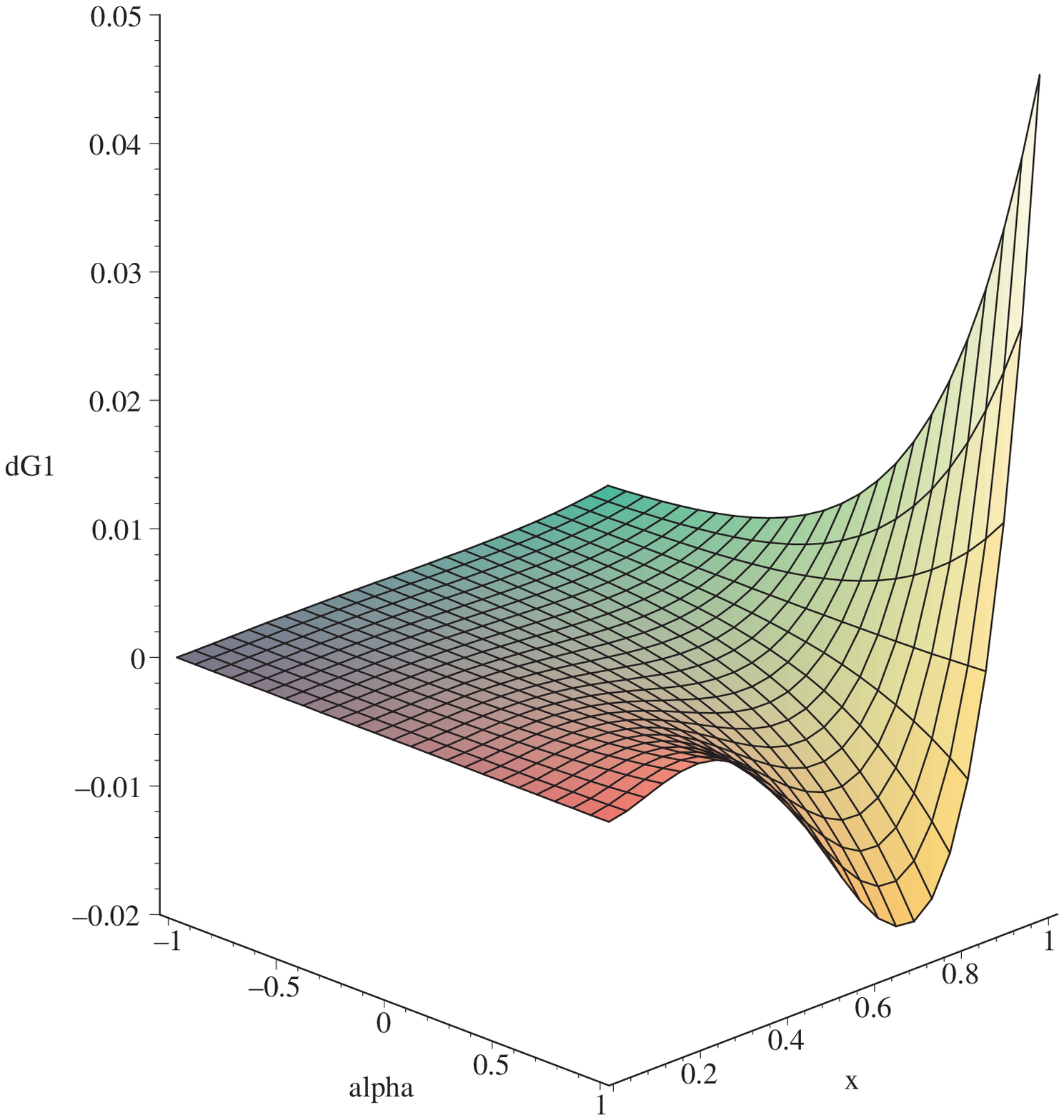}}}
\end{minipage}%
\hspace{0.5cm}
\begin{minipage}[t]{0.3 \textwidth}
\centering
\vspace{-3.5cm}
\caption{\label{fig:3d-Michel-r=4e-11} Three-dimensional plot of the Michel spectrum for varying $\al$ and $x$, with $M = 4 \times 10^{-11}$~cm.
The minimum in Fig.~\ref{fig:3d-curv+ni-Michel-r=4e-11} due to curvature and non-inertial effects becomes more pronounced,
with a correspondingly larger net effect found in Fig.~\ref{fig:3d-rel-Michel-r=4e-11}.
\vspace{1mm}}
\end{minipage}
\end{figure*}
\begin{figure*}
\psfrag{x}[tc][][1.8][0]{\Large $x$}
\psfrag{alpha}[tc][][1.8][0]{\Large $\al$}
\psfrag{dG}[bc][][1.8][90]{\Large ${1 \over \Gm_0} \, \hat{\Gm}(x)$}
\psfrag{dG1}[bc][][1.8][90]{\Large ${1 \over \Gm_0} \, \Dl \hat{\Gm}(x)$}
\begin{minipage}[t]{0.3 \textwidth}
\centering
\subfigure[\hspace{0.2cm} $M = 2 \times 10^{-11}$ cm]{
\label{fig:3d-curv+ni-Michel-r=2e-11}
\rotatebox{0}{\includegraphics[width = 5.0cm, height = 3.5cm, scale = 1]{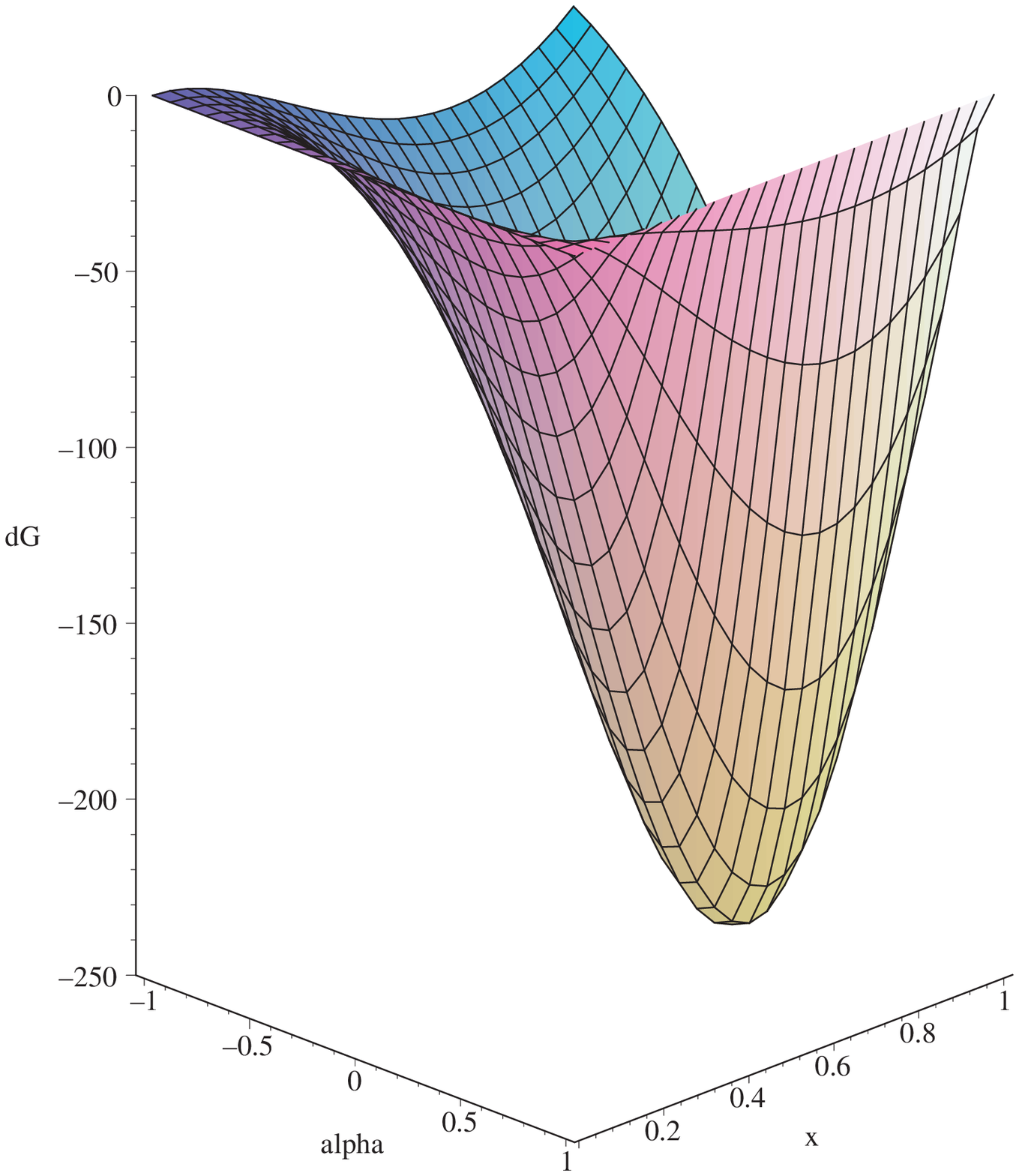}}}
\end{minipage}%
\hspace{0.5cm}
\begin{minipage}[t]{0.3 \textwidth}
\centering
\subfigure[\hspace{0.2cm} $M = 2 \times 10^{-11}$ cm]{
\label{fig:3d-rel-Michel-r=2e-11}
\rotatebox{0}{\includegraphics[width = 5.0cm, height = 3.5cm, scale = 1]{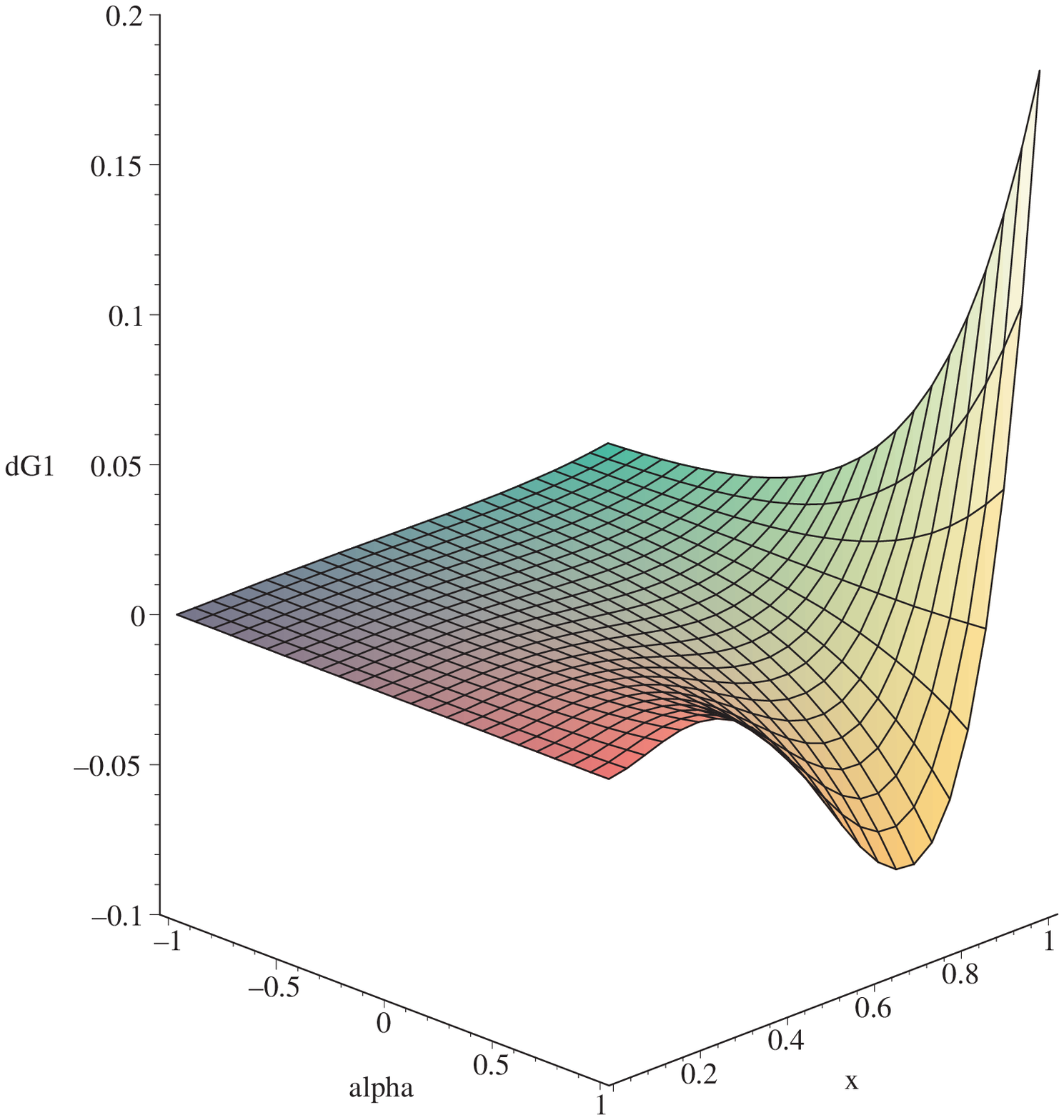}}}
\end{minipage}%
\hspace{0.5cm}
\begin{minipage}[t]{0.3 \textwidth}
\centering
\vspace{-3.5cm}
\caption{\label{fig:3d-Michel-r=2e-11} Three-dimensional plot of the Michel spectrum for varying $\al$ and $x$, with $M = 2 \times 10^{-11}$~cm.
As $M$ becomes progressively smaller, curvature effects on the plot strongly determines the plot's structure and magnitude.
\vspace{1mm}}
\end{minipage}
\end{figure*}

To determine the effect of varying $\al$ on the Michel spectrum, Figures~\ref{fig:3d-Michel-r=2e-10}--\ref{fig:3d-Michel-r=2e-11}
display three-dimensional plots of $\hat{\Gm}(x)$ as functions of $\alpha$ and $x$ for
$M = 2 \times 10^{-10}$ cm, $M = 4 \times 10^{-11}$ cm, and $M = 2 \times 10^{-11}$ cm, respectively.
Figs.~\ref{fig:3d-curv+ni-Michel-r=2e-10}, \ref{fig:3d-curv+ni-Michel-r=4e-11}, and \ref{fig:3d-curv+ni-Michel-r=2e-11}
have the full contributions due to curvature and $\Rvec$, while
Figs.~\ref{fig:3d-rel-Michel-r=2e-10}, \ref{fig:3d-rel-Michel-r=4e-11}, and \ref{fig:3d-rel-Michel-r=2e-11}
reflect the net spectra after subtracting off the contribution due to curvature alone, according to
\be
\Dl \hat{\Gm}(x) & \equiv & \hat{\Gm}(x) - \hat{\Gm}(x)_{\rm Curvature} \, .
\label{Dl-Gm2=}
\ee
Comparison between Fig.~\ref{fig:3d-curv+ni-Michel-r=2e-10} with Fig.~\ref{fig:3d-curv+ni-Michel-r=4e-11} illustrates
the gradual emergence of space-time curvature for $r_0 \sim 2 \times 10^{-10}$ cm, where the first plot just begins to develop a minimum
at around $\al = 0.40$ and $x = 0.65$, while the minimum in the second plot becomes strongly pronounced in the same region of parameter space,
at around $\al = 0.30$ and $x = 0.70$.
This becomes even more evident when they are compared to Fig.~\ref{fig:3d-curv+ni-Michel-r=2e-10}.
From examining Figs.~\ref{fig:3d-rel-Michel-r=2e-10} and \ref{fig:3d-rel-Michel-r=4e-11}, it is apparent that the region of $-1/4 \leq \al \lesssim 0$
is dominated by curvature contributions alone, while the rapid increase
corresponding to $0 \lesssim \al \leq 1$ and large $x$ is due to the mixed coupling of curvature with $\Rvec$.
This trend continues for Figure~\ref{fig:3d-Michel-r=2e-11}, indicating a minimum in Fig.~\ref{fig:3d-curv+ni-Michel-r=2e-11}
whose magnitude becomes four times larger than for Fig.~\ref{fig:3d-curv+ni-Michel-r=4e-11}, with a
similar outcome for Figs.~\ref{fig:3d-rel-Michel-r=2e-10} and \ref{fig:3d-rel-Michel-r=2e-11}.
\begin{figure}
\psfrag{M}[cc][][1.8][0]{\Large $M$ (cm)}
\psfrag{dG}[bc][][1.8][0]{\Large ${1 \over \Gm_0} \, \lt|\Dl \Gm(M) \rt|$}
\begin{minipage}[t]{0.3 \textwidth}
\centering
\subfigure[\hspace{0.2cm} ($N_0 = 1/\sqrt{2} \, , \ \al = 0$)]{
\label{fig:Decay-Rate-N=inv-sqrt-2-al=000}
\rotatebox{0}{\includegraphics[width = 6.0cm, height = 4.0cm, scale = 1]{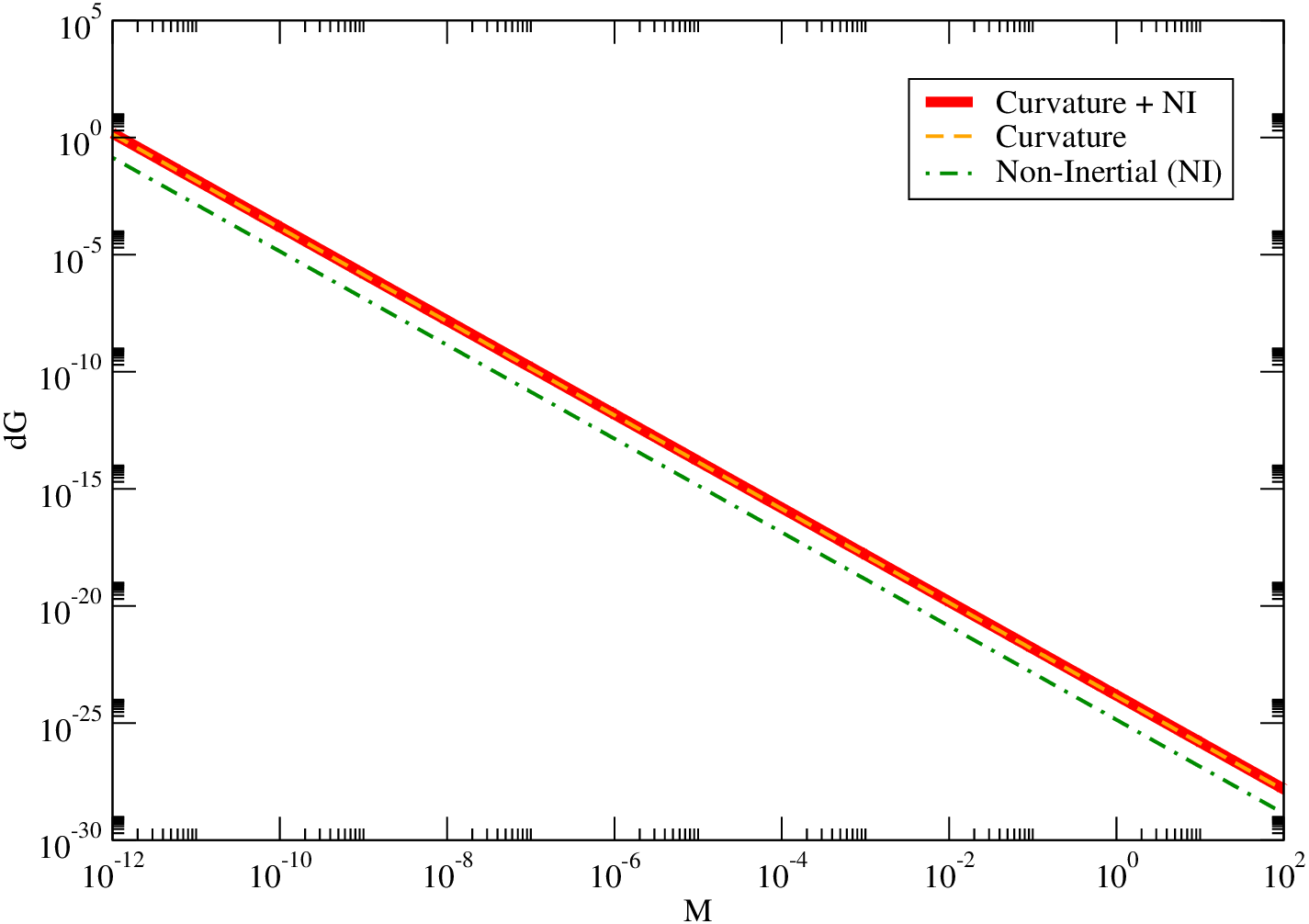}}}
\vspace{5mm}
\end{minipage}
\begin{minipage}[t]{0.3 \textwidth}
\centering
\subfigure[\hspace{0.2cm} ($N_0 = 10^{-2} \, , \ \al = 0$)]{
\label{fig:Decay-Rate-N=1e-2-al=000}
\rotatebox{0}{\includegraphics[width = 6.0cm, height = 4.0cm, scale = 1]{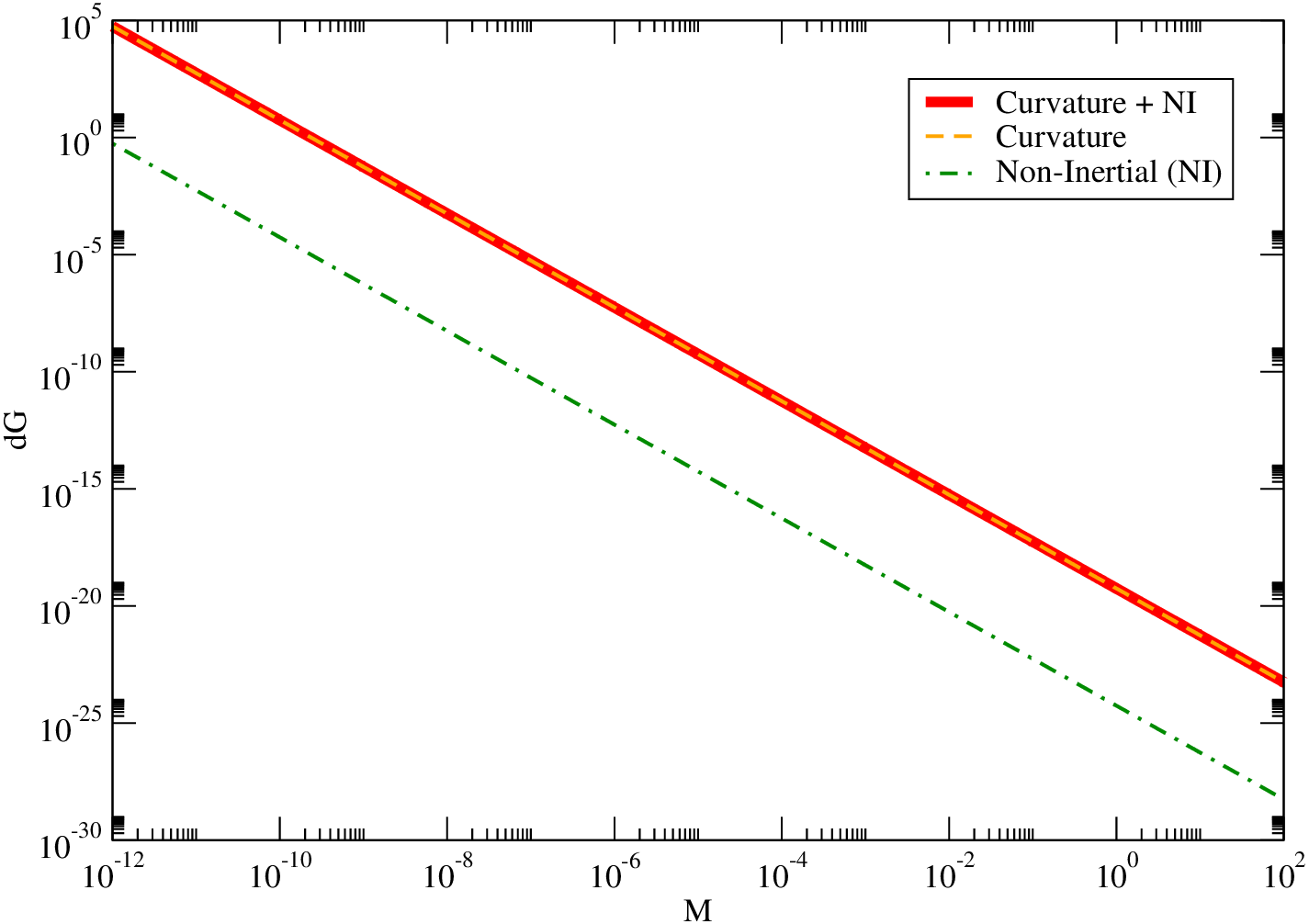}}}
\end{minipage}
\caption{\label{fig:Decay-Rate-al=000} Relative total decay rate as a function of $M$ for a Schwarzschild black hole $(\al = 0)$.
Fig.~\ref{fig:Decay-Rate-N=inv-sqrt-2-al=000} corresponds to $r_0 = 6 \, M$ for $N_0 = 1/\sqrt{2}$, while
Fig.~\ref{fig:Decay-Rate-N=1e-2-al=000} implies that $r_0 \sim 3 \, M$ for $N_0 = 10^{-2}$.
It is evident that curvature strongly enhances the lifetime of the muon as $r_0 \rightarrow 2 \, M$, the Schwarzschild radius,
while the contribution due to $\Rvec$ indicates strong enhancement for $r_0 \sim 10^{-12}$ cm, the Compton wavelength
scale for the muon.}
\end{figure}

\begin{figure}
\psfrag{M}[cc][][1.8][0]{\Large $M$ (cm)}
\psfrag{dG}[bc][][1.8][0]{\Large ${1 \over \Gm_0} \, \lt|\Dl \Gm(M) \rt|$}
\begin{minipage}[t]{0.3 \textwidth}
\centering
\subfigure[\hspace{0.2cm} ($N_0 = 10^{-2} \, , \ \al = 1$)]{
\label{fig:Decay-Rate-al=+100}
\rotatebox{0}{\includegraphics[width = 6.0cm, height = 4.0cm, scale = 1]{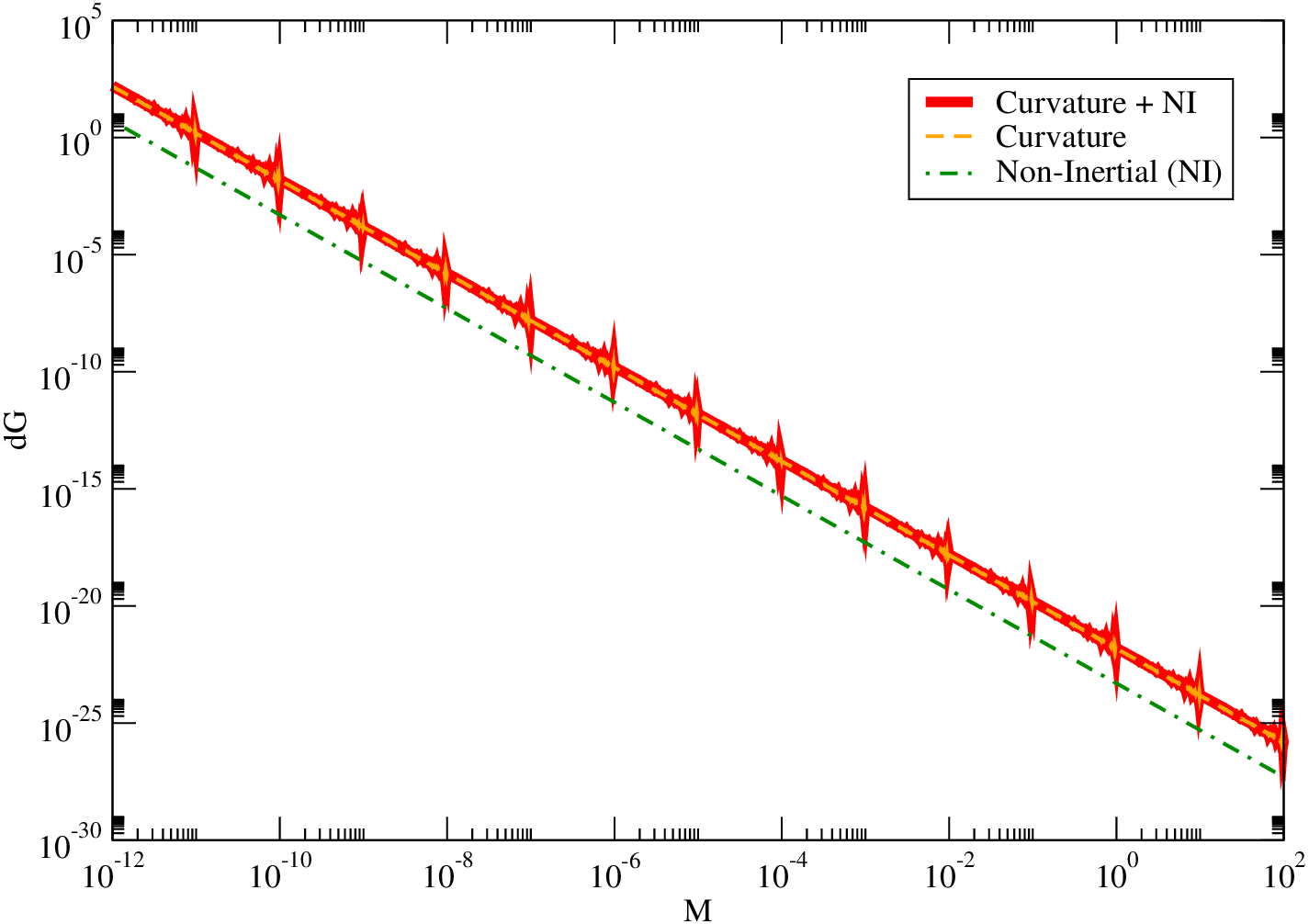}}}
\vspace{5mm}
\end{minipage}
\begin{minipage}[t]{0.3 \textwidth}
\centering
\subfigure[\hspace{0.2cm} ($N_0 = 10^{-2} \, , \ \al = -1/4$)]{
\label{fig:Decay-Rate-al=-025}
\rotatebox{0}{\includegraphics[width = 6.0cm, height = 4.0cm, scale = 1]{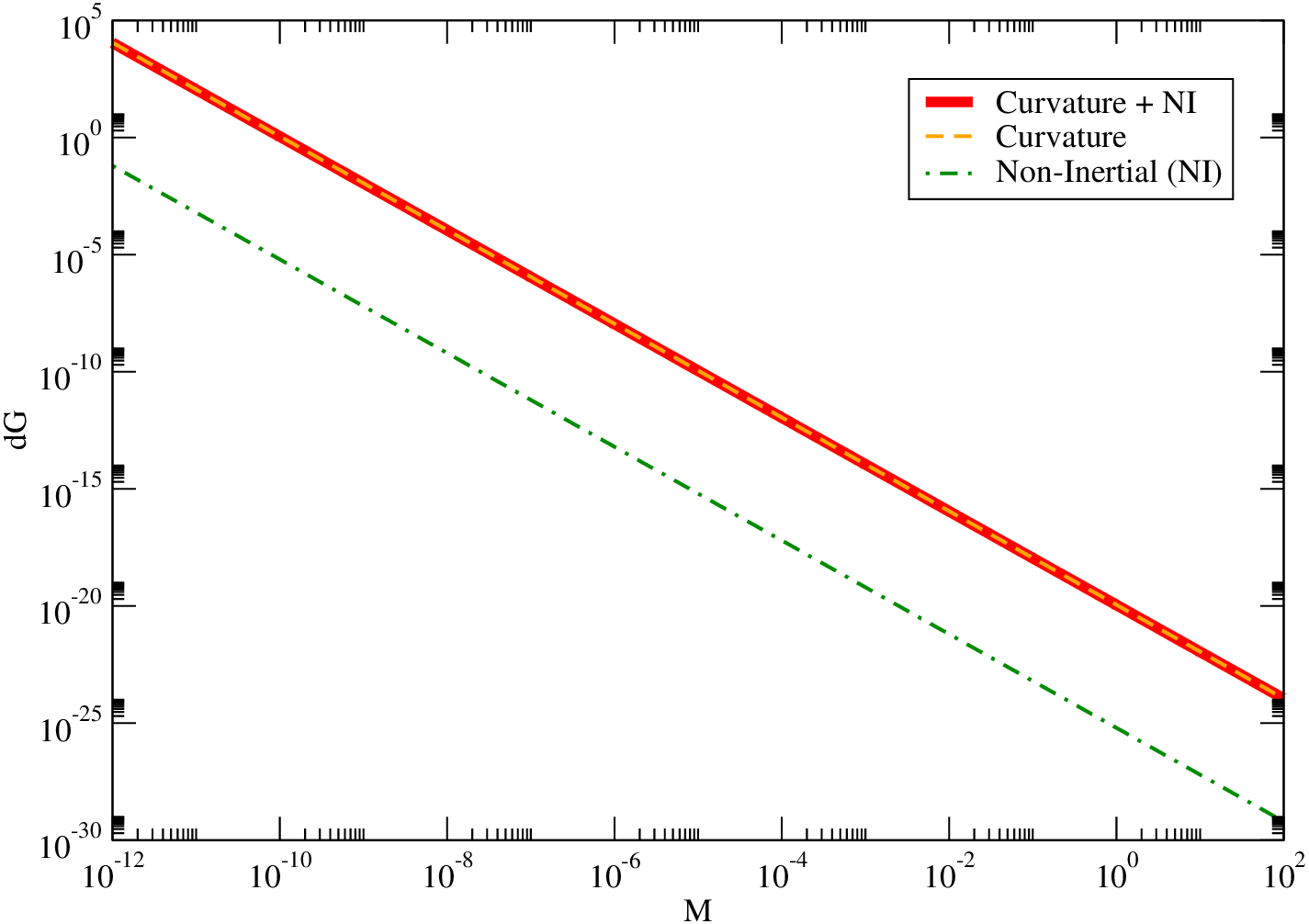}}}
\end{minipage}
\caption{\label{fig:Decay-Rate-al=extreme} Relative total decay rate as a function of $M$ for an extreme Kerr black hole.
Fig.~\ref{fig:Decay-Rate-al=+100} denotes $r_0 \sim M$ for a co-rotating black hole $(\al = 1)$ with respect to the muon's orbit, while
Fig.~\ref{fig:Decay-Rate-al=-025} refers to $r_0 \sim 4 \, M$ for a counter-rotating black hole $(\al = -1/4)$.
The spikes on the main curve shown in Fig.~\ref{fig:Decay-Rate-al=+100}, along with the apparent upward translation of the
curve due to $\Rvec$, are numerical artifacts that should be discounted.}
\end{figure}

A final integration of (\ref{Michel-defn}) over $x$ leads to the muon's total decay rate $\Gamma$.
Since $M$ sets the energy scale for determining the curvature and non-inertial contributions to the decay rate,
$\Gamma$ can be expressed in terms of
\be
\Dl \Gm(M) & \equiv & \Gm(M) - \Gm_0 \, .
\label{Dl-Gm(M)=}
\ee
It is very interesting to report, based on Figures~\ref{fig:Decay-Rate-al=000} and \ref{fig:Decay-Rate-al=extreme}, that
(\ref{Dl-Gm(M)=}) is {\em negative-valued}, which indicates that the muon's lifetime is enhanced due to curvature
and non-inertial effects due to $\Rvec$.
This agrees with the main result presented earlier \cite{Singh-Mobed}, following an approximated approach in terms of
cylindrical co-ordinates, but now determined from a more careful treatment of the problem that includes
contributions due to curved space-time.

Figure~\ref{fig:Decay-Rate-al=000} displays $|\Dl \Gm(M)|/\Gm_0$ for a Schwarzschild black hole $(\al = 0)$, where
in Fig.~\ref{fig:Decay-Rate-N=inv-sqrt-2-al=000}, $N_0 = 1/\sqrt{2}$ is chosen in accordance with (\ref{r0=}) to obtain
$r_0 = 6 \, M$, the innermost stable circular orbit \cite{Chandra}.
For comparison, Fig.~\ref{fig:Decay-Rate-N=1e-2-al=000} is the corresponding expression in terms of
$r_0 \sim 3 \, M$, the (unstable) photon orbit \cite{Chandra}.
It is self-evident that the total decay rate is very sensitive to changes in space-time curvature as $r_0 \rightarrow 2 \, M$,
the Schwarzschild radius, given that the total contribution shown in Fig.~\ref{fig:Decay-Rate-N=1e-2-al=000}
is five orders of magnitude larger than for its corresponding curve in Fig.~\ref{fig:Decay-Rate-N=inv-sqrt-2-al=000}.
Furthermore, it is predicted that the decay rate goes to zero for $r_0 \sim 10^{-9}$~cm, the atomic scale.
In the absence of curvature, both plots indicate that the contribution to $|\Dl \Gm(M)|/\Gm_0$ due to the non-inertial dipole operator
leads to a strong enhancement of the muon's lifetime at around its Compton wavelength of $r_0 = 1.18 \times 10^{-12}$ cm.

The case of an extreme Kerr black hole is presented in Figure~\ref{fig:Decay-Rate-al=extreme}, where
comparison is made between a co-rotating $(\al = 1)$ and a counter-rotating extreme black hole $(\al = -1/4)$, as shown
by Figs.~\ref{fig:Decay-Rate-al=+100} and \ref{fig:Decay-Rate-al=-025}, respectively.
In this instance, it is clear that the dominant contributions in the muon's lifetime enhancement are due to
curvature components dependent on odd functions of $\al$, especially since $r_0 \sim 4 \, M$ in Fig.~\ref{fig:Decay-Rate-al=-025} compared to
both $r_0 \sim M$ in Fig.~\ref{fig:Decay-Rate-al=+100} and $r_0 = 6 \, M$ in Fig.~\ref{fig:Decay-Rate-N=inv-sqrt-2-al=000}.
A possible physical explanation is that the muon is extracting energy from the extra inertia of space-time it must overcome
to orbit a black hole with $a < 0$, which then serves to reduce its decay rate.
While it is too early to make any definitive statements at present, this outcome roughly coincides with the Penrose process for
classical particles \cite{Chandra}.
However, it should be noted that this result also challenges the established understanding that super-radiance, the
phenomenon analogous to the Penrose process for quantum fields in a Kerr background, {\em does not} occur for
spin-1/2 fields \cite{Chandra}.
It is entirely possible that the machinery employed for the differential cross-section given by (\ref{diff-cross-section=}) may
be inadequate when applied to muons within a Compton wavelength of the black hole event horizon.
If plane wave solutions do not correctly approximate the asymptotic states, then appropriate measures are required
to properly address this technical challenge, which may then resolve the apparent mismatch between the predictions in
Figures~\ref{fig:Decay-Rate-al=000} and \ref{fig:Decay-Rate-al=extreme} and the lack of spin-1/2 field superradiance effects.
In addition, it is certainly reasonable to envision other non-trivial effects emerging at the Compton wavelength scale
unanticipated by this approach that would have to be investigated thoroughly.

Irrespective of this matter, the results in Figures~\ref{fig:Decay-Rate-al=000} and \ref{fig:Decay-Rate-al=extreme}
suggest that the muon's total decay rate becomes {\em negative-valued} for sufficiently small $r_0$, already identified
as a matter of concern in a previous paper \cite{Singh-Mobed} when considering non-inertial effects alone.
It is very clear that curvature effects heighten this possibility rather than mitigate against it, and while
electromagnetic bremmstrahlung radiation may have a role in avoiding this possibility for muon decay,
this would likely have little to no bearing on the behaviour of unstable particles with no electrical charge.
Though unresolved issues still need to be thoroughly examined first, this may be further evidence of new physical phenomena
emerging at the intersection between quantum mechanics and general relativity \cite{Singh-Mobed}.

\section{Potential Contributions due to Noncommutative Geometry}
\label{sec:noncommutative-geometry}

One recent development in quantum gravity research that has gained a considerable following is
noncommutative geometry \cite{Connes}, which proposes that space-time co-ordinates
no longer commute at some scale $\kbar$ with dimensions of area, but rather satisfy 
\be
\lt[\Xvec^\mu , \Xvec^\nu\rt] & = & i \kbar \, \Jvec^{\mu \nu}(\Xvec^\al) \, ,
\label{noncomm}
\ee
where the $\Xvec^\mu$ are Hermitian co-ordinate operators and $\Jvec^{\mu \nu}$ is a dimensionless antisymmetric tensor.
In fact, this idea was proposed several decades ago \cite{Snyder}, but in the context of finding
a mechanism to truncate ultraviolet divergences in quantum field theory prior to the
advent of renormalization and regularization.
The $\Jvec^{\mu \nu}$ then resemble the generators of orbital angular momentum satisfying an SU(2) Lie algebra \cite{Chaichian0}.

While it is certainly logical to propose this route for determining quantum gravity,
its viability compared to the other approaches can be confirmed only if it produces
an unambiguous signature in relation to an identifiable background measurement.
Based on present knowledge, it seems that $\kbar$ can be treated as a free parameter,
whose magnitude may be identified empirically.
However, a recent paper based on noncommutative QED suggests an upper bound of
$\sqrt{\kbar} \ \sim \ 10^{-18}$~cm \cite{Chaichian}.
For the purposes of this paper, the question of whether $\kbar$ can be derived from first principles
is a matter of speculation not addressed here.

It so happens that the formalism of this paper allows for the possibility of finding nontrivial contributions
due to noncommutative geometry that would otherwise be set to zero if space-time co-ordinates always commute.
The possibility of this identification is based on the following consideration of the metric tensor in Fermi normal co-ordinates.
According to (\ref{F-g}), a symmetric metric tensor ${}^F{}g_{\mu \nu}(X)$ implies that the Riemann tensor is
also symmetric in the middle two indices.
However, this symmetry {\em is not} an inherent property of the Riemann tensor, but rather imposed upon due to the contraction of
$\d X^\mu \, \d X^\nu$ in (\ref{ds^2-symm}).

This is a crucial observation because, by relaxing this requirement for ${}^F{}g_{\mu \nu}(X)$, it allows for the possibility of the
{\em antisymmetric} combination ${}^F{}R_{\mu [\al \bt] \nu}(T) =
{1 \over 2} \lt[{}^F{}R_{\mu \al \bt \nu}(T) - {}^F{}R_{\mu \bt \al \nu}(T)\rt]$ to appear, such that terms of the form
\be
{}^F{}R_{\mu [\al \bt] \nu}(T) \, X^\mu \, X^\nu & = & -{1 \over 4} \, {}^F{}R_{\al \bt \mu \nu}(T) \lt[X^\mu , X^\nu\rt] \qquad
\label{Riemann-antisymmetric}
\ee
are identifiable.
If the $X^\mu$ are ordinary c-numbers, then (\ref{Riemann-antisymmetric}) automatically vanishes,
but if $X^\mu \rightarrow \Xvec^\mu$, then (\ref{Riemann-antisymmetric}) is nonzero according to (\ref{noncomm})
and needs to be retained.
It then follows naturally that
\be
\d s^2 & \rightarrow &
{1 \over 2} \, {}^F{}g_{(\mu \nu)}(\Xvec) \lt( \d \Xvec^\mu \otimes \d \Xvec^\nu + \d \Xvec^\nu \otimes \d \Xvec^\mu \rt)
\nn
&  &{} + {1 \over 2} \, {}^F{}g_{[\mu \nu]}(\Xvec) \, \d \Xvec^\mu \wedge \d \Xvec^\nu \, ,
\label{ds^2-asymm}
\ee
%
where the antisymmetric part of (\ref{ds^2-asymm}) emerges at the $\kbar$-scale only.


From closer inspection of the second terms in (\ref{Spin-Connection-j-S}) and (\ref{Spin-Connection-0-T}),
respectively, it becomes evident that contributions due to noncommutative geometry emerge
at the $\kbar$-scale, since it follows from invoking (\ref{noncomm}) and (\ref{Riemann-antisymmetric})
that
\begin{subequations}
\label{spin-connection-noncomm}
\be
\bar{\Gmvec}^{(\rm S)}_{\hat{\jmath}} & = & - {1 \over 2} {}^F{}R_{j00m}(T) \, X^m + {1 \over 3} {}^F{}R_{l(j0)m,0}(T) \, X^l \, X^m
\nn
&  &{} - {i \kbar \over 12} \, {}^F{}R_{j0lm,0}(T) \, \Jvec^{lm} \, , \qquad
\label{Spin-Connection-j-S-noncomm}
\nl
\bar{\Gmvec}^{(\rm T)}_{\ze[\hat{l}\hat{m}]} & = &
{1 \over 2} \, {}^F{}R_{lm0k}(T) \, X^k + {i \kbar \over 48} \, {}^F{}R_{lmjk,0}(T) \, \Jvec^{jk} \, . \qquad \hspace{2mm}
\label{Spin-Connection-0-T-noncomm}
\ee
\end{subequations}

Given (\ref{Spin-Connection-j-S-noncomm}) and (\ref{Spin-Connection-0-T-noncomm}), it becomes obvious that
the presence of noncommutative geometry formally appears in both the Casimir spin scalar $\Wvec^{\hat{\al}} \, \Wvec_{\hat{\al}}$
and the polarization vector for muon decay.
However, when applied to the specific example of muon decay near a Kerr black hole, it turns out that
the anticipated contributions {\em vanish} in the time average over a complete cycle.
This outcome is very surprising, with no clear explanation on a conceptual level for why this occurs.
%

One clue that may be responsible for the null result 
comes from the azimuthal symmetry of the Kerr background.
This suggests a possibility for extracting the noncommutative part by introducing an asymmetric perturbation in the orbital
plane, such as an electromagnetic potential along the $x$-axis in Figure~\ref{fig:black-hole-orbit}, for example.
By then solving the Einstein-Maxwell field equations for a perturbed orthonormal tetrad $\lm^{\bar{\mu}}{}_\al$
that is biased for some value of $\phi_0$, the integration over quantum fluctuations should then yield
$\kbar$-dependent contributions when time averaged over a complete cycle.


\section{Discussion}
\label{sec:discussion}

Given the considerable detail shown for identifying curvature and non-inertial contributions to the muon decay spectrum
in a microscopic Kerr background, it is nonetheless appropriate to critically assess the likelihood for observation of these phenomena
within a realistic setting.
It is widely accepted that any formation of microscopic black holes will rapidly and spontaneously decay
into a spectrum of massless particles as Hawking radiation \cite{Hawking,Page}.
Furthermore, current scenarios predicting microscopic black hole production at the Large Hadron Collider (LHC) \cite{Dimopoulos}
are critically dependent on the existence of both Hawking radiation and large extra dimensions in order to
attain a signature at the TeV energy scale.
It is debatable whether the intensity of collisions performed at the LHC can achieve a high enough energy density
for quantum mechanical gravitational collapse to occur, or whether this process can happen at all
due to unforeseen modifications of general relativity at sufficiently small length scales.
For these and other reasons, it is difficult to ascertain whether the analysis presented here, as it concerns the curvature effects,
has any value beyond purely theoretical considerations.

This question becomes more interesting, however, when considering the observation of the non-inertial dipole operator $\Rvec$ within muon decay,
since this a contribution that is independent of space-time curvature and purely a byproduct of non-inertial motion due to rotation.
As noted earlier \cite{Singh-Mobed}, the strength of $\Rvec$ becomes large when the local radius of curvature for the muon's orbit
approaches the Compton wavelength scale.
While it is not realistic to test for this effect using circulating muon beams in macroscopic storage rings,
for muonic atoms the likelihood becomes more promising, since they already exist and are relatively long-lived objects.
In these respects, the test for non-inertial effects in muon decay within this context becomes a realistic possibility
and worthy of a more detailed examination.

The extension of this analysis to incorporate more sophisticated details is certainly possible.
For example, it is already shown that electromagnetic interactions can be introduced in a straightforward
and natural way into the formalism.
The possibility of curvature-induced pair production is definitely an important factor to consider,
as well as back reaction effects due to the self-energy of the decaying muon.
For the case of classical particles propagating in curved space-time \cite{Poisson}, this is an area of
active research, so due consideration for quantum particles must also be taken into account for a more sophisticated treatment
of the problem.
Furthermore, this overall approach is not necessarily limited to leptons, and it so should be possible to apply
this formalism to an accelerated neutron undergoing beta decay, for example.
However, the possibility of identifying the curvature and non-inertial contributions likely becomes much more difficult,
due to the finite size of the neutron and inherent complexity of hadronic processes that would effectively mask any signature
that may result from the decay process.
As well, the formalism is not limited to spin-1/2 particles, as the Casimir scalar properties for higher-order spin systems
can be examined systematically \cite{Singh-Mobed1}, with particularly interesting applications for the spin-3/2 gravitino in supersymmetry
and the spin-2 graviton.

One issue not examined here in detail but needs to be acknowledged is the presence of frame-based singularities
due to performing the computations in the muon's local rest frame.
Specifically, there exist terms in the differential cross section that grow singular in the limit as
$|\Pvec| = \sqrt{-\Pvec^{\hat{\jmath}} \, \Pvec_{\hat{\jmath}}} \rightarrow 0$.
However, it is reassuring to note that all such terms are coupled to the gravitational field and so they make no contribution in the
absence of space-time curvature.
In addition, there exist non-zero finite terms in the limit of $|\Pvec| \rightarrow 0$
that are also coupled to the gravitational field, along with a second set of non-zero finite terms that come exclusively from $\Rvec$.
If the weak equivalence principle is to be satisfied in the rest frame of the muon, then it is reasonable
to surmise that the singular terms and the non-zero finite terms coupled to the gravitational field must somehow cancel
each other.
Specifically, the relevant expression for consideration is
\be
\lim_{|\Pvec| \rightarrow 0} G \lt({\lm_0 \over |\Pvec|} \, f_0 + \lm_1 \, f_1 \rt) & \equiv & K \, ,
\label{frame-singularity}
\ee
where $G$ is the (observed) universal gravitational constant, $f_i$ are numerical coefficients associated with the
singular and non-zero finite terms, $\lm_i$ are their respective coupling parameters, and $K$ is some parameter in units of $M/r_0$.
The value of $K$ may possibly be determined empirically by comparison to observation, but must equal zero to satisfy the weak equivalence principle.
With $K = 0$ assumed throughout this analysis, all plots presented are well behaved without any indication of
numerical difficulties associated with this assumption.
It, therefore, seems possible that (\ref{frame-singularity}) is indicative of a relationship between the coupling parameters $\lm_0$
and $\lm_1$ that is analogous to the renormalization flow concept for running coupling constants as functions of energy.
This is also a matter worthy of further study at a later date.

\section{Conclusion}
\label{sec:conclusion}
This paper contains a detailed study of the Pauli-Lubanski spin vector $\Wvec$
for spin-1/2 particles when applied to non-inertial motion in curved space-time, as described by Fermi normal co-ordinates.
It confirms an earlier result obtained in flat space-time \cite{Singh-Mobed}, which shows the existence of the
non-inertial dipole operator $\Rvec$, that leads to a predicted quantum violation of Lorentz invariance for
$\Wvec^{\hat{\al}} \, \Wvec_{\hat{\al}}$, the Casimir scalar for spin.
When the Pauli-Lubanski vector is applied to the study of muon decay in a gravitational background,
it follows that curvature and non-inertial corrections become present in the differential decay rate,
that has theoretically observable consequences when applied to decays near the event horizon of a microscopic Kerr black hole.
This paper also puts forward the possibility that the formalism can be extended to incorporate noncommutative geometry
upon considering a nonsymmetric space-time metric in the Fermi frame.
Interestingly, while contributions due to noncommutative geometry formally exist in muon decay, they integrate
to zero for the specific case of circular motion in a Kerr background.
The reasons for this outcome are unknown at present.

It seems apparent that the results presented here have potentially broad implications for identifying
possible directions for future quantum gravity research.
In particular, if the predictions of non-inertial effects in muon decay due to $\Rvec$ become verified,
then its presence must be properly accounted for in any future developments towards attaining a self-consistent
theory of quantum gravity, whose full range of consequences are yet to be determined.
Nonetheless, potentially outstanding issues like the applicability of plane wave asymptotic states
in the decay rate calculation, for example, as well as other unanticipated effects at the Compton wavelength
scale will have to be addressed in greater detail before a more definitive conclusion can be stated.


\appendix
\section{Derivation of the Momentum Commutator}
\label{sec:momentum-comm}

The derivation of (\ref{P-commutator}) begins with (\ref{Pvec})--(\ref{Ovec}), leading to
\be
\lefteqn{i \lt[\Pvec_{\hat{\al}} , \Pvec_{\hat{\bt}}\rt] \ = \ i \lt\{
(i \hbar)^2 \, \lt[\nabvec_{\hat{\al}} , \nabvec_{\hat{\bt}} \rt]
+ 2 i \hbar \, \nabvec_{[\hat{\al}} \, \Om_{\hat{\bt}]} \rt\} }
\nn
&& = \ (i \hbar)^2 \lt\{i \lt[\nabvec_{\hat{\al}} , \nabvec_{\hat{\bt}} \rt]
\lt[1 +  \ln \lt(\lm^{\hat{1}}(u) \, \lm^{\hat{2}}(u) \, \lm^{\hat{3}}(u)\rt)^{1/2} \rt] \rt\} \, .
\nn
\label{A1}
\ee
By using the orthonormal vierbein set $\lt\{e^\sg{}_{\hat{\al}} \rt\}$ to describe
$\nabvec_{\hat{\al}} = e^\sg{}_{\hat{\al}} \, \nabla_\sg$ in terms of Fermi normal co-ordinates,
it follows that
\be
\lt[\nabvec_{\hat{\al}} , \nabvec_{\hat{\bt}} \rt] & = &
- 2 \lt(\nabla_\lm \, e^\sg{}_{[\hat{\al}} \rt) \, e^\lm{}_{\hat{\bt}]} \, \nabla_\sg \, ,
\label{A2}
\ee
where
\be
\lefteqn{ \lt(\nabla_\lm \, e^\sg{}_{[\hat{\al}} \rt) \, e^\lm{}_{\hat{\bt}]} \, \nabla_\sg
\ = \ \lt({\partial \over \partial U^\lm} \, \bar{e}^\gm{}_{[\hat{\al}} \rt) e^\lm{}_{\hat{\bt}]}
\lt({\partial U^\sg \over \partial X^\gm} \, {\partial \over \partial U^\sg}\rt)} \qquad \qquad
\nn
&&{} + \bar{e}^\gm{}_{[\hat{\al}} \, \bar{e}^\lm{}_{\hat{\bt}]} \,
\lt[{\partial \over \partial X^\lm} \lt(\partial U^\sg \over \partial X^\gm\rt)\rt] {\partial \over \partial U^\sg} \, .
\qquad \qquad
\label{A3}
\ee
The first term of (\ref{A3}) corresponds to $\hbar \, C^{\hat{\mu}}{}_{\hat{\al}\hat{\bt}} \, \Pvec_{\hat{\mu}}$
in (\ref{P-commutator}), while the second term contains the non-inertial contribution and deserves closer inspection.

The key to this derivation lies in the identification of
\be
{\partial \over \partial X^\gm} & = & {\partial U^\sg \over \partial X^\gm} \, {\partial \over \partial U^\sg}
\label{A4}
\ee
with
\be
\nabvec_{\hat{\gm}} & = & {1 \over \lm^{(\gm)}} \, {\partial \over \partial U^\gm} \, ,
\label{A5}
\ee
which is always possible by a suitable rotation of the orthonormal frame.
If the Fermi normal co-ordinates are strictly Cartesian, it is self-evident that the mixed partial
derivatives in the second term of (\ref{A3}) will commute, leading to a net contribution of zero when contracted by
the antisymmetric vierbein combination $\bar{e}^\gm{}_{[\hat{\al}} \, \bar{e}^\lm{}_{\hat{\bt}]}$.
However, this is no longer true when using general curvilinear co-ordinates, since
the second term of (\ref{A3}) described with respect to (\ref{A5}) results in
\be
\lt[{\partial \over \partial X^\lm}
\lt(\partial U^\sg \over \partial X^\gm\rt)\rt] {\partial \over \partial U^\sg}
& \rightarrow & \lt[\nabvec_{\hat{\lm}} \, \lt(\lm^{(\gm)}\rt)^{-1}\rt] \, {\partial \over \partial U^\gm}
\nn
& = & - \lt[\nabvec_{\hat{\lm}} \, \ln \, \lm^{(\gm)}\rt] \nabvec_{\hat{\gm}} \, ,
\label{A6}
\ee
which is generally nonzero.
It follows that substitution of (\ref{A2}), (\ref{A3}), and (\ref{A6}) into (\ref{A1}) leads to (\ref{P-commutator}).


\section{Cross Section Plots for Various Scattering Angles}
\label{sec:diff-cross-section-plots}

\begin{figure*}
\psfrag{x}[tc][][1.8][0]{\Large $x$}
\psfrag{dG}[bc][][1.8][0]{\Large ${1 \over \Gm_0} \, \hat{\Gm}\lt(x,\cos \Th\rt)$}
\begin{minipage}[t]{0.3 \textwidth}
\centering
\subfigure[\hspace{0.2cm} ($\Th = 120^\circ \, , \ \al = 1$)]{
\label{fig:dG-th=120-al=100}
\rotatebox{0}{\includegraphics[width = 6.6cm, height = 5.0cm, scale = 1]{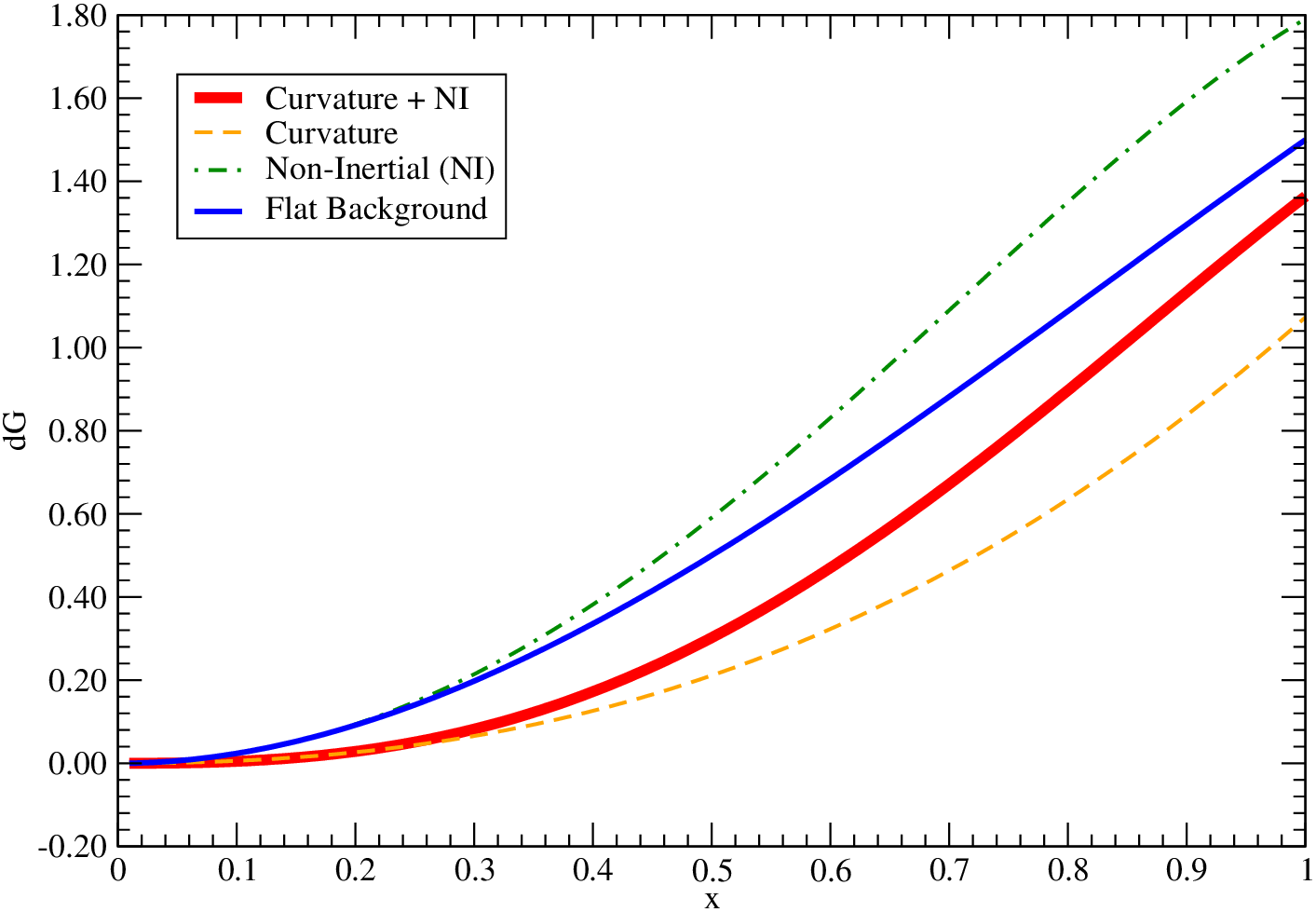}}}
\end{minipage}%
\hspace{2.0cm}
\begin{minipage}[t]{0.3 \textwidth}
\centering
\subfigure[\hspace{0.2cm} ($\Th = 120^\circ \, , \ \al = 0.99$)]{
\label{fig:dG-th=120-al=099}
\rotatebox{0}{\includegraphics[width = 6.6cm, height = 5.0cm, scale = 1]{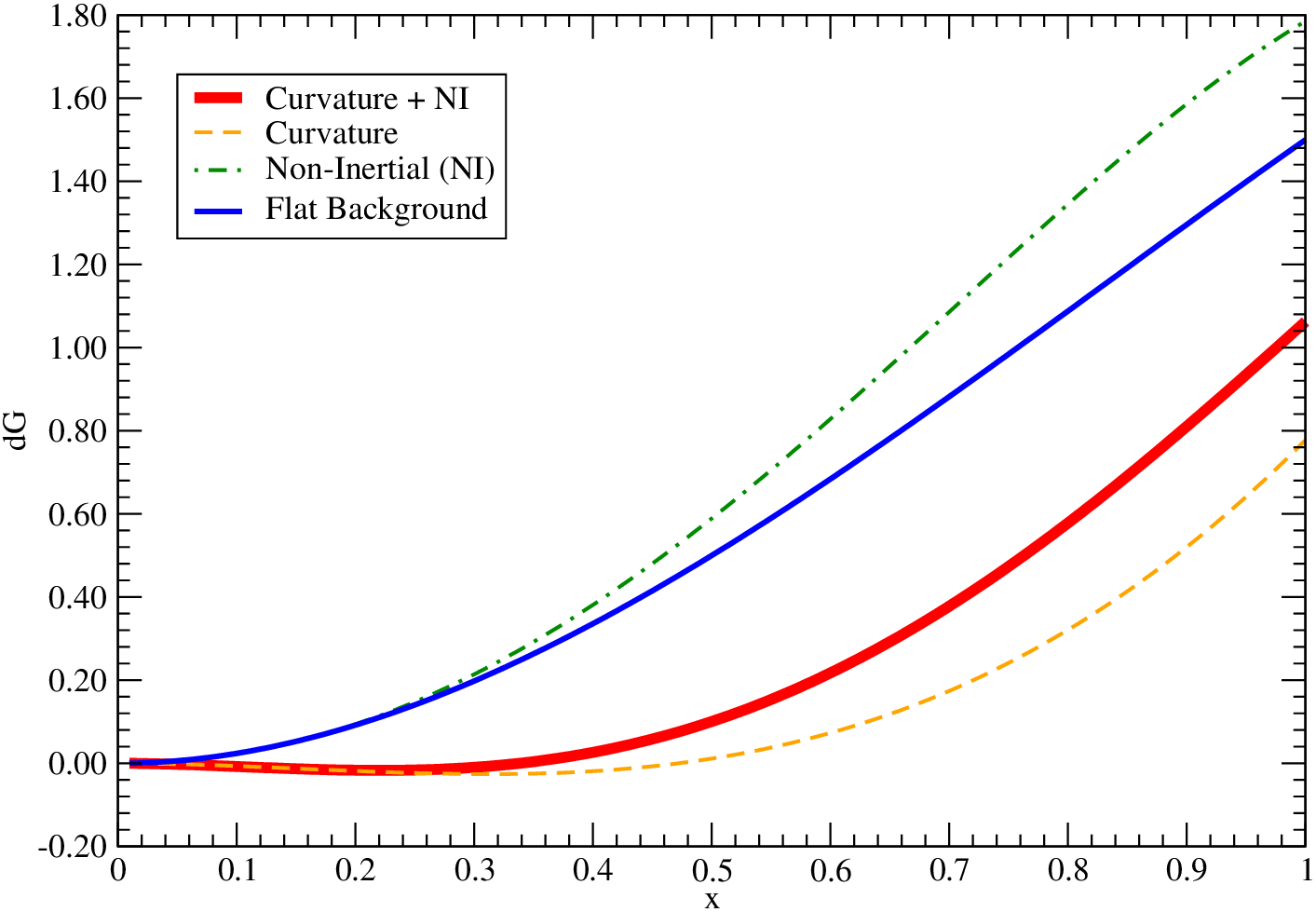}}}
\end{minipage} \\
\vspace{0.8cm}
\begin{minipage}[t]{0.3 \textwidth}
\centering
\subfigure[\hspace{0.2cm} ($\Th = 150^\circ \, , \ \al = 1$)]{
\label{fig:dG-th=150-al=100}
\rotatebox{0}{\includegraphics[width = 6.6cm, height = 5.0cm, scale = 1]{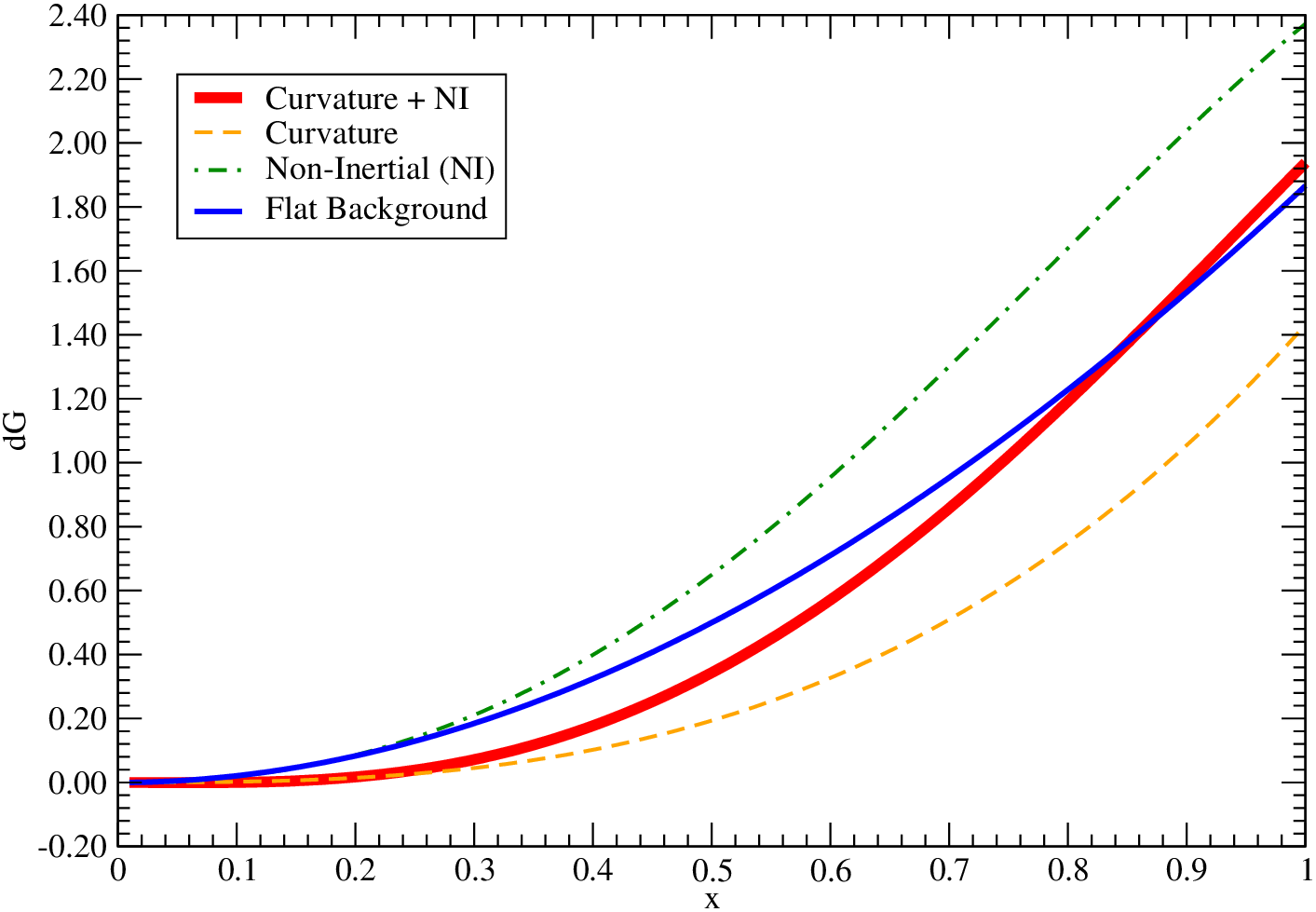}}}
\end{minipage}%
\hspace{2.0cm}
\begin{minipage}[t]{0.3 \textwidth}
\centering
\subfigure[\hspace{0.2cm} ($\Th = 150^\circ \, , \ \al = 0.99$)]{
\label{fig:dG-th=150-al=099}
\rotatebox{0}{\includegraphics[width = 6.6cm, height = 5.0cm, scale = 1]{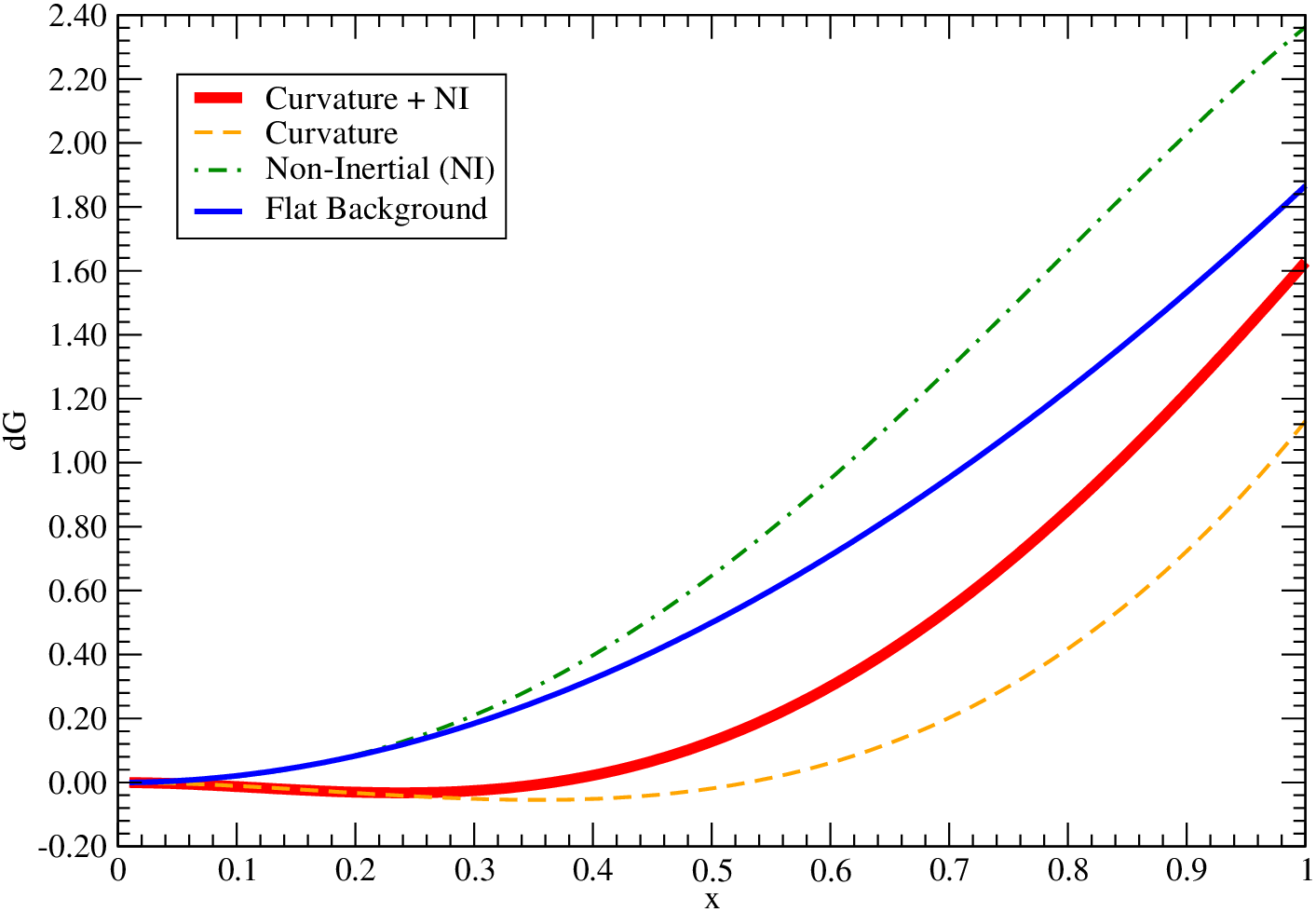}}}
\end{minipage} \\
\vspace{0.8cm}
\begin{minipage}[t]{0.3 \textwidth}
\centering
\subfigure[\hspace{0.2cm} ($\Th = 180^\circ \, , \ \al = 1$)]{
\label{fig:dG-th=180-al=100}
\rotatebox{0}{\includegraphics[width = 6.6cm, height = 5.0cm, scale = 1]{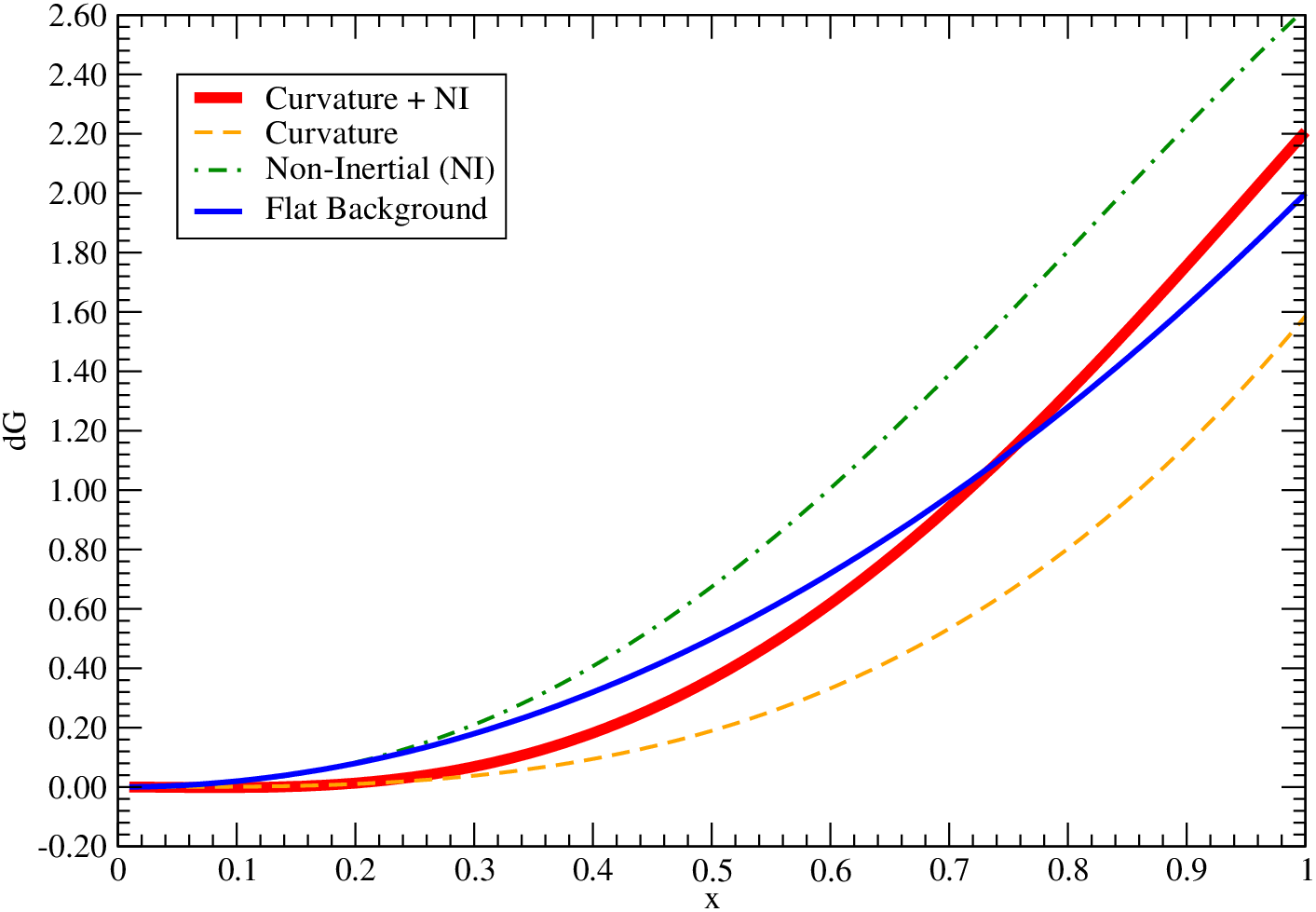}}}
\end{minipage}%
\hspace{2.0cm}
\begin{minipage}[t]{0.3 \textwidth}
\centering
\subfigure[\hspace{0.2cm} ($\Th = 180^\circ \, , \ \al = 0.99$)]{
\label{fig:dG-th=180-al=099}
\rotatebox{0}{\includegraphics[width = 6.6cm, height = 5.0cm, scale = 1]{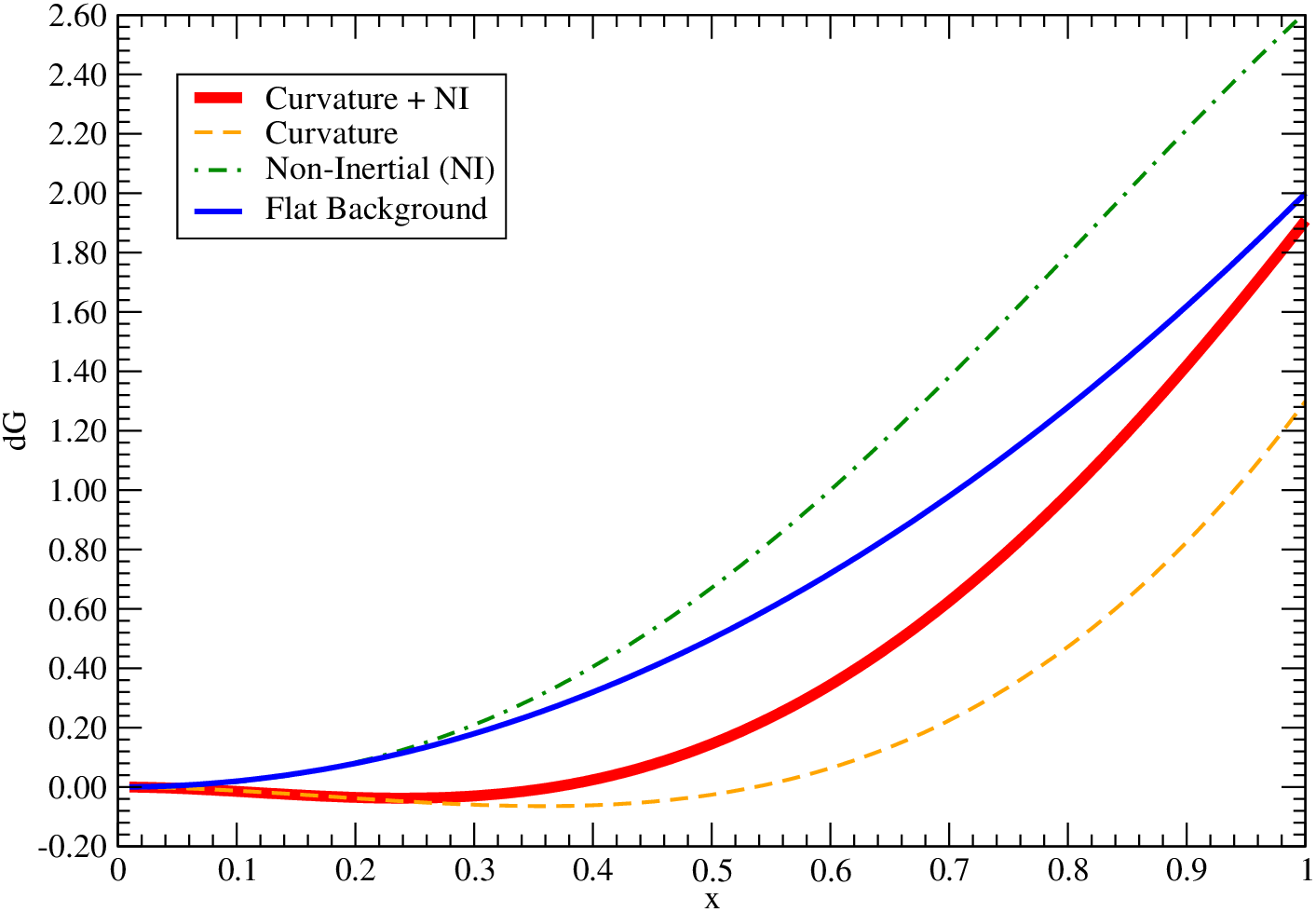}}}
\end{minipage}
\caption{\label{fig:dG-th-backward} Differential decay rate displaying contributions due to curvature and the
non-inertial dipole operator (backward emission).
}
\end{figure*}

\begin{figure*}
\psfrag{x}[tc][][1.8][0]{\Large $x$}
\psfrag{dG}[bc][][1.8][0]{\Large ${1 \over \Gm_0} \, \Dl \hat{\Gm}\lt(x,\cos \Th\rt)$}
\begin{minipage}[t]{0.3 \textwidth}
\centering
\subfigure[\hspace{0.2cm} ($\Th = 30^\circ \, , \ \al = 1$)]{
\label{fig:rel-dG-th=030-al=100}
\rotatebox{0}{\includegraphics[width = 6.6cm, height = 5.0cm, scale = 1]{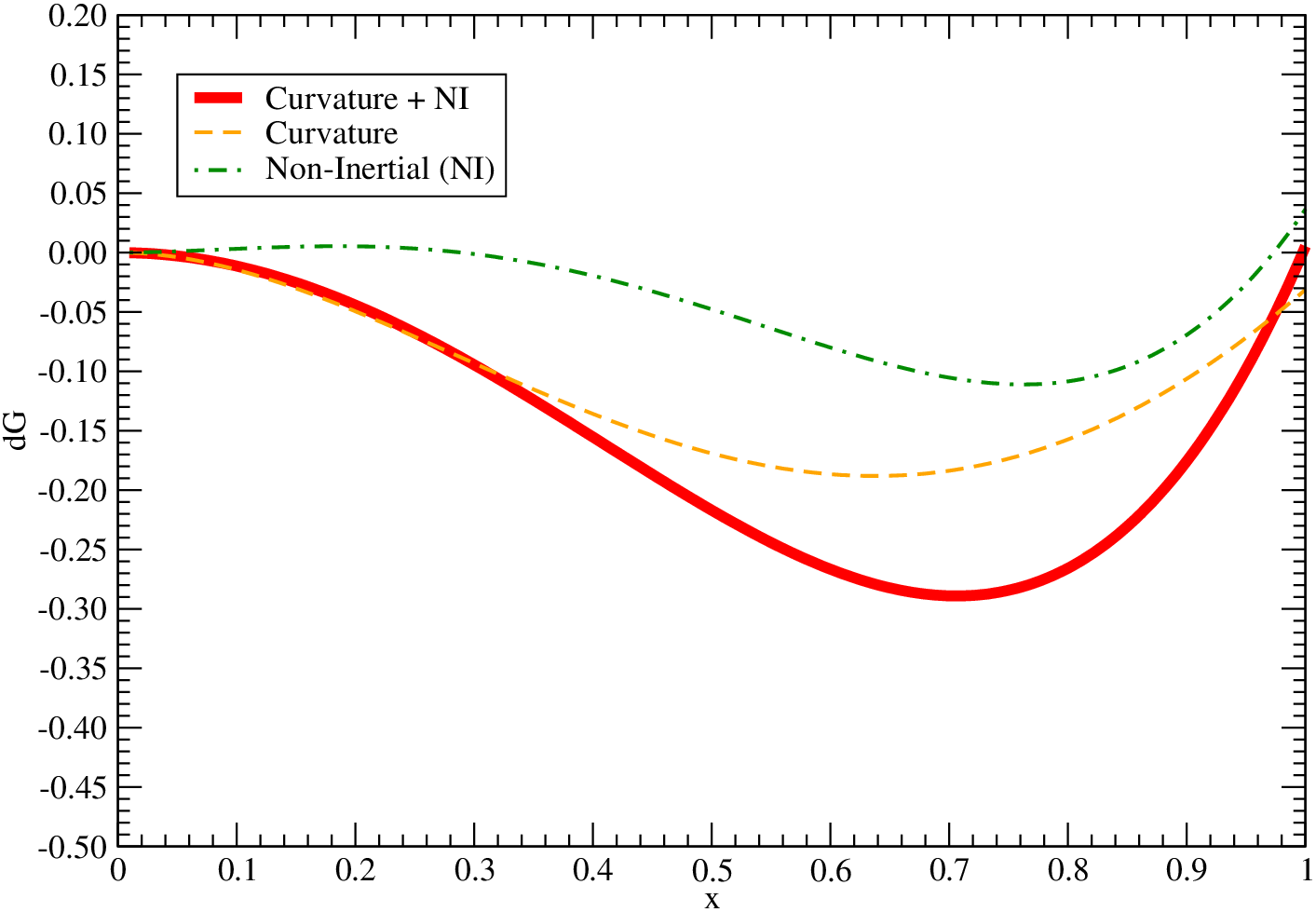}}}
\end{minipage}%
\hspace{2.0cm}
\begin{minipage}[t]{0.3 \textwidth}
\centering
\subfigure[\hspace{0.2cm} ($\Th = 30^\circ \, , \ \al = 0.99$)]{
\label{fig:rel-dG-th=030-al=099}
\rotatebox{0}{\includegraphics[width = 6.6cm, height = 5.0cm, scale = 1]{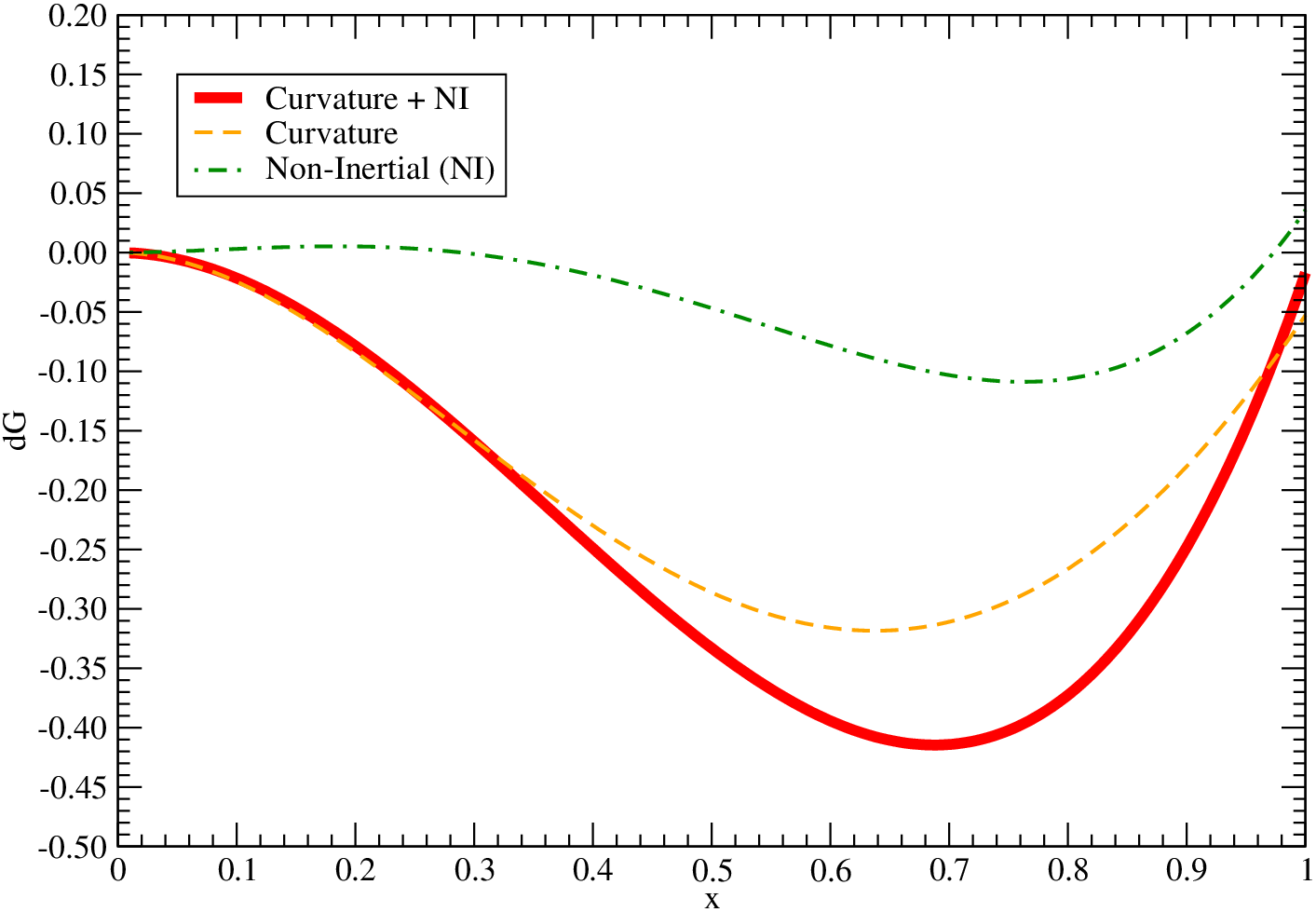}}}
\end{minipage} \\
\vspace{0.8cm}
\begin{minipage}[t]{0.3 \textwidth}
\centering
\subfigure[\hspace{0.2cm} ($\Th = 60^\circ \, , \ \al = 1$)]{
\label{fig:rel-dG-th=060-al=100}
\rotatebox{0}{\includegraphics[width = 6.6cm, height = 5.0cm, scale = 1]{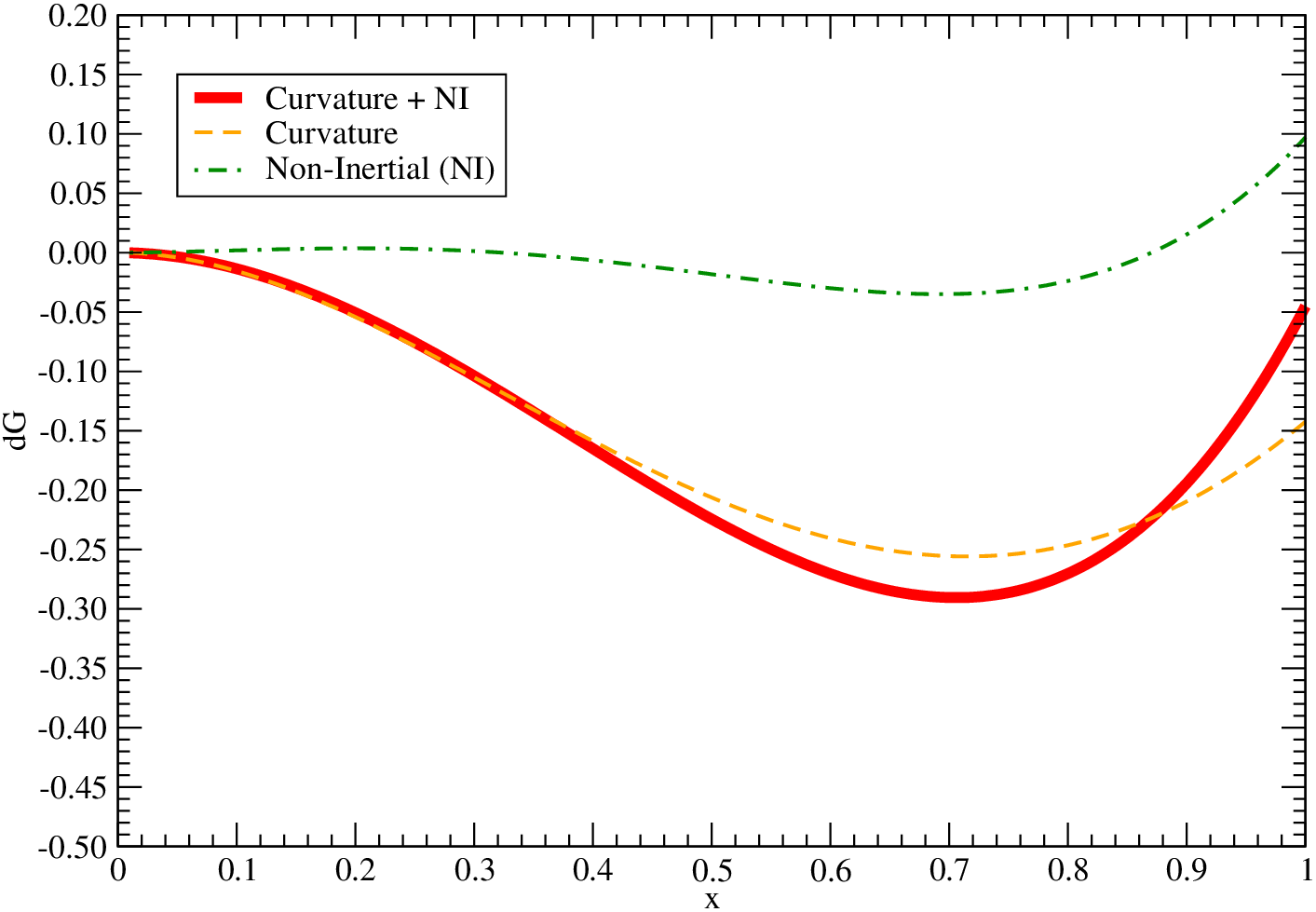}}}
\end{minipage}%
\hspace{2.0cm}
\begin{minipage}[t]{0.3 \textwidth}
\centering
\subfigure[\hspace{0.2cm} ($\Th = 60^\circ \, , \ \al = 0.99$)]{
\label{fig:rel-dG-th=060-al=099}
\rotatebox{0}{\includegraphics[width = 6.6cm, height = 5.0cm, scale = 1]{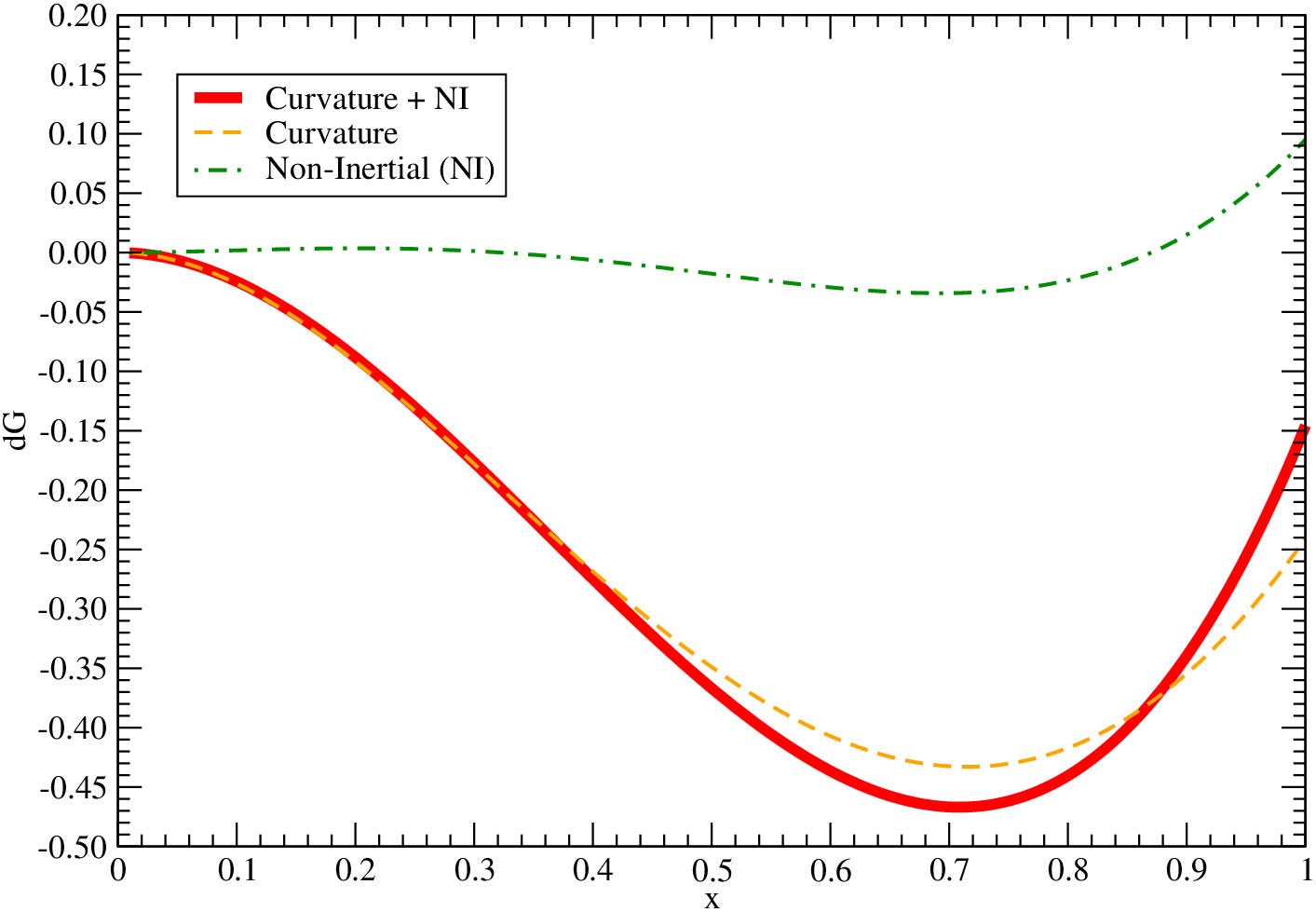}}}
\end{minipage} \\
\vspace{0.8cm}
\begin{minipage}[t]{0.3 \textwidth}
\centering
\subfigure[\hspace{0.2cm} ($\Th = 90^\circ \, , \ \al = 1$)]{
\label{fig:rel-dG-th=090-al=100}
\rotatebox{0}{\includegraphics[width = 6.6cm, height = 5.0cm, scale = 1]{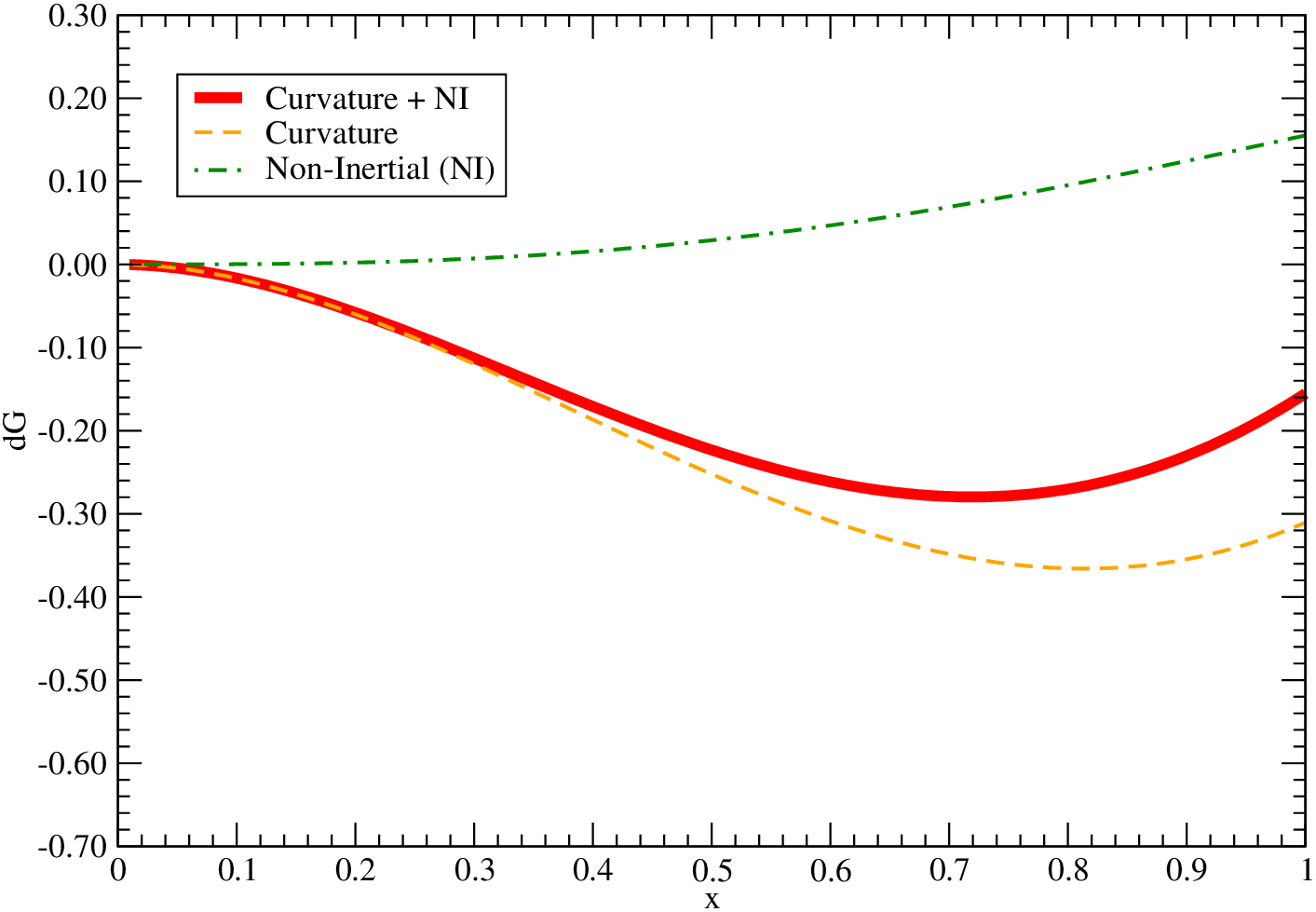}}}
\end{minipage}%
\hspace{2.0cm}
\begin{minipage}[t]{0.3 \textwidth}
\centering
\subfigure[\hspace{0.2cm} ($\Th = 90^\circ \, , \ \al = 0.99$)]{
\label{fig:rel-dG-th=090-al=099}
\rotatebox{0}{\includegraphics[width = 6.6cm, height = 5.0cm, scale = 1]{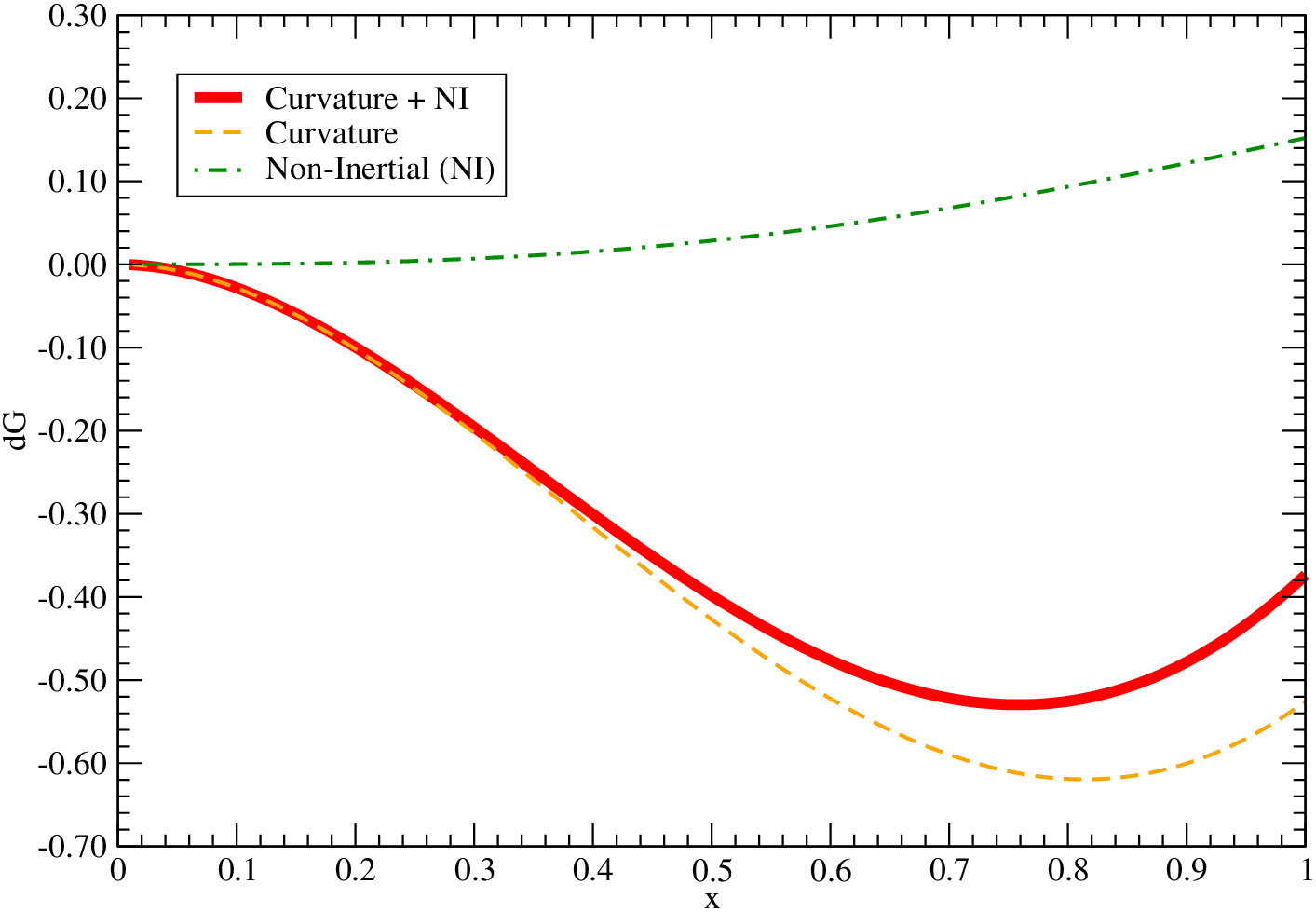}}}
\end{minipage}
\caption{\label{fig:rel-dG-th-forward} Net differential decay rate due to curvature and the non-inertial dipole operator
after subtracting the flat space-time background contribution (forward emission).
}
\end{figure*}

\begin{figure*}
\psfrag{x}[tc][][1.8][0]{\Large $x$}
\psfrag{dG}[bc][][1.8][0]{\Large ${1 \over \Gm_0} \, \Dl \hat{\Gm}\lt(x,\cos \Th\rt)$}
\begin{minipage}[t]{0.3 \textwidth}
\centering
\subfigure[\hspace{0.2cm} ($\Th = 120^\circ \, , \ \al = 1$)]{
\label{fig:rel-dG-th=120-al=100}
\rotatebox{0}{\includegraphics[width = 6.6cm, height = 5.0cm, scale = 1]{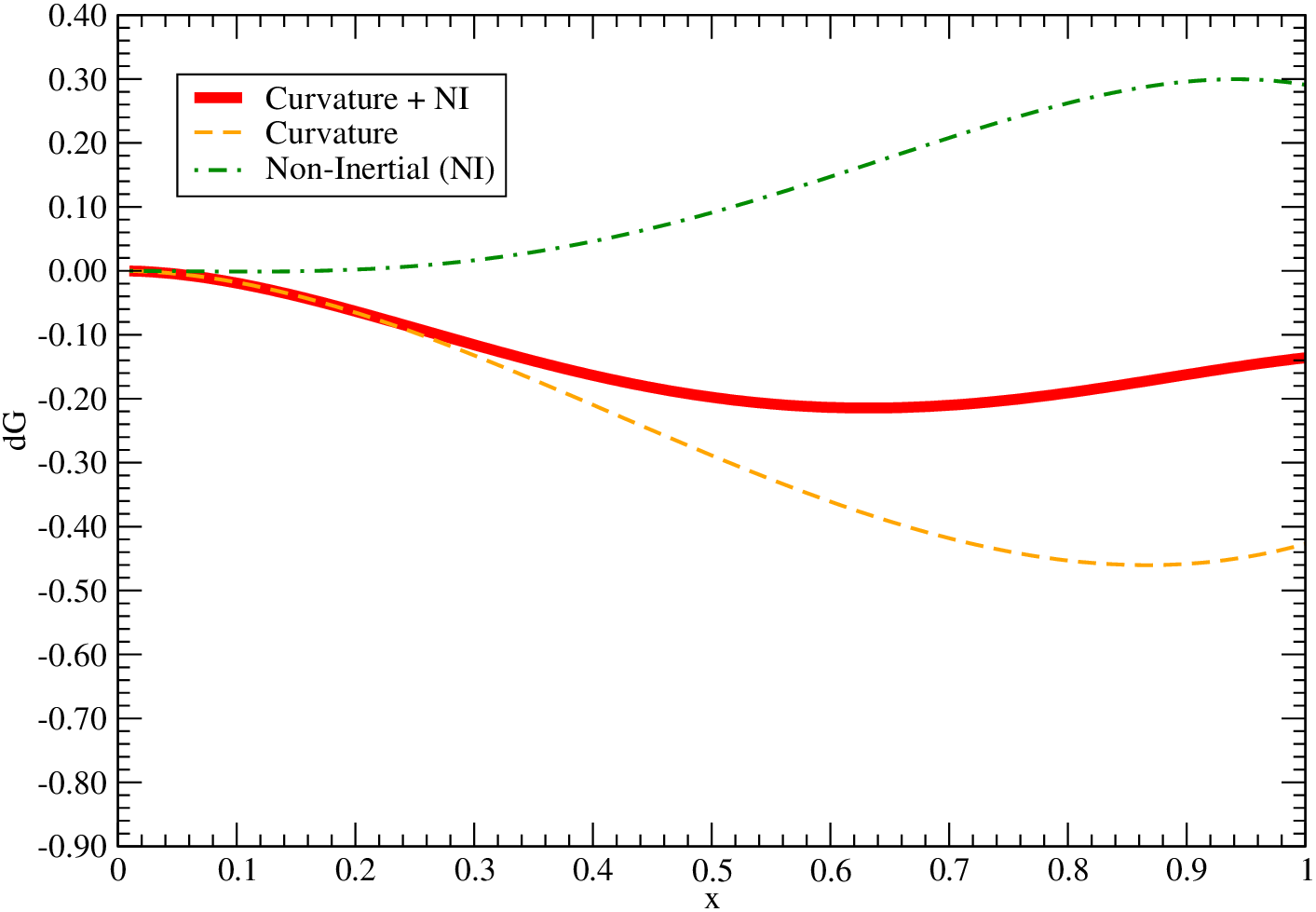}}}
\end{minipage}%
\hspace{2.0cm}
\begin{minipage}[t]{0.3 \textwidth}
\centering
\subfigure[\hspace{0.2cm} ($\Th = 120^\circ \, , \ \al = 0.99$)]{
\label{fig:rel-dG-th=120-al=099}
\rotatebox{0}{\includegraphics[width = 6.6cm, height = 5.0cm, scale = 1]{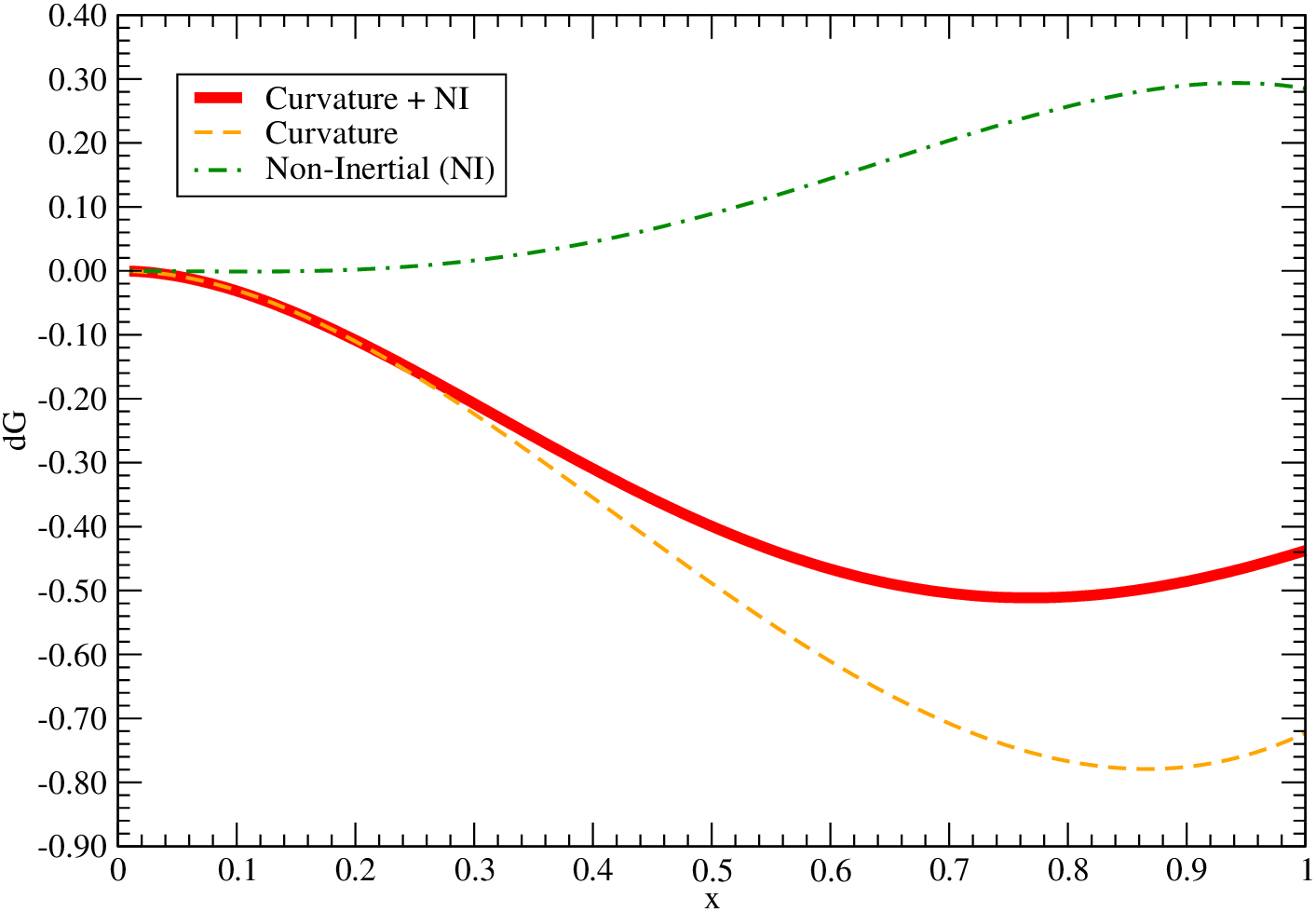}}}
\end{minipage} \\
\vspace{0.8cm}
\begin{minipage}[t]{0.3 \textwidth}
\centering
\subfigure[\hspace{0.2cm} ($\Th = 150^\circ \, , \ \al = 1$)]{
\label{fig:rel-dG-th=150-al=100}
\rotatebox{0}{\includegraphics[width = 6.6cm, height = 5.0cm, scale = 1]{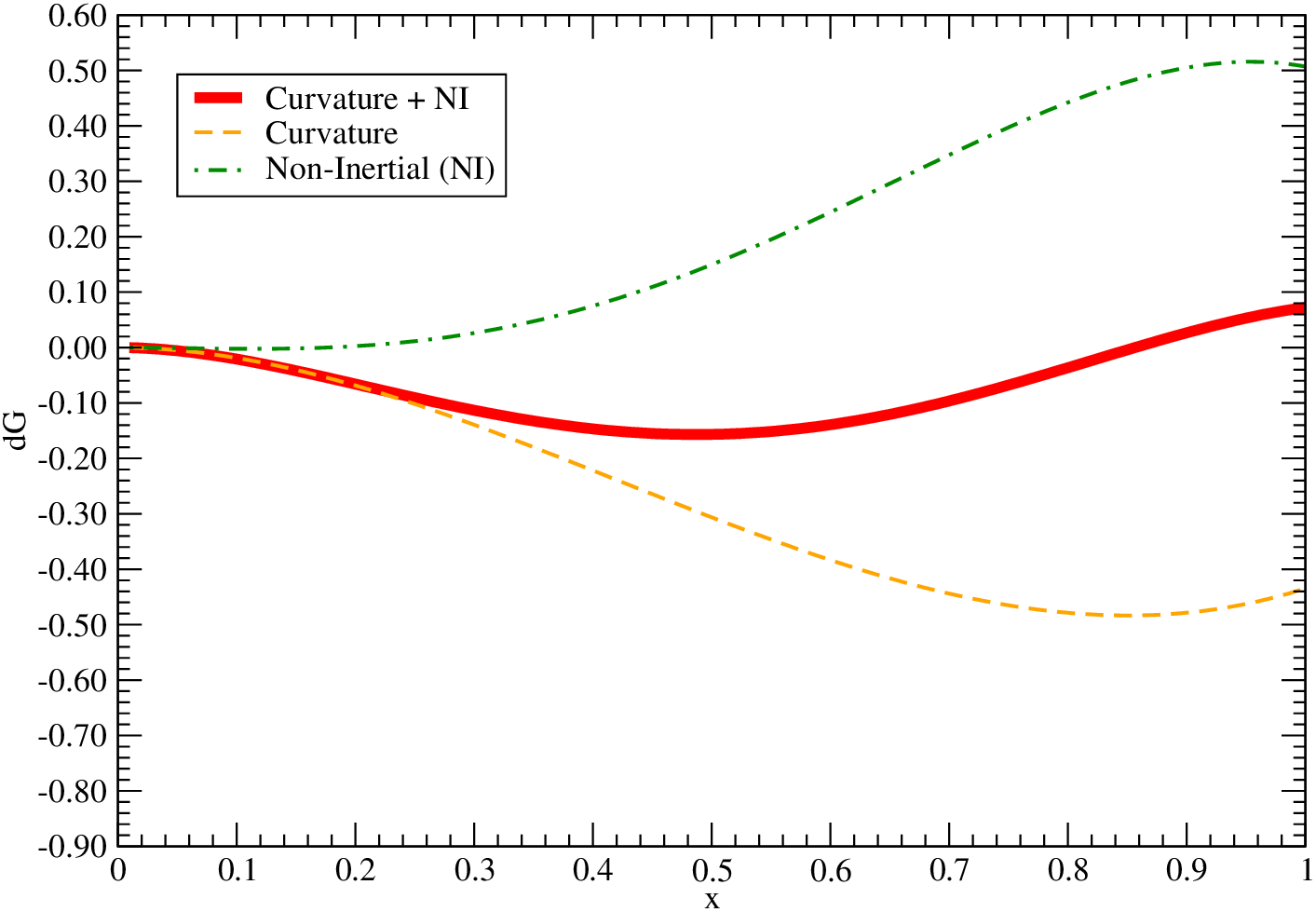}}}
\end{minipage}%
\hspace{2.0cm}
\begin{minipage}[t]{0.3 \textwidth}
\centering
\subfigure[\hspace{0.2cm} ($\Th = 150^\circ \, , \ \al = 0.99$)]{
\label{fig:rel-dG-th=150-al=099}
\rotatebox{0}{\includegraphics[width = 6.6cm, height = 5.0cm, scale = 1]{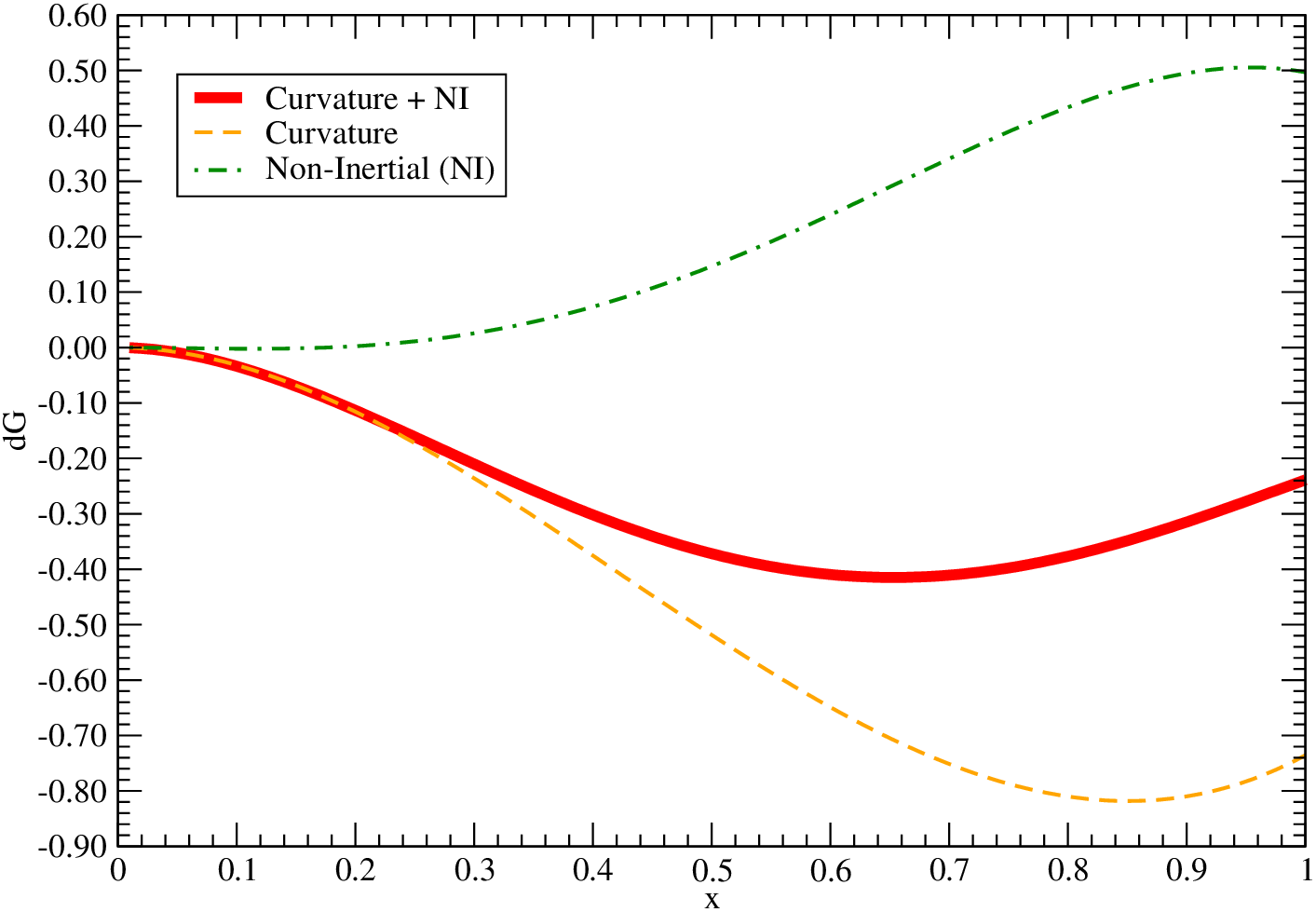}}}
\end{minipage} \\
\vspace{0.8cm}
\begin{minipage}[t]{0.3 \textwidth}
\centering
\subfigure[\hspace{0.2cm} ($\Th = 180^\circ \, , \ \al = 1$)]{
\label{fig:rel-dG-th=180-al=100}
\rotatebox{0}{\includegraphics[width = 6.6cm, height = 5.0cm, scale = 1]{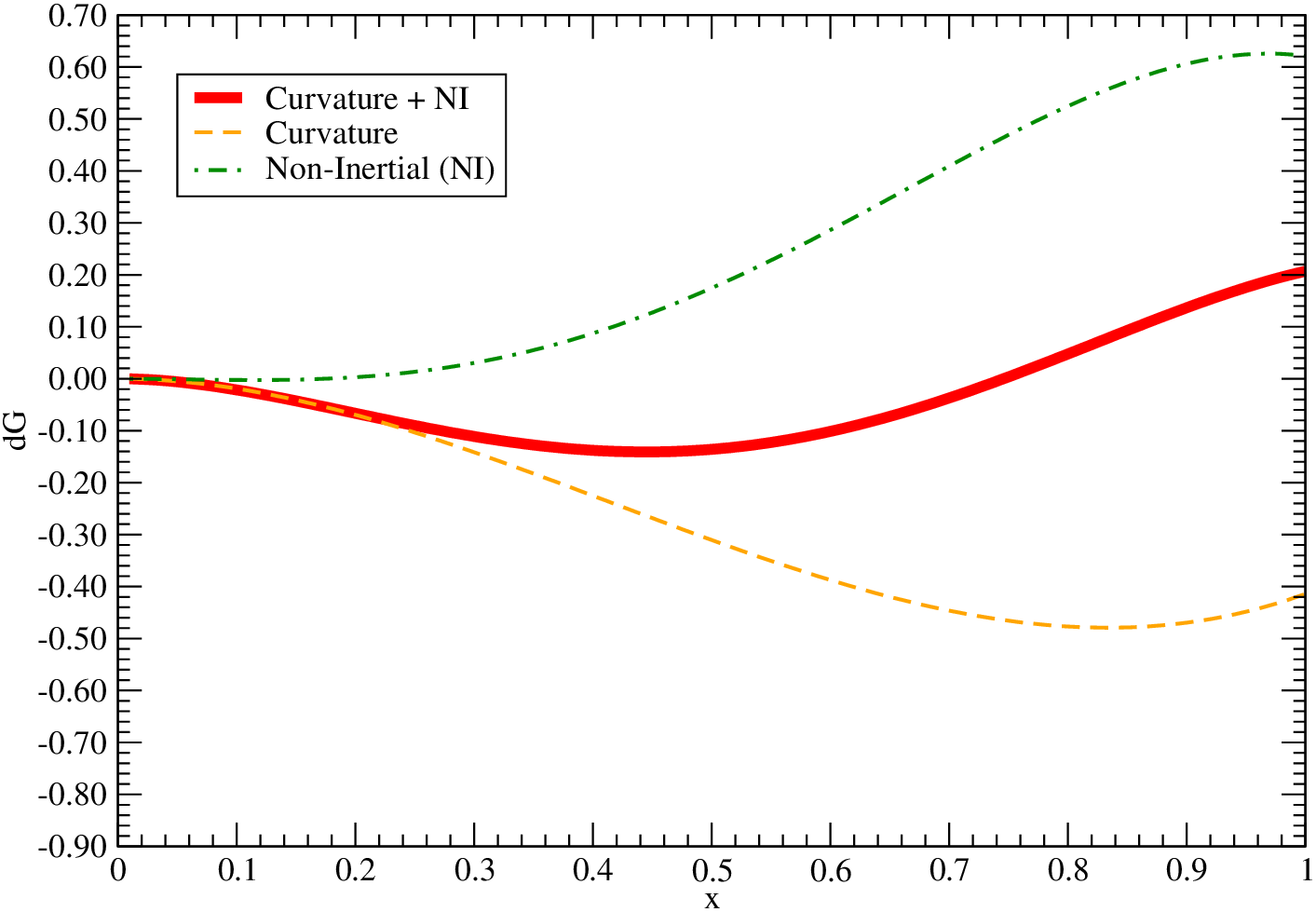}}}
\end{minipage}%
\hspace{2.0cm}
\begin{minipage}[t]{0.3 \textwidth}
\centering
\subfigure[\hspace{0.2cm} ($\Th = 180^\circ \, , \ \al = 0.99$)]{
\label{fig:rel-dG-th=180-al=099}
\rotatebox{0}{\includegraphics[width = 6.6cm, height = 5.0cm, scale = 1]{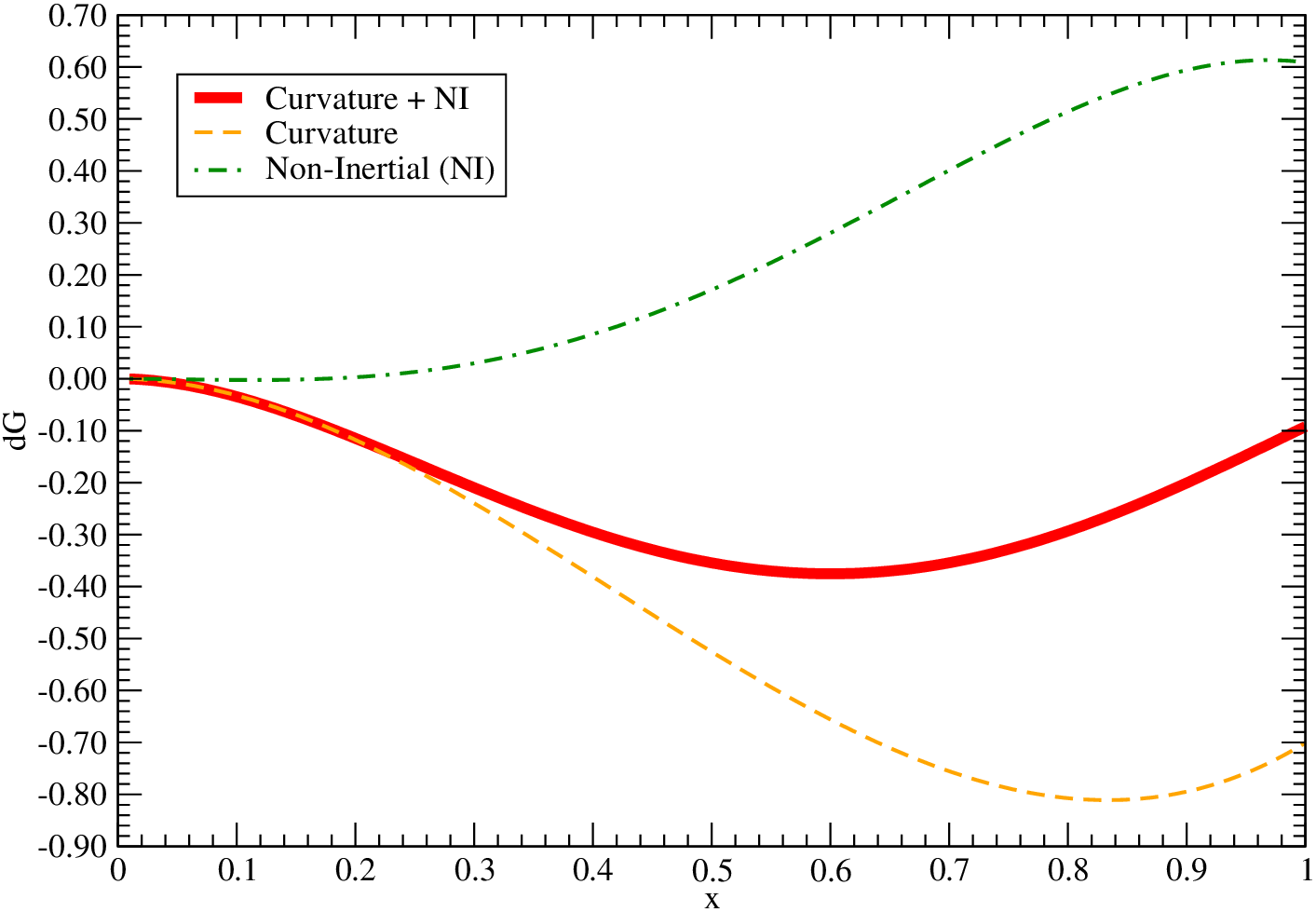}}}
\end{minipage}
\caption{\label{fig:rel-dG-th-backward} Net differential decay rate due to curvature and the non-inertial dipole operator
after subtracting the flat space-time background contribution (backward emission).
}
\end{figure*}

Figures~\ref{fig:dG-th-forward}--\ref{fig:rel-dG-th-backward} comprise
a list of differential cross section plots for the full range of scattering angles $\Th$ in $30^\circ$
intervals:

\begin{figure*}
\psfrag{x}[tc][][1.8][0]{\Large $x$}
\psfrag{dG}[bc][][1.8][0]{\Large ${1 \over \Gm_0} \, \hat{\Gm}\lt(x,\cos \Th\rt)$}
\begin{minipage}[t]{0.3 \textwidth}
\centering
\subfigure[\hspace{0.2cm} ($\Th = 30^\circ \, , \ \al = 1$)]{
\label{fig:dG-th=030-al=100}
\rotatebox{0}{\includegraphics[width = 6.6cm, height = 5.0cm, scale = 1]{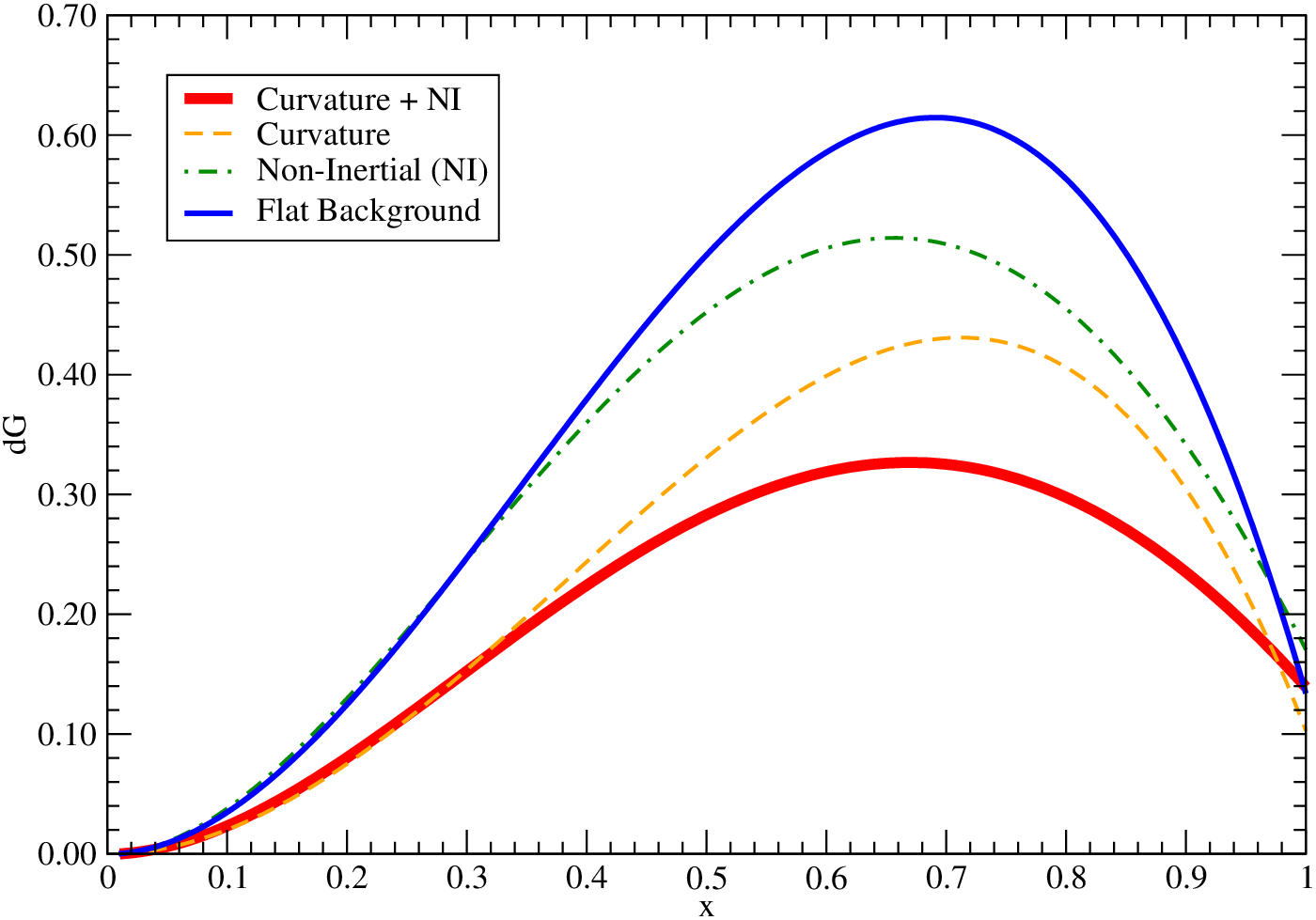}}}
\end{minipage}%
\hspace{2.0cm}
\begin{minipage}[t]{0.3 \textwidth}
\centering
\subfigure[\hspace{0.2cm} ($\Th = 30^\circ \, , \ \al = 0.99$)]{
\label{fig:dG-th=030-al=099}
\rotatebox{0}{\includegraphics[width = 6.6cm, height = 5.0cm, scale = 1]{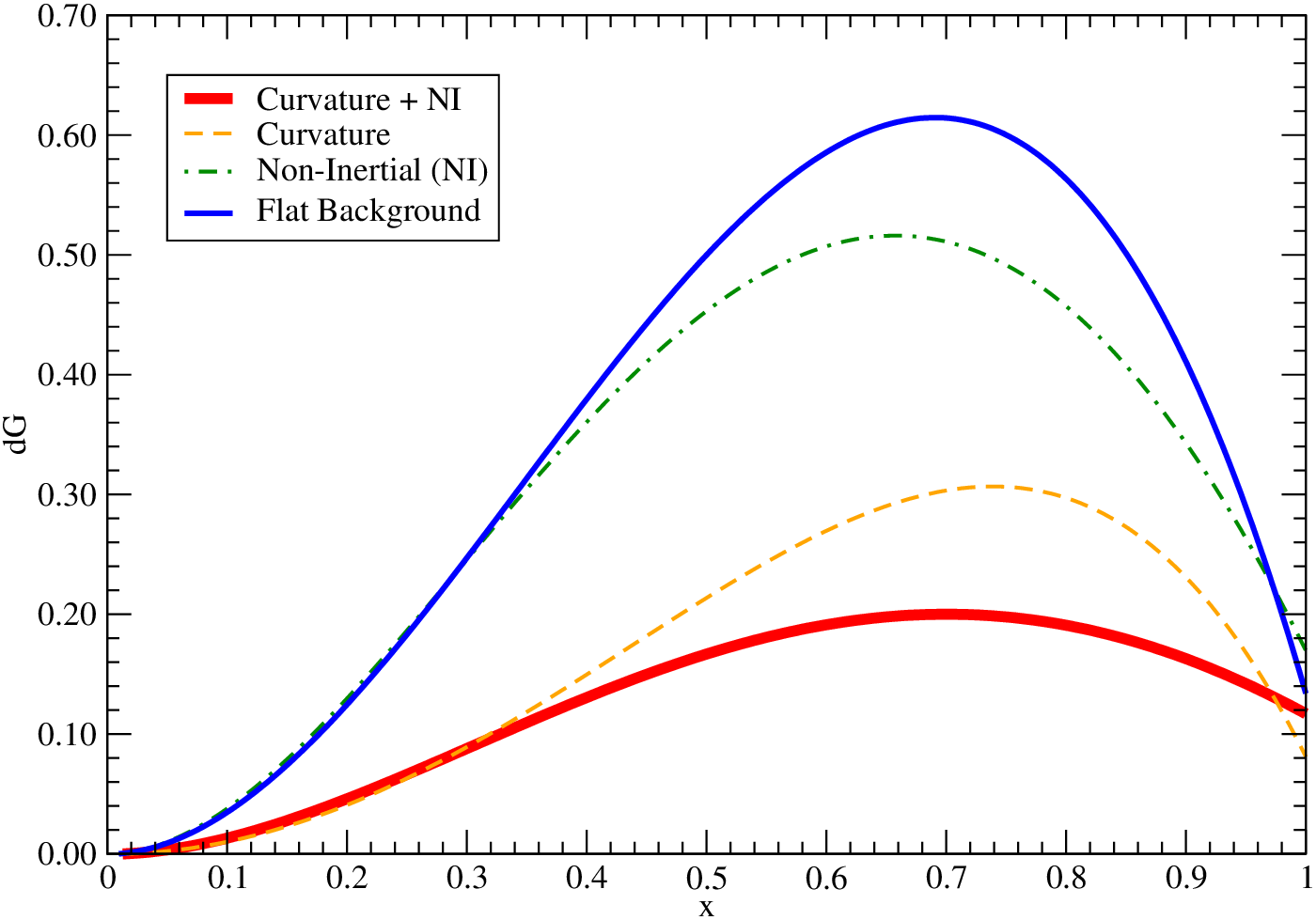}}}
\end{minipage} \\
\vspace{0.8cm}
\begin{minipage}[t]{0.3 \textwidth}
\centering
\subfigure[\hspace{0.2cm} ($\Th = 60^\circ \, , \ \al = 1$)]{
\label{fig:dG-th=060-al=100}
\rotatebox{0}{\includegraphics[width = 6.6cm, height = 5.0cm, scale = 1]{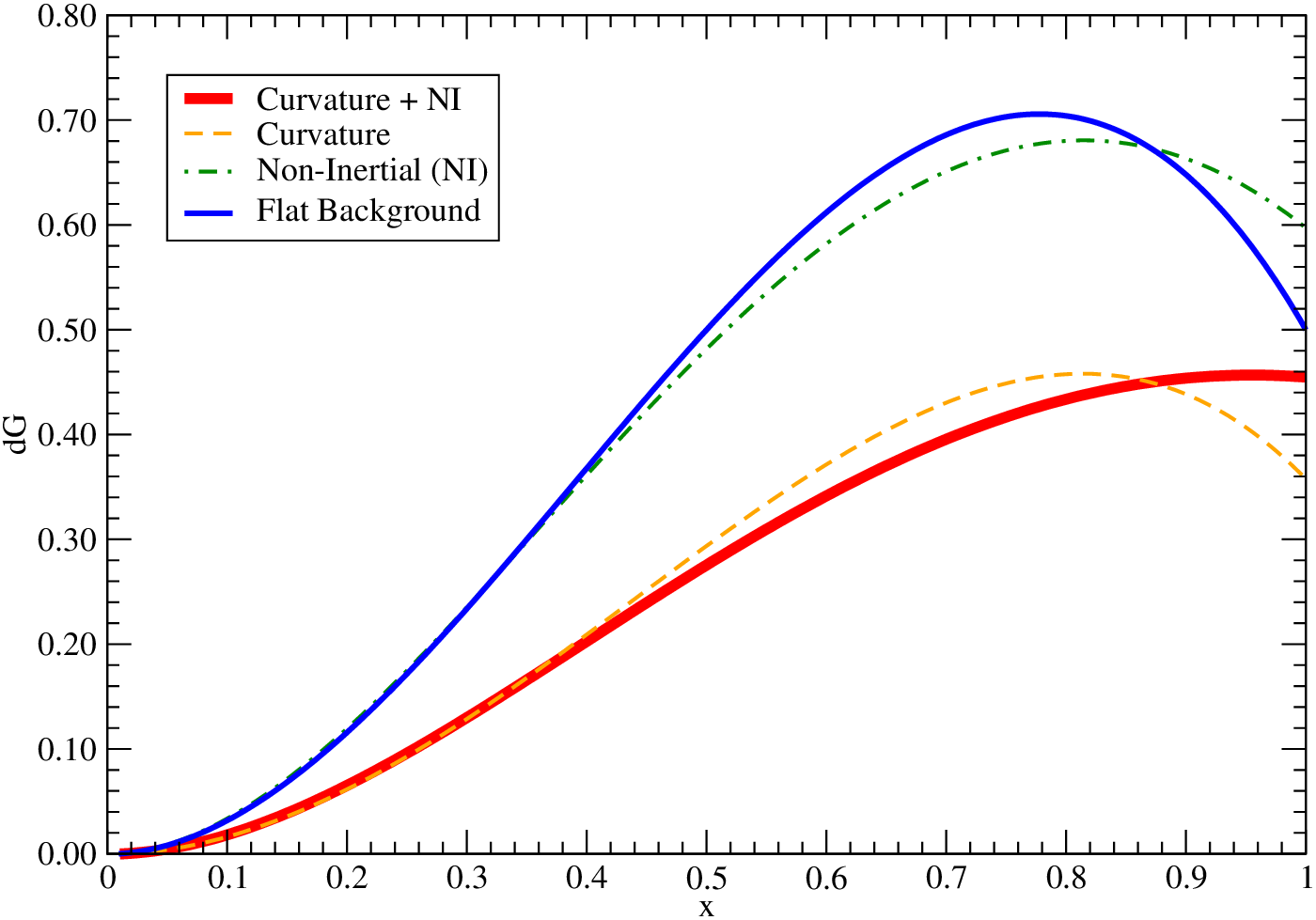}}}
\end{minipage}%
\hspace{2.0cm}
\begin{minipage}[t]{0.3 \textwidth}
\centering
\subfigure[\hspace{0.2cm} ($\Th = 60^\circ \, , \ \al = 0.99$)]{
\label{fig:dG-th=060-al=099}
\rotatebox{0}{\includegraphics[width = 6.6cm, height = 5.0cm, scale = 1]{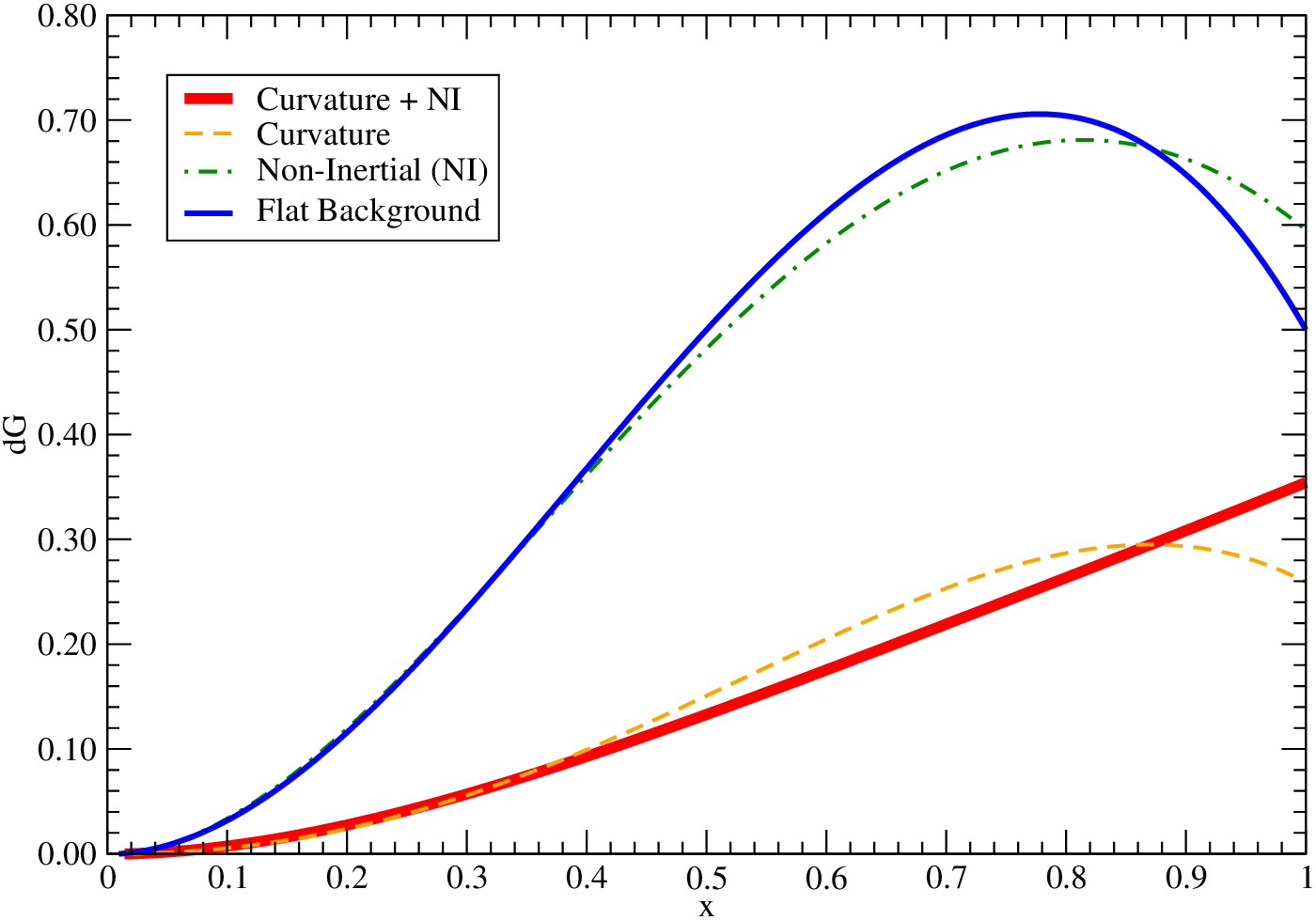}}}
\end{minipage} \\
\vspace{0.8cm}
\begin{minipage}[t]{0.3 \textwidth}
\centering
\subfigure[\hspace{0.2cm} ($\Th = 90^\circ \, , \ \al = 1$)]{
\label{fig:dG-th=090-al=100}
\rotatebox{0}{\includegraphics[width = 6.6cm, height = 5.0cm, scale = 1]{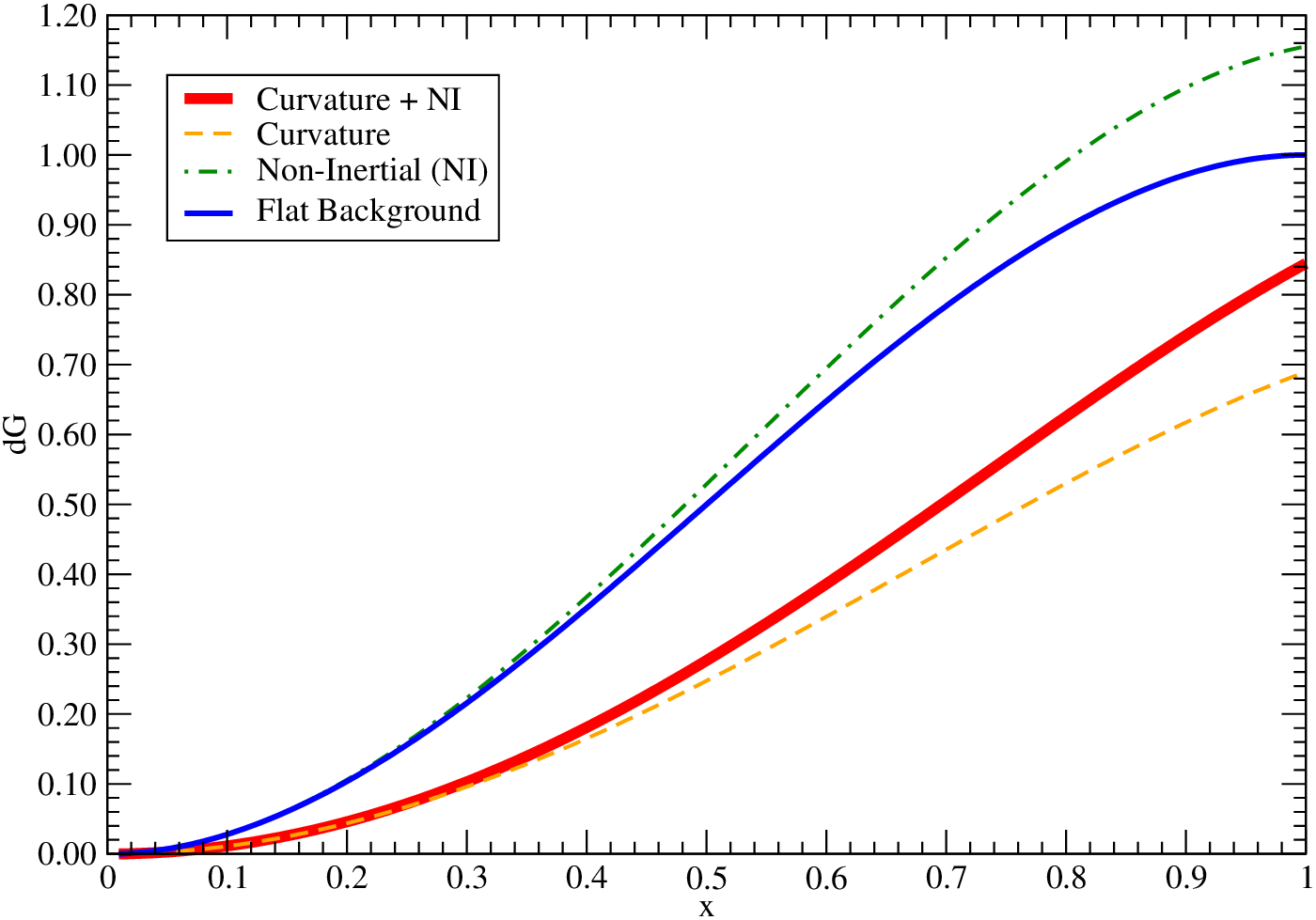}}}
\end{minipage}%
\hspace{2.0cm}
\begin{minipage}[t]{0.3 \textwidth}
\centering
\subfigure[\hspace{0.2cm} ($\Th = 90^\circ \, , \ \al = 0.99$)]{
\label{fig:dG-th=090-al=099}
\rotatebox{0}{\includegraphics[width = 6.6cm, height = 5.0cm, scale = 1]{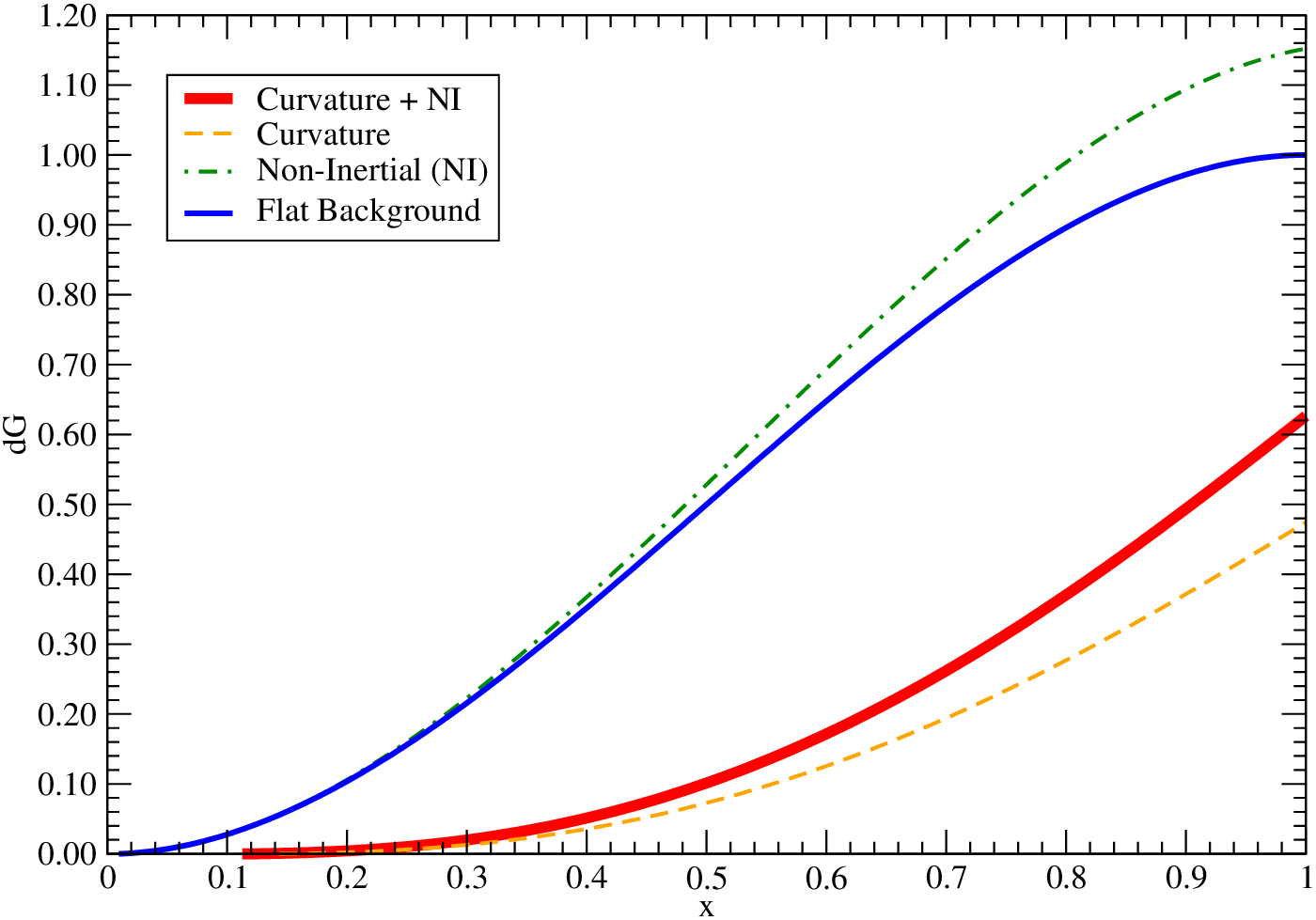}}}
\end{minipage}
\caption{\label{fig:dG-th-forward} Differential decay rate displaying contributions due to curvature and the
non-inertial dipole operator (forward emission).
}
\end{figure*}

\end{document}